\documentclass[11pt,letterpaper,fleqn]{article}

\usepackage{amsthm}
\usepackage{mathtools}
\usepackage{sidecap}
\usepackage{dsfont}
\usepackage{amssymb}
\usepackage{wasysym,graphicx}
\usepackage{yhmath}
\usepackage{fontawesome5}
\usepackage{enumitem}
\setitemize{noitemsep,topsep=0pt,parsep=0pt,partopsep=0pt,leftmargin=*}
\setenumerate{noitemsep,topsep=0pt,parsep=0pt,partopsep=0pt,leftmargin=*}
\setdescription{noitemsep,topsep=0pt,parsep=0pt,partopsep=0pt,leftmargin=*}
\usepackage[flushleft]{threeparttable} 

\usepackage{amsmath}
\usepackage{tikz}
\usepackage{xspace}
\usepackage{tcolorbox}
\tcbuselibrary{skins} 
\usepackage[margin=1in]{geometry}
\usepackage{hyperref}
\usepackage[font=footnotesize]{caption}
\usepackage[nosort,capitalise]{cleveref}
\usepackage{thm-restate}
\usepackage{thmtools}

\newcounter{strategycount}
\newcounter{algocount}

\crefname{strategycount}{Strategy}{Strategies}
\Crefname{strategycount}{Strategy}{Strategies}
\crefname{equation}{Eq.}{Eqs.}
\crefname{figure}{Figure}{Figures}

\newtcolorbox{strategyboxlayout}[1]{
	colback=blue!5!white,
	colframe=blue!65!black,
	fonttitle=\bfseries,
	title={Strategy \thestrategycount: #1},
	arc=4pt, 
	boxrule=1pt,
	enhanced
}

\newtcolorbox{algoboxlayout}[1]{
	colback=blue!5!white,
	colframe=blue!65!black,
	fonttitle=\bfseries,
	title={#1},
	arc=4pt, 
	boxrule=1pt,
	enhanced
}

\newenvironment{strategybox}[2]{%
	\refstepcounter{strategycount}
	\label{#2}
	\begin{strategyboxlayout}{#1}
	}{%
	\end{strategyboxlayout}
}

\newtcolorbox{assumptionboxlayout}[1]{
	colback=blue!5!white,
	colframe=blue!65!black,
	fonttitle=\bfseries,
	title={#1},
	arc=4pt, 
	boxrule=1pt,
	enhanced
}

\DeclareRobustCommand{\pizza}{%
	\tikz[baseline=-0.75ex]{
		\fill (0,0) circle (5.2pt);
		\foreach \angle in {45, 135, 225, 315} {
			\fill[white] (\angle:2.5pt) circle (0.75pt);
		}
	}\xspace%
}
\DeclareRobustCommand{\emptypizza}{%
	\tikz[baseline=-0.75ex]{
		\draw[line width=0.8pt] (0,0) circle (5.2pt); 
	}\xspace%
}
\DeclareRobustCommand{\slice}{%
	\tikz[baseline=-0.2ex, scale=0.15, rotate=40]{
		\fill (0,0) -- (2,0.8) arc (21.8:-24:2.15) -- cycle;
		\fill[white] (1.4, 0) circle (5pt); 
	}\xspace%
}
\DeclareRobustCommand{\emptyslice}{%
	\tikz[baseline=-0.2ex, scale=0.15, rotate=40]{
		\draw[line width=0.8pt] (0,0) -- (2,0.8) arc (21.8:-24:2.15) -- cycle;
	}\xspace%
}
\DeclareRobustCommand{\templateslice}{%
	\tikz[baseline=-0.2ex, scale=0.15, rotate=40]{
		\draw[line width=0.8pt] (0,0) -- (2,0.8) arc (21.8:-24:2.15) -- cycle;
		\node[scale=0.55, font=\bfseries\itshape] at (1.4, 0.1) {t};
	}\xspace%
}

\newcommand{\NNDG}{\ensuremath{\Delta G(S) = \sum_{l\in S} \Delta G(l) + (c-1)\Delta G^{\textrm{assoc}}  + k_\mathrm{B} T \log R_{\text{max}}(S)}}
\newcommand{\PF}{partition function\xspace} \newcommand{\pf}{\PF}

\usepackage{booktabs}
\usepackage{makecell}
\usepackage{multirow}
\usepackage{array}
\usepackage{arydshln}
\usepackage{stmaryrd} 
\usepackage{makecell} 
\usepackage{multirow}
\usepackage{arydshln} 

\theoremstyle{definition}
\newtheorem{theorem}{Theorem}
\newtheorem{lemma}[theorem]{Lemma}
\newtheorem{definition}[theorem]{Definition}
\newtheorem{note}[theorem]{Note} 
\newtheorem{remark}[theorem]{Remark}
\newtheorem{example}[theorem]{Example}
\newtheorem{definition2}[theorem]{Definition} 
\crefname{definition2}{Definition}{Definitions} 

\newcommand{\PolySpi}{\ensuremath{\mathrm{Poly}(S,\pi)}\xspace}
\newcommand{\PolySpiPrime}{\ensuremath{\mathrm{Poly}(S,\pi')}\xspace}

\newcommand{\Polypi}{\ensuremath{\mathrm{Poly}(\emptypizza,\pi)}\xspace}

\newcommand{\distinguishable}[1]{\ensuremath{\overline{#1}}\xspace}
\newcommand{\indistinguishable}[1]{\ensuremath{#1}\xspace}

\newcommand{\DeltaGD}{\distinguishable{\Delta G}}

\newcommand{\DeltaGI}{\indistinguishable{\Delta G}\xspace}

\newcommand{\OmegaD}{\distinguishable{\Omega}}
\newcommand{\OmegaDS}{\OmegaD}
\newcommand{\OmegaI}{\indistinguishable{\Omega}}
\newcommand{\OmegaIS}{\OmegaI}

\newcommand{\PiD}{\distinguishable{\Pi}}
\newcommand{\PiI}{\indistinguishable{\Pi}}

\newcommand{\QD}{\distinguishable{Q}}
\newcommand{\QI}{\indistinguishable{Q}}
\newcommand{\QalphaD}{\distinguishable{Q}_{\alpha}}
\newcommand{\QalphaI}{\indistinguishable{Q_{\alpha}}}
\newcommand{\OmegaDpi}{\distinguishable{\Omega}(\pi)\xspace}
\newcommand{\OmegaDpiS}{\OmegaDpi}
\newcommand{\OmegaIpiS}{\indistinguishable{\Omega}(\pi)\xspace}
\newcommand{\OmegaIpi}{\OmegaIpiS}

\newcommand{\OmegaDRpi}{\OmegaD(\pi, R)\xspace}
\newcommand{\OmegaDtwopi}{\OmegaD(\pi, 2)\xspace}

\title{One partition function call to rule them all: 
Efficiently computing the equilibrium energy landscape of multiple nucleic acid strands}

\author{} 
\date{}

\usepackage{algorithm,algpseudocode,xcolor}
\usepackage{arydshln}

\usepackage{tikz}
\usetikzlibrary{arrows.meta, bending, positioning, calc}
\definecolor{darkergreen}{rgb}{0.0, 0.65, 0.0}

\definecolor{vlightgray}{gray}{0.98}
\usepackage[textsize=tiny, color=lightgray 
]{todonotes}

\newif\ifdwsfinal

\usepackage{xspace}

\newcommand{\ssELbase}{SSEL}
\newcommand{\ssEL}{\ensuremath{\#\mathrm{\ssELbase}}\xspace}
\newcommand{\ssELtext}{$\#$\ssELbase\xspace}

\newcommand{\base}[1]{\ensuremath{\mathrm{#1}}\xspace}
\newcommand{\baseA}{\base{A}}
\newcommand{\baseT}{\base{T}}
\newcommand{\baseG}{\base{G}}
\newcommand{\baseC}{\base{C}}
\newcommand{\baseU}{\base{U}}

\author{Ahmed Shalaby\thanks{Hamilton Institute and Department of Computer Science, Maynooth University, Ireland. Research supported by Science Foundation Ireland (SFI) under grant numbers 20/FFP-P/8843 and 18/ERCS/5746 and the European Union's European Research Council (ERC, Active-DNA, no 772766); European Innovation Council  (EIC, DISCO, No 101115422). Views and opinions expressed are however those of the author(s) only and do not necessarily reflect those of the European Union, European Research Council,  European Innovation Council or Science Foundation Ireland. Neither the European Union nor the granting authority can be held responsible for them.} 
	\and Damien Woods\footnotemark[1]}

\date{}

\begin{document}

\maketitle


\begin{abstract}
Throughout physics, chemistry, machine learning and discrete mathematics, the partition function is a normalisation factor for probability assignment.  As a sum over all system states, it is often, but not always, intractable depending on the context.  In the nearest neighbour  model of DNA/RNA strand binding, the partition function is a sum of Boltzmann-weighted free energies of the exponentially large set of secondary structures---discrete base-pairing states.  Dirks, Bois, Schaeffer, Winfree and Pierce [SIAM review, 2007] gave a polynomial time partition function algorithm for multiple ($\mathcal{O}(1)$) strands, generalising  McCaskill's single-stranded algorithm [Biopolymers, 1990].  Despite the intuition that partition function may somehow encode and unify properties over the entire state space, thus far, different approaches were required to solve various different energy-landscape prediction problems.  For instance: finding the most favoured, or minimum free energy (MFE), structure(s) for multiple strands was recently solved [ICALP, 2025], but required new techniques beyond those used for partition function.

We asked whether there is a generalised {\em magnifiable} partition function algorithm [DNA31, 2025], which we answer here in the affirmative.  
We give an $\mathcal{O}(N^3(c-1)!)$ time and $\mathcal{O}(N^2)$ space for magnifiable partition function, for $c=\mathcal{O}(1)$ strands and $N$ bases, the same overhead as Dirks et al's (non-magnifiable) partition function algorithm.  However, we go on to show that magnification allows for efficient decoding of the entire energy landscape, concretely:  
a single function call to the {\em magnifiable} partition function efficiently solves and improves upon a large number of problems, giving:
\begin{itemize}
\item Partition function in $\mathcal{O}(N^3(c-1)!)$ time and $\mathcal{O}(N^2)$ space.
\item MFE in $\mathcal{O}(N^3(c-1)!)$ time, $\mathcal{O}(N^2)$ space; a considerable improvement over our $\mathcal{O}(N^4(c-1)!)$ time, $\mathcal{O}(N^4)$ space result  [ICALP 2025]. 
\item  The first polynomial time algorithm for the multi-stranded {\em density of states} problem, 
also in  time, space $\mathcal{O}(N^3(c-1)!), \mathcal{O}(N^2)$. 
\item  The first polynomial time algorithm outputting the entire list of energy levels occupied by at least one structure (in time, space $\mathcal{O}(N^3(c-1)!), \mathcal{O}(N^2)$), and the first algorithms for various other energy landscape and counting problems.
\end{itemize}
Our approach unifies the field of energy landscape prediction showing that a single,  general, algorithmic framework neatly unifies a large collection of problems that practitioners care about, and that hitherto required distinct techniques with worse time bounds.  
\end{abstract}

\section{Introduction}\label{sec:typesetting-summary}

The almost-optimal information encoding capabilities of DNA and RNA are harnessed by biology to encode myriad structures and complex behaviours, and  by scientists and engineers to create bespoke nanostructures~\cite{rothemund2006folding} and molecular computers~\cite{algoSST}. 
Consequently, predicting the structures formed by DNA and RNA strands is of rather wide interest. 
Mathematical models are deep enough that theorists have found interesting mathematical structure, and in some cases have designed useful prediction algorithms for practitioners~\cite{santa,NUPACK,viennaRNA}, 
while needing to know very little about biology/chemistry.    
In the past few decades, DNA's proven computational abilities provide further motivation:   
molecular programmers have designed collections of DNA strands that form nanostructures~\cite{rothemund2006folding} and 
embed computation (Boolean circuits)~\cite{rothemund2000program,SoloveichikWinfree,meunier2014intrinsic,pumpability,IUSA,wang2018effective,nikitin2023non} 
leading to P-hard or NP-hard prediction problems~\cite{algoSST}, even at equilibrium~\cite{TBNs,TBNexp,SDCexp}. 
 
The most used and theoretically interesting model of DNA/RNA secondary structure is the nearest neighbour model.  
Its history begins in statistical mechanics~\cite{zimm1959theory,crothers1964theory,crothers1965melting,tinoco},
and predicting structures in that model
fits with the  long tradition~\cite{MooreMertens} of algorithmic prediction for statistical mechanics models with state spaces of exponential size in system description~\cite{JerrumSinclair}.  
When algorithmically predicting statistical mechanical systems important considerations are: 
What kinds of combinatorial  structures populate the state space? 
And how are probabilities assigned to those structures? 
What is the energy landscape, or probability distribution?

\subsection{Background: the nearest neighbour model of DNA binding}\label{sec: intro NN and rot sym}
\begin{figure}[t]
\centering\includegraphics[width=\textwidth]{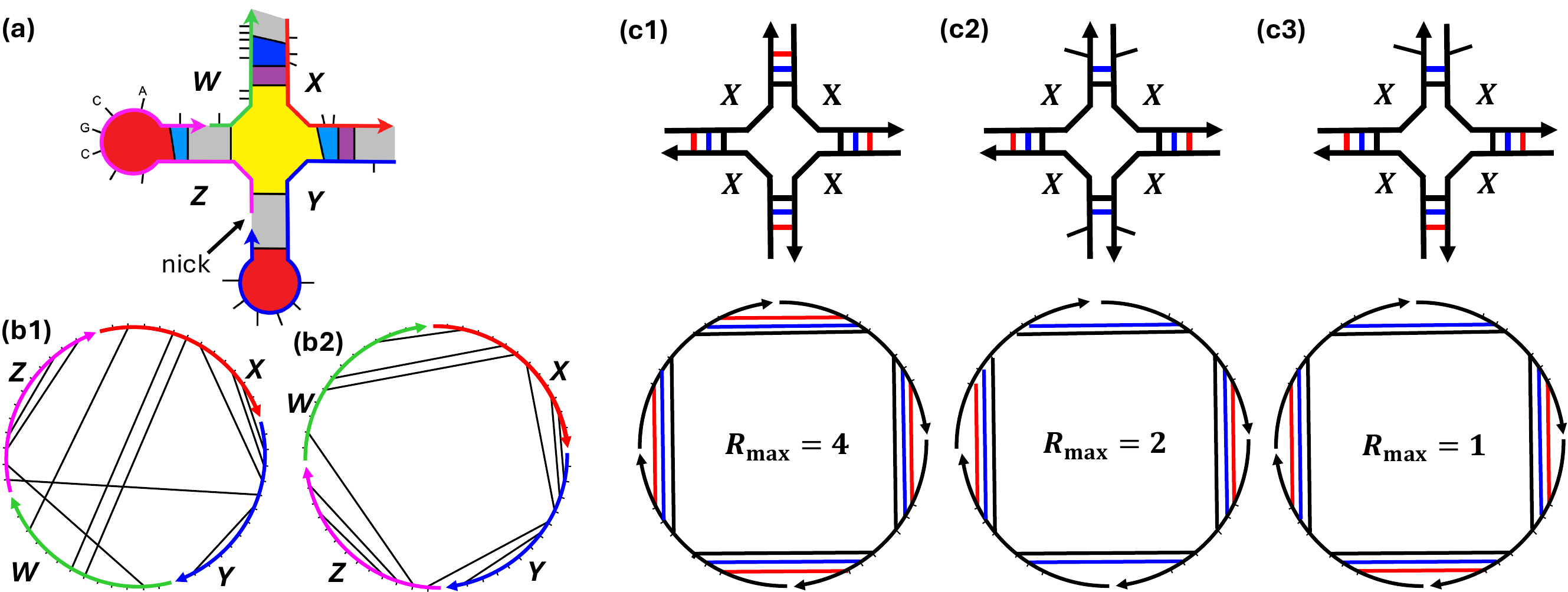}
\vspace{-1.2\baselineskip}
\caption{%
	(a)~Illustration of one of the many possible secondary structures for $c=4$ DNA strands $W,X,Y,Z$. 
	Short black ticks:  
	DNA bases (a few are shown $\ldots \baseC, \baseG, \baseC, \baseA \ldots$).   Black lines: base pairs (drawing not to scale). 
	Loops are colour-coded:  stack (purple), multiloop (yellow), hairpin (red), bulge (light blue), internal (dark blue), external (grey).  
	The nick separates two strands. 
	Formally, the secondary structure is a set of base pairs: $S=\{ (1_Z, 5_Y), (2_Z,2_W) \ldots, (6_Y,12_Y) \}$. 
	(b1)~Polymer graph for strand ordering $\pi' = WZXY$, denoted \PolySpiPrime, with base-pair crossings. 
	(b2)~By  reordering to $\pi = ZWXY$ we get another polymer graph \PolySpi for $S$, without crossings, hence  $S$ is unpseudoknotted. 
	{(c1--3)}: Three secondary structures with their associated polymer graphs,
	each with different maximum symmetry degree $R_\text{max}$, and each composed of the same four indistinguishable (identical; same DNA sequence) strands of type $X$.  
	(c1) $R_\text{max}$ = 4 (rotation by $90^{\circ}$ or $180^{\circ}$  gives the same secondary structure). 
	(c2) $R_\text{max}$ = 2 (rotation by $180^{\circ}$  gives the same secondary structure). 
	(c3) $R_\text{max}$ = 1 (asymmetric secondary structure). Each   structure has $4$ nests on the top, right, bottom, and left of their polymer graphs.}
\label{fig:1}
\end{figure}

Here, we study the {\em nearest neighbour model} of DNA/RNA binding~\cite{santa,NUPACK,viennaRNA}. 
The {\em primary structure} of a DNA strand is a word over the alphabet of {\em bases} $\{\baseA, \baseC, \baseG, \baseT\}$, with $\baseU$ instead of $\baseT$ for RNA.  
Bases bond in pairs, \baseA-\baseT and \baseC-\baseG. 
A {\em secondary structure} (\cref{fig:1}) of one or more strands is a set of base pairs.\footnote{In this paper, we are concerned  only about DNA/RNA secondary structures, which are lists of base pairs with associated free energies/probabilities.  We are not concerned with 3D molecular structure, nor any other chemical nor biological details.} 
The set of all secondary structures of a multieset of strands is called the ensemble, $\Omega$,    
and has size exponential in the total number of bases---since each base can be bound or not (assuming a non-trivial system). 
$\Omega$ is a combinatorial state space, but what we care about are which states are more or less likely, or how the entire probability distribution of states looks---i.e.~we assign probabilities to states via free-energy and the Boltzmann distribution. 

A secondary structure $S$ has a free energy $\Delta G(S) \in \mathbb{R}$, where more negative means more favourable~\cite{tinoco}. 
$\Delta G(S)$ is linear-time computable in the number of bases in $S$:  
simply sum over the {\em loops} (features) defined by base pairing (\cref{fig:1,eq:NNDG}): 
adjacent base pairs or stacks are favourable (more negative), 
but single-stranded regions confer an unfavourable (more positive)  polymer-chain entropy of logarithmic cost in loop length. 
A key, non-local, contribution to $\Delta G(S)$ is called {\em rotational symmetry}, which arises when multiple indistinguishable strands (they have the same sequence) are in the same structure.
If the structure has certain symmetries, it means that multiple configurations (arrangement of strands) are actually the same structure 
hence these ``extra'' symmetries need to be factored out to avoid erroneous overcounting of structures when computing probabilities. 
   The higher the degree of symmetry $R \geq 1$ of the secondary structure the greater its free energy (by positive $O(\log R)$), 
which lowers its probability under the Boltzmann distribution.\footnote{DNA strands with the same DNA sequence are indistinguishable in the nearest neighbour model and, for all intents and purposes, experimentally  (in statistical mechanics, two identical particles are indistinguishable).  Hence there can be several ``configurations'' (rearrangements of strands) of a secondary structure $S$ that are indistinguishable. Specifically, if identical strands can be rearranged according to a circular permutation without changing $S$, there are multiple identical/indistinguishable  configurations of $S$.  When there are symmetries in $S$ (i.e.~$R>1$), it turns out that including a positive $O(\log R)$ term in $\Delta G(S)$ makes it so that the probability distributions under indistinguishable and distinguishable model match~\cite{ShalabyPhD} (Chapter 2)~\cite{dirks}.  This is formalised using  cyclic permutations in \cref{sec:symmetry}, and indistinguishable and distinguishable models are respectively called the physical and algorithmic models in \cref{sec:dis}. } 
	Other features of the nearest neighbour model are local, in the sense they are amenable to decomposition by dynamic programming, but rotational symmetry is non-local meaning that decomposed subproblems may form part of rotationally symmetric structures or not---the subproblem itself does not encode this information leading to a history of algorithmic challenges~\cite{dirks,ShalabyWoods}.

Under the Boltzmann distribution the probability of a secondary structure $S$ at equilibrium is 
\begin{align}\label{eq:pf intro}
	\textrm{Prob}[S] = \frac{1}{Q} \cdot \mathrm{e}^{- \Delta G(S)/k_\mathrm{B}T} & & \textrm{where} & &	Q  = \sum_{S\in\Omega} \mathrm{e}^{- \Delta G(S)/k_\mathrm{B}T} 
\end{align} 
is a real number called the partition function used as the normalisation factor in $\mathrm{Prob}[S]$, 
and where  $k_\mathrm{B},T$ are constants that can be ignored, or set to 1, in this work (Boltzmann's constant and  temperature in Kelvin).

\subsection{The challenge of handling rotational symmetry in partition function \& MFE}\label{sec:intro:challenge of rot sym}
Since the partition function is a sum over the entire state space $\Omega$, 
intuition suggests that this single real number might somehow encode information about the entire energy landscape. 
Is this information obfuscated or algorithmically accessible? 
Can one extract the MFE from the \pf? 
What other  energy landscape information, or thermodynamic predictions, can be extracted via this intuition?

\begin{table}[t]\centering
	\begin{threeparttable}
		\begin{tabular}{>{\raggedright\arraybackslash}p{5cm}||>{\raggedright\arraybackslash}p{3.5cm}|>{\raggedright\arraybackslash}p{3.3cm}|>{\raggedright\arraybackslash}p{3cm}}
			Input type  & Partition function   & MFE  & \#SSEL \\ \hline\hline
			\multirow{1}{*}{Single strand}  & \multirow{1}{*}{\makecell[lt]{$\mathcal{O}(N^3)$~~~~~~\cite{mccaskill1990equilibrium} \protect\footref{ft:N3}}}  
			&   \makecell[lt]{$\mathcal{O}(N^3)$ \\ \cite{zukeroptimal,zukerrna, nussinov1978algorithms, nussinov1980fast,waterman1986rapid}\protect\footref{ft:N3}}   &   \makecell[lt]{$\mathcal{O}(N^3)$ \\ {\bf [\cref{thm:SSEL}]}\tnote{***}} \\  \hline
			Multiple strands, bounded,\newline i.e.~$c= \mathcal{O}(1)$ strands  &  \makecell[lt]{$\mathcal{O}(N^3 (c-1)!)$ \\ \cite{dirks}\footref{ft:N3} \bf[{\cref{thm:main}]}\tnote{*}}  &  \makecell[lt]{$\mathcal{O}(N^3 (c-1)!)$ \\ {\bf [\cref{thm:MFE}]}\tnote{**}}  &   \makecell[lt]{$\mathcal{O}(N^3 (c-1)!)$ \\ {\bf[\cref{thm:SSEL}]}\tnote{****}}  \\ \hline
			Multiple strands, unbounded,\newline i.e.~$c= \mathcal{O}(N)$ strands  &  \multirow{1}{*}{\makecell[lt]{\#P hard  \cite{ducloz}}} 
			&  \multirow{1}{*}{\makecell[lt]{APX hard  \cite{condon2021predicting}}}  &  \multirow{1}{*}{\makecell[lt]{\#P complete  \cite{ducloz}}} 
			\\ \hline
		\end{tabular}
		\begin{tablenotes}[ resume = tobecont, ,topsep = 0pt, partopsep = 0pt]
		\footnotesize
			\item[*] Same runtime as \cite{dirks}, but {\em magnifiable}, leads to new results:~\cref{thm:MFE,thm:SSEL,thm:counting,th:SSELMFEalgo,lem:high} 
			\item[**] Improvement over $\mathcal{O}(N^4 (c-1)!)$ in Ref.~\cite{ShalabyWoods}.  
			\item[***]  Improvement over $\mathcal{O}(N^5)$ in Ref.~\cite{cupal1997density}.
			\item[****] First polynomial time algorithm.  
		\end{tablenotes}
		\vspace{-0.1\baselineskip}
	\end{threeparttable}
	\caption{\label{table}
		Run-time bounds for algorithms  computing partition function,  MFE and \#SSEL for unpseudoknotted\protect\footref{ft:pseudoknot short} nucleic acid systems, and hardness results.
		$N$ is the sum of all strand lengths, or number of DNA/RNA bases.
		Positive results are proven in the nearest neighbour model studied here, 
		negative/hardness results in the simpler base pair matching (BPM) model~\cite{condon2021predicting,ducloz},
		and all results assume unpseddoknottedness since  these problems are NP-hard or \#P-hard with pseudoknots~\cite{ducloz}. 
		}
\end{table}

To date, however,  intuition has not been fortified with proof, 
in the sense that   somewhat distinct techniques have been used to solve \pf versus other prediction problems in the nearest neighbour model.  
The main blockage to a unifying framework has been rotational symmetry, which, as a global property of a secondary structure does not seem amenable to the kind of dynamic programming (DP) algorithms used to date for \pf. 
I.e. small pieces of a secondary structure may be part of several structures, some having and some not having, rotational symmetry. 
To see why this matters, we take  the MFE  problem as an example since, until this paper, it has required different algorithmic techniques from partition function. 
The  \pf algorithm of Dirks, Bois, Schaeffer, Winfree and Pierce~\cite{dirks} 
uses 
McCaskill-style dynamic programming~\cite{mccaskill1990equilibrium}, 
with novel characterisation theorems, 
to correctly count  structures
so that a single, final, algebraic step can correct for rotation symmetry penalties. 
In other words, rotational symmetry is not handled during dynamic programming recursions. 
That approach, of algebraically handling rotational symmetry outside of dynamic programming did not automatically carry over to the  MFE problem---since MFE uses minimisation, instead of summation, there was no obvious way to algebraically correct a multitude of candidate solutions for differing rotational symmetry inaccuracies in a  single, final step. 
Ref.~\cite{dirks} left MFE open,   
solved recently~\cite{ShalabyWoods} by first computing a (possibly incorrect) MFE that ignores rotational symmetry, 
but then proving a characterisation lemma about rotational symmetry that yields a quadratic bound on how far the algorithm needs to then backtrack~\cite{wuchty1999complete} to find the  true MFE. 
Unfortunately, backtracking meant expensive time and space bounds of $O(N^4 (c-1)!)$ and $O(N^4)$---i.e. factor-$N$ and factor-$N^2$ in time and space respectively, worse than Dirks,  Bois, Schaeffer, Winfree and  
Pierce~\cite{dirks}. 

There are many other problems one can ask about the energy landscape, one of the most all-encompassing is {\em density of states} (DOS, or \#SSEL~\cite{ducloz}) where we seek an efficient description of the entire energy landscape, specifically the set of energy levels that are occupied and the number of structures per level.  
DOS is especially useful for simulations of the dynamics of DNA/RNA folding, since transition rates can be derived from that description, and regions dense in state may indicate (so called) kinetic folding traps~\cite{machadoLima2008}.  
To date, there is no polynomial time algorithm for DOS, other than in the single stranded setting~\cite{cupal1997density} ($O(N^5$ time, \cref{table}). Any multi-stranded algorithm would need to simultaneously handle a multitude of rotational symmetry penalties over the entire state space, something previous  algorithms are not suited to (and anyway, they merely solve \pf and MFE in the multi-stranded setting). 

Here, we take a new approach, by having the algorithm decompose the \pf problem in terms of its rotational symmetry structure~\cite{boehmer2025rna}. 
For that we need a more general \pf algorithm.


\subsubsection{A magnifiable partition function algorithm}   
Ducloz, Shalaby and Woods~\cite{ducloz} identified and defined  {\em magnifiability} as a desirable property of a \pf algorithm\footnote{Ref.~\cite{ducloz} used the term ``magnification adaptable'' that we shorten here.} (\cref{fig:mag}), meaning that it computes over $\alpha$-scaled states:    
something previous algorithms for the nearest neighbour model~\cite{dirks,ShalabyWoods}
can not explicitly handle:
\begin{equation}\label{eq:Qa intro}
Q_\alpha  = \sum_{S\in\Omega} \mathrm{e}^{- \alpha \cdot \Delta G(S)/k_\mathrm{B}T} 
\end{equation}
for any $\alpha \in \mathrm{poly}(N)$, and where other variables are defined in \cref{eq:pf intro}.  
The magnified \pf $Q_\alpha$ generalises the usual \pf $Q$, since  $\alpha=1$ implies $Q_1 =Q$. 

Ducloz, Shalaby and Woods~\cite{ducloz} showed that the simplified base pair matching/stacking models {\em that ignore  rotational symmetry}
 have a magnifiable \pf algorithm (which they used to show hardness results in those simple models,   \cref{table}).   
But they left open the problem of whether there is a magnifiable \pf algorithm for the most-studied and applied nearest neighbour model. 
The key challenge being rotational symmetry: 
how to efficiently handle the potentially exponentially-sized set of rotationally symmetric secondary structures under $\alpha$-magnification?  
Rotational symmetry is a global property of a secondary structure, and previous  local (subproblem) decomposition techniques fail to address this question.

\begin{figure}[h]
	\centering
	\begin{tikzpicture}
		
		\tikzset{
			scale=0.8, transform shape
		}

		\draw[->, thick] (0,0) -- (0,-4.5);

		\foreach \y/\label in {0/0, -1/-5, -2/-10, -3/-15, -4/-20} {
			
			\draw (-0.1,\y) -- (0.1,\y); 
			
			\node[left] at (-0.1,\y) {\label}; 
			
		}
		
		\node[rotate=90, anchor=south, yshift=1em] at (-0.5,-2.25) {$\Delta G$ (kcal/mol)};
		
		\foreach \y in {0,-2,-1.2,-1.7} {
			
			\draw (1,\y) -- (3,\y); 
			
		}

		\newcommand{\placeNodes}[3]{
			
			\pgfmathsetmacro{\N}{#1}
			
			\pgfmathsetmacro{\a}{1}
			
			\pgfmathsetmacro{\b}{3}
			
			\pgfmathsetmacro{\step}{(\b - \a)/(\N + 1)}
			
			\foreach \i in {1,...,\numexpr\N} {
				
				\pgfmathsetmacro{\x}{\a + \i * \step}
				
				\filldraw[black] (\x, #2) circle (2pt);
				
			}
			
		}

		\placeNodes{2}{0};   
		
		\placeNodes{6}{-1.7};  
		
		\filldraw[black] (2, -2) circle (2pt);
		
		\filldraw[black] (2, -1.2) circle (2pt);
		
		\draw (2,-1) node {$\Delta G(S)$};
		
		\draw (8.2,-2.15) node {$\alpha \cdot \Delta G(S)$};

		\shade[left color=white, right color=blue!60] (3.3,0) -- (6.9,0) -- (6.9,-4) -- (3.3,-2) -- cycle;

		\draw[->,dashed, line width=1.2pt] (3.8,-1.5) -- (6.4,-2.5)
		
		node[above,pos=0.5, yshift= 2em] {Magnification}
		
		node[above,pos=0.5, yshift=1em] {by $\alpha = 2$}; 

		\draw[->, thick] (10.2,0) -- (10.2,-4.5);

		\foreach \y/\label in {0/0, -1/-5, -2/-10, -3/-15, -4/-20} {
			
			\draw (10.3,\y) -- (10.1,\y); 
			
			\node[right] at (10.3,\y) {\label}; 
			
		}
		
		\node[rotate= 90, anchor=south, yshift=1em] at (12.2,-2.25) {$\alpha \cdot \Delta G$ (kcal/mol)};

		\foreach \y in {0,-2,-1.2,-1.7} {
			
			\pgfmathsetmacro{\newy}{2*\y}
			
			\draw (7.2,\newy) -- (9.2,\newy); 
			
		}

		\newcommand{\placeNodesright}[3]{
			
			\pgfmathsetmacro{\N}{#1}
			
			\pgfmathsetmacro{\a}{7.2}
			
			\pgfmathsetmacro{\b}{9.2}
			
			\pgfmathsetmacro{\step}{(\b - \a)/(\N + 1)}
			
			\foreach \i in {1,...,\numexpr\N} {
				
				\pgfmathsetmacro{\x}{\a + \i * \step}
				
				\filldraw[blue!80] (\x, #2) circle (2pt);
				
			}
			
		}

		\placeNodesright{2}{0};   
		
		\placeNodesright{6}{-1.7*2};  
		
		\filldraw[blue!80] (8.2, -2*2) circle (2pt);
		
		\filldraw[blue!80] (8.2, -1.2*2) circle (2pt);
	\end{tikzpicture} 
	\caption{Magnification. 
		Each node is a secondary structure $S$ on an energy level, with nodes on the left having free energy $\Delta G(S)$, and right having magnified free energy $\alpha\cdot \Delta G(S)$.
		Magnification increases the absolute value of energy levels, without changing the distribution of secondary structures per level. From~\cite{ducloz}.}\label{fig:mag}
\end{figure}

\subsection{Main results}

In this paper, we give a magnifiable \pf algorithm (\cref{eq:Qa intro}) 
with the same runtime as the \pf algorithm of Dirks,  Bois, Schaeffer, Winfree and Pierce~\cite{dirks} (\cref{eq:pf intro}), but that 
(a) explicitly handles rotational symmetry in the problem decomposition, 
in order to (b) unite several common prediction problems under the umbrella of \pf,  which in turn will yield (c) more efficient algorithms for those problems.  Our main results are: 

\begin{restatable}{theorem}{main} 
\label{thm:main}
There is an $\mathcal{O}(N^3(c-1)!)$ time and $\mathcal{O}(N^2)$ space magnifiable \pf algorithm   
for  $c = \mathcal{O}(1)$ DNA/RNA strands with  $N$ bases, in the unpseudoknotted\footnote{All results in this paper assume unpseudoknotted secondary structures, since with pseudoknots essentially all of the problems discussed here (\pf, MFE, \#SSEL) are known to be NP-hard or \#P-hard. 
See, for example, the introduction of Ref.~\cite{ShalabyWoods} for an explanation and  discussion of pseudoknots.\label{ft:pseudoknot short}} nearest neighbour model.\footnote{We note that in the original literature~\cite{mccaskill1990equilibrium, dirks} that established these results, the polynomial is to the power 4 due to interior loop contributions (i.e. $\mathcal{O}(N^4)$ for single-stranded and $\mathcal{O}(N^4(c-1)!)$) for multistranded). The subsequent improvement in the exponent from 4 to 3 is achieved using techniques	in Dirks and Pierce (2003)~\cite{dirks2003partition} to handle the energy contributions of interior loops more efficiently in dynamic programming.\label{ft:N3}} 
\end{restatable}
\noindent The proof spans \cref{sec:dis,sec:chunks,sec:prosym,sec:PFMAalgo}, with  overviews in \cref{sec:paper structure,sec:strategy mag PF}. 
Setting $\alpha=1$ gives the (usual) \pf with the same run time and space usage of Ref.~\cite{dirks}. 

With larger $\alpha \in \Theta(N)$,\footnote{Results in this paper use only three values  $\alpha =0$ ,  $\alpha =1$,  $\alpha = k_\mathrm{B}TN/\varepsilon = O(N)$ (we assume $k_\mathrm{B},T,\varepsilon=O(1)$).} a single call to our magnifiable \pf algorithm solves MFE (improving on the $\mathcal{O}(N^4(c-1)!)$ time and $\mathcal{O}(N^4)$ space bounds of~\cite{ShalabyWoods}, proof in \cref{sec:MFE from finite alpha theorem and algorithm}): 
\begin{restatable}{theorem}{MFE} 
\label{thm:MFE}
There is an $\mathcal{O}(N^3(c-1)!)$ time and $\mathcal{O}(N^2)$  space algorithm for the 
Minimum Free Energy nearest neighbour unpseudoknotted secondary structure prediction problem, 
for  $c = \mathcal{O}(1)$ DNA/RNA strands of total length $N$ bases. 
\end{restatable}
Likewise, a single call to our magnifiable \pf algorithm, with the same value for $\alpha \in \Theta(N)$ as before, yields a description of the entire energy landscape, known as density of states~\cite{cupal1997density,cupal1996dynamic,lou2010thermodynamics,wang2001determining,ducloz} (proof in \cref{sec:SSEL occupied}):  
\begin{restatable}{theorem}{SSEL} 
\label{thm:SSEL}
There is an $\mathcal{O}(N^3(c-1)!)$ time and $\mathcal{O}(N^2)$  space algorithm for the 
\#SSEL (density of states, or DOS) nearest neighbour unpseudoknotted secondary structure prediction problem
for a set of $c = \mathcal{O}(1)$ DNA/RNA strands of  total length  $N$ bases. 
\end{restatable}
To further showcase the benefits of a magnifiable \pf, with a single call with  $\alpha \in \Theta(N)$, and thus the same time/space bounds as before,  
we get various other results  that further characterise various properties of the full energy landscape 
such as:
\begin{itemize} 
\item 
The total number of secondary structures (\cref{sec:one call} shows how this is a corollary of \cref{thm:SSEL} when $\alpha \in \Theta(N)$, 
although one can more easily see that setting $\alpha=0$ in \cref{thm:main} gives $Q_\alpha= Q_0$ in \cref{eq:pf intro} which is the desired result). 
\item \cref{thm:counting}: the number of structures with various symmetry properties, 
\item \cref{th:SSELMFEalgo}: the number of MFE structures for a given permutation of strands, 
\item \cref{lem:high}: the number of structures of a given energy level, strand permutation \& symmetry degree. 
\end{itemize} 

\subsection{Paper structure, and the key technical and conceptual ideas}\label{sec:paper structure}

 We give an overview of key  technical and conceptual ideas, but first the paper structure: 
 Definitions are in \cref{sec:def}.  
 \cref{sec:dis} discusses  distinguishability and the mathematical relationship between the algorithmic  ($\OmegaD$) and physical ($\OmegaI$) ensembles, or sets, of secondary structures.   
 \cref{sec:chunks} summarises the strategy to prove \cref{thm:main} (in more detail than this section) and includes lemmas on converting from the algorithmic to physical \pf. 
\cref{sec:prosym} describes structural/combinatorial properties of secondary structures.  
\cref{sec:PFMAalgo} gives the the Master Chef and Assistant Chef algorithms, the core algorithms for  \cref{thm:main}.  
\cref{sec:Applications} provides multiple applications of our main result by essentially showing that the entire landscape (many of its properties), can be efficiently decoded from the magnified \pf value computed in  \cref{thm:main}.

\subsubsection{The algorithmic and physical ensembles (sets) of secondary structures} 
\cref{sec: intro NN and rot sym,fig:1} describe the intuition behind rotational symmetry, and  \cref{sec:intro:challenge of rot sym} gives one challenge in handling it efficiently, which we next elaborate on.

Consider a multi-stranded system of  where all $c \geq 1$ strands, even if some have identical sequence,  are considered to be distinct  (we simply pretend they are distinct).  The {\em algorithmic ensemble} of secondary structures denoted by $\OmegaD$ contains all connected unpseudoknotted secondary structures $S$. 
Since strands are distinct, there is no rotational symmetry to consider ($R=1$ in $\Delta G(S)$, \cref{eq:NNDG}).
 State of the art algorithms~\cite{dirks,ShalabyWoods} use a dynamic programming paradigm that assumes (pretends) strands are distinct, by operating on the algorithmic ensemble (see \cref{sec:dis} for details). 
However, the {\em  physical ensemble} denoted by $\OmegaI \subseteq \OmegaD$, is  a maximal subset of distinct secondary structures $S$ 
(where we consider identical-sequence strands to actually be identical/indistinguishable, in other words we count only one of their many indistinguishable
structures ($1 \leq R \leq c$ in $\Delta G(S)$, \cref{eq:NNDG})). 
The physical ensemble, $\Omega$, is what we care about, since the statistical mechanics underlying the nearest neighbour model, and experiments, involve indistinguishable strands. 
In prior multi-strand work, computed algorithmic ensemble quantities were algebraically post-processed~\cite{dirks}, 
or backtracked over (a form of post-processing)~\cite{ShalabyWoods}, 
to obtain desired physical ensemble quantities. 
Our \pf algorithm will decompose by distinct symmetry degrees $R$ in order to more directly convert from the algorithmic to the physical ensemble during processing.

\subsubsection{Key technical challenges for the main result: \cref{thm:main}}
To correctly handle computation of rotational symmetry penalties, we decomposed the problem into all possible rotational symmetries $R$:  
essentially computing a partition function over ``slices'' (sectors) of the polymer graph (\cref{fig:1}), for each relevant symmetry value of $R$.  
The goal is then to simply  
combine these magnified slice \pf values to get the correct global \pf.  
Several things can go wrong if one attempts to do this naively. 

\begin{enumerate}
\item {\em The same secondary structure erroneously appears multiple times}:   
One immediate problem we are faced with is that the traditional McCaskill~\cite{mccaskill1990equilibrium} recursions, also used by \cite{dirks}, are carefully crafted to work on full (circular) secondary structures and naive application to $N/R$-sized slices (sectors) gives incorrect \pf values due to erroneously overcounting secondary structures.
\cref{lem:criterion} gives a combinatorial criterion telling us how to develop  a new  dynamic programming recursion scheme that modifies the recursions of \cite{mccaskill1990equilibrium,dirks} to work on slices by forbidding certain base pairs, which prevents erroneous overcounting. 
We call this algorithm the ``Master chef'', see overview in \cref{sec:PFMAalgo} and details  in \cref{sec:overcounting,sec:master}. 

\item {\em Slices (structures): correct handling of  connectedness,   $\alpha$-scaling, and symmetry:~~} 
Our  dynamic programming recursion on slices (segments of polymer graphs) is required to ensure a slice is connected  by base pairs.  
Also, for each loop  (\cref{fig:1}) formed by adding a base pair  in a slice during dynamic programming, 
the correct $\alpha$ scaling should be applied and the the correct repetition (for $R$-fold symmetry).  
The algorithm that enforces these conditions, and others, is called the ``Assistant chef'', see overview in \cref{sec:PFMAalgo} and details in \cref{lem:assistant,sec:assistant}.

\item {\em Erroneous overcounting of symmetry degrees:~~}For example, if a structure has maximum symmetry degree $R_\mathrm{max}=8 \leq c$ it also has symmetry degrees of all other divisors of $R_\mathrm{max}$, i.e. $R=4$, $R=2$ and $R=1$), and including all of these would lead to overestimation of the \pf. 
\cref{lem:conv1} uses a kind of inclusion-exclusion principle to separate these merged symmetry $R$ classes (\cref{eq:rec-proof}) into disjoint maximum ($R_\mathrm{max}$) symmetry degree classes. See \cref{sec:chunks} for details.

\item {\em Running time should be no worse than~\cite{dirks}:~~}  
For the special case of all distinct strands ($R=1$), our algorithm degenerates to that of~\cite{dirks}, hence the running time is lower bounded by~\cite{dirks}. 
Our algorithm also ensures the asymptotic running time is not worse than~\cite{dirks}, even for arbitrary (but constant) symmetry degree $R \leq c$.
Secondly, to get all of the applications of our theorem, one call is sufficient (see below), which might have interest to practitioners  (software packages~\cite{NUPACK,viennaRNA} typically run separate algorithms for the various problems discussed in this paper that actually have algorithms, \pf, MFE, \#SSEL on single strands, etc.)

\end{enumerate}
This is not intended to be a complete list of the challenges, and it includes only steps 1 and 2 of the 4-step overview  (\cref{strategy1}) in \cref{sec:chunks}.

\subsubsection{Applications: key ideas} 
In summary, \cref{sec:Applications} shows that to get all algorithmic results in the paper we need only call the magnified \pf with a single magnification parameter value $\alpha = k_\mathrm{B} T N / \varepsilon = \Theta(N)$, and then do some simple (algebraic) post-processing to get the (usual) \pf $Q$ in \cref{eq:pf intro}, and \cref{thm:MFE,thm:SSEL,thm:counting,th:SSELMFEalgo,lem:high}.

However, it is  easier to first understand what is returned by calling the magnified \pf  under the various $\alpha$ values:  
\begin{itemize}
\item $\alpha=1$ (returns \pf, since $Q_\alpha = Q_1 = Q$ in \cref{eq:pf intro,eq:Qa intro}). 
\item $\alpha= 0$ (returns the total number of distinguishable secondary structures;  a trivial consequence of setting $\alpha=0$ in \cref{eq:Qa intro}.   
\item $\alpha = k_\mathrm{B} T N / \varepsilon = \Theta(N)$ (in this case the returned number, $Q_\alpha$, contains a description of the entire energy landscape: $Q_\alpha$ can be efficiently decoded with no additional asymptotic time/space overhead to give \cref{thm:MFE,thm:SSEL,thm:counting,th:SSELMFEalgo,lem:high} and the previous two bullets).
\end{itemize}

In order to get these applications,  \cref{sec:lim alpha is MFE,thm:lim} first show that in the limit as  $\alpha \rightarrow \infty$, the magnified partition function is in fact the MFE---it is a property of the exponential function that the most favoured (MFE) energies get blown up enough to dominate all energies in the magnified \pf.  
 Of course, algorithmically we desire a finite  $\alpha$, and \cref{sec:MFE from finite alpha theorem and algorithm} proves \cref{thm:MFE}, showing that merely linear $\alpha = k_\mathrm{B} T N / \varepsilon = \Theta(N)$ (for constant precision $\varepsilon$) is sufficient to extract the MFE from $Q_\alpha$ with a few algebraic operations.  
 
Perhaps surprisingly, in \cref{lem:dom} (``Hierarchical dominance of energy levels'') it is shown that the same $\alpha = k_\mathrm{B} T N / \varepsilon = \Theta(N)$ value (and hence the same call to $Q_\alpha$), can be used to extract/separate any energy level, that has structures at that level, from the set of all energy levels; 
which sets the scene for  \cref{sec:SSEL occupied} which contains the proof of \cref{thm:SSEL} (and \cref{algo:onion}):  
first compute $Q_\alpha$, then subtract the MFE value (times the number of structures with that value), then iteratively repeat the same level by level. Essentially, ``peeling'' each energy level off, one at a time, from lowest to highest (there are at most a polynomial number of levels).  
This structure on the energy landscape 
can also be unravelled in the opposite direction (from least to most favourable energy), by flipping the sign of $\alpha$ using the same procedure  (see \cref{algo:onion}).

Finally \cref{sec:one call} gives a relatively simple algorithm showing all of the previous-mentioned results can be obtained by a single magnified \pf call.

\subsection{Future work}

\begin{enumerate}
\item  Our MFE algorithm gives the MFE value but not the MFE secondary structure(s). 
Can the algorithm be modified to return the MFE structure with the same time/space bound?
 More broadly, in the context of  reductions from problem $A$ to $B$, where we  have only a dynamic programming algorithm for $B$, 
 can we use the matrices for $B$ to construct the optimal solution of $A$?

\item A single call to our \pf algorithm returns a number that can be quickly analysed to reveal the shape of the entire landscape. However, it does not encode paths through that landscape, hence: 
is there a polynomial time algorithm outputting an efficient description of all pathways through the energy landscape? 
Such a description would somehow efficiently encode an exponential size graph, where each edge is the formation or breakage of a base pair.

\item Checking all cyclic permutations of $c$ strands yields the $(c-1)!$ factor in the
multi-stranded running times in \cref{table}. 
Boehmer, Berkemer, Will, and Ponty~\cite{boehmer2025rna}  reduce that to $3^c$, but for the simple BPM model\footref{foot:BPM}. 
Can we make such improvements in the nearest neighbour model? 

\item Can we increase the number of strands $c$ beyond $O(1)$ and still have a polynomial time \pf or MFE algorithm? We know “not by much”, since for $c = \Theta(N)$ 
 \pf is \#P-hard~\cite{ducloz} and MFE is NP-complete~\cite{condon2021predicting} in the simple BPM model.
\end{enumerate}


\section{Definition of multi-stranded DNA systems}\label{sec:def}

Formally, a DNA strand\footnote{Chemically, a DNA strand $s$ is a sequence of nucleotide bases connected by \textit{covalent bonds} which together make up the backbone of $s$, the left end of $s$ corresponds to the $5'$ end of $s$ and the right end corresponding to the $3'$ end. When drawing $s$ we label  the $3'$ end with an arrow which also shows the strand directionality, see \cref{fig:1}.}\label{foot:cov} $s$ is a word over the alphabet of DNA {\em bases} $\{\mathrm{A},\mathrm{T},\mathrm{G},\mathrm{C}\}$ (RNA uses $\mathrm{U}$ instead of $\mathrm{T}$), indexed from 1 to $|s|$, where $|s|$ denotes the length of $s$.
A base pair\footnote{Base pairs are \textit{hydrogen bonds} formed between Watson-Crick pairs, namely C–G and A–T.} is a tuple $(i, j)$ such that $i<j$. 
For any $c$ strands, we assign to each of them a unique distinct identifier in $\{1, . . . ,c\}$. 
Each base is specified by a strand identifier and a position on that strand; $i_s$ denotes the base of index $i$ of strand $s$.
In this paper we consider multi-stranded systems   $\mathcal{S}$, i.e.~a multiset of either DNA or RNA strands, and $N \in \{1,2,3,\ldots\}$, called the system {\em length}, denotes the sum of the number of bases in all strands. Throughout this paper all logarithms are to the base $\mathrm{e}$, meaning $\log = \ln$.     

\subsection{Connected unpseudoknotted secondary structures and  polymer graphs}\label{sub:SS}

\begin{definition}[Secondary structure $S$]\label{def:sec struct}
	For any multi-stranded system $\mathcal{S}$ of $c$ interacting strands, a secondary structure $S$ is a set of base pairs such that each base appears in at most one pair, i.e.~if $(i_n, j_m)\in S$ and $(k_q, l_r)\in S$ then $i_n,j_m,k_q,l_r$ are all distinct.
\end{definition}
The {\em graph representation of a  secondary structure} $S$
is the graph $G=(V,E)$, where $V$ is the set of bases of each strand $s \in \{1, . . . ,c\}$, and $E = E_v \cup E_b$, where $E_v$ is the set of \emph{covalent backbone bonds} connecting base $i_s$ with base $(i+1)_s$ for all bases $i = 1,2, ..., |s|-1$ on all strands $s \in \{1, . . . ,c\}$, and $E_b = S$ is the set of base pairs in $S$. 

The set of circular permutations, $\Pi$, of $c$ strands has size $(c-1)!$~\cite{brualdi1977introductory} (e.g., for the three strands $\{A, B, C\}$, $\Pi = \{A B C, ACB\}$), 
since the orderings $ABC$, $BCA$, and $CAB$ are the same on a circle. 
Next, we define a  polymer graph for each ordering $\pi \in \Pi$, see also  \cref{fig:1}.

\begin{definition}[Polymer graph]
	For any secondary structure $S$, and any strand ordering $\pi$ of its $c$ strands, the polymer graph representation of $S$,  denoted  \PolySpi, is a graph representation of $S$, embedded in the unit disk from $\mathbb{R}^2$, where the $c$ strands are placed in succession from their $5'$ to $3'$ ends around the circumference of the circle in the order given by $\pi$, the bases, $V$, are spaced evenly around the circle circumference, each element of $E_v$ is represented by an arc on the circumference between covalently-bonded bases, and each element of $E_b$ is represented by a chord between two different bases. See \cref{fig:1}.
\end{definition}

\begin{definition}[Unpseudoknotted secondary structure]
	A secondary structure $S$ is unpseudoknotted if there exists at least one circular permutation $\pi \in \Pi$ such that $\PolySpi$ is planar (none of the chords cross), otherwise $S$ is pseudoknotted. 
	An example is shown in \cref{fig:1}.
\end{definition} 

A secondary structure $S$ is connected if the graph representation of $S$ is a connected graph. 
In this work, we only consider \textbf{connected unpseudoknotted} secondary structures of multi-stranded systems  $\mathcal{S}$ 
with $c = \mathcal{O}(1)$ strands and $N$ bases, i.e. $c$ is a constant independent of $N$ which is consider to be the input size.  
 And following recent 
 literature~\cite{zukerrna,mccaskill1990equilibrium, dirks2003partition,dirks2004algorithm,dirks,viennaRNA}, our algorithms are analysed on the unit cost RAM model~\cite{cormen2022introduction}. 

\subsection{The nearest neighbour energy model}\label{sec:BG}

Simple temperature-independent (or more accurately, temperature-ignorant) energy models for nucleic acid secondary structures, called the Base Pair Matching (BPM) model, and the Base Pair Stacking (BPS) model were proposed in literature, were used to prove hardness results~\cite{condon2021predicting,ducloz,Lyngso2004}.\footnote{In the BPM model, the free energy of a secondary  structure $S$ is simply the number of base pairs in $S$, and in BPS it is the number of stacks in $S$. Although these models are not an accurate reflection of the underlying statistical mechanics since they ignore multiple important features that matter for prediction (temperature, single-stranded loops, rotational symmetry), hardness results in such simple settings intuitively suggest hardness should carry over to the more realistic  nearest neighbour model.\label{foot:BPM}}
Beyond temperature dependence, simple models do not account for a number of features of DNA and RNA that provide significant free energy contributions: single-stranded regions and rotational symmetry. This motivates the most realistic energy model in physics which is called the nearest neighbour energy model for any secondary structure $S \in \Omega$, where $\Omega$ is the set of all connected unpseudoknotted secondary structures of multi-stranded systems $\mathcal{S}$.

\begin{definition2}[The Nearest Neighbour (NN) energy model]\label{def:NN}
	Let $S$ be an unpseudoknotted connected 
	secondary structure over a multi-stranded system $\mathcal{S}$ of $c\geq 1$ strands,  
	and where strands with the same sequence are treated as indistinguishable.  
	$S$ is decomposed into loops, 
	each of one the types: 
		hairpin, interior, exterior, stack, bulge, or multiloop, as described in \cref{fig:1}. 
		Then, the free energy of any secondary structure $S \in \Omega$,
		$\Delta G(S,T)$ or simply $\Delta G(S)$, when temperature $T$ is clear in the context, is defined as the sum of the following three terms: 
		\begin{equation}\label{eq:NNDG}
			\NNDG
		\end{equation}
		
		\begin{itemize}
			\item The first is the sum of the 
			free energies $\Delta G(l)$ of $S$'s constituent loops~\cite{dirks}, where the free energy for each loop $\Delta G(l)$ is defined with respect to the free energy of the unpaired reference state~\cite{santa}.
			\item $\Delta G^{\textrm{assoc}} $ is the helix association~\cite{dirks} penalty applied for each of the $c-1$ strands added to the first strand to form a complex of $c$ strands to guarantee the connectedness of $S$.
			\item $R_{\text{max}}(S)$ is the maximum degree of rotational symmetry of $S$, defined in \cref{sec:symmetry}. In particular, since favourable free energies are usually negative, the term  $k_\mathrm{B} T \log R_{\text{max}}(S) \geq 0$ corresponds to the reduction in the entropic contribution of $S$, as any $R_{\text{max}}$-fold rotationally symmetric secondary structure has a corresponding $R_{\text{max}}$-fold reduction in its distinguishable conformational space~\cite{dirks}, see \cref{fig:1,rem:VS}, in \cref{sec:symmetry}.
		\end{itemize}
	\end{definition2}

\subsection{Multi-stranded rotational symmetry}\label{sec:symmetry}

Intuitively, in \cref{sub:SS} we assigned each one of the $c$ strands a unique identifier, dealing with them as \textit{distinct} strands even if two or more have the same sequence. 
However, 
in statistical mechanics, strands with the same sequence are \emph{indistinguishable}  in the sense that they behave identically with respect to relevant measurable quantities~\cite{dirks}.
Mathematically, we say that {\em two strands are  indistinguishable} if they have the same sequence. 
Also, \emph{two secondary structures are indistinguishable} if their polymer graphs can be rotated so that all strands are mapped onto indistinguishable strands, all base pairs are mapped onto base pairs, and all unpaired bases are mapped onto unpaired bases; otherwise the structures are {\em distinct}~\cite{dirks,NUPACK}.

For any $c$ strands, not necessarily distinct, they consist of $k \leq c$ \emph{strand types}, usually denoted by uppercase English letters $X,Y,\ldots$\footnote{In contrast with some of the literature. we exclude using $\{\mathrm{A},\mathrm{T},\mathrm{G},\mathrm{C}\}$ for strand types; to avoid any confusion between strand and base types.}
A {\em multi-stranded DNA system} 
$\mathcal{S} = \{(t_1,n_1),(t_2,n_2), ..,(t_k,n_k)\}$,  
is a \emph{multiset} of $k$ strand types $t_1,  ..., t_k$ with repetition numbers 
$n_1,  ..., n_k \in \mathbb{N}$ such that $n_1+ ...+n_k = c$.\footnote{It is known~\cite{sawada2003fast} how to efficiently reduce the circular permutation space by getting rid of  circular permutations that are redundant due to indistinguishable strands, which is important to consider when computing the  partition function to avoid overcounting, but not needed for MFE.} 
For such a multiset $\mathcal{S}$
we can think of each circular permutation $\pi$  as a string over strand types such that each strand type $t_i$ appears exactly $n_i$ times (e.g., $\mathcal{S} = \{(X,6),(Z,3)\}$ one valid $\pi$ is $XZXXZXXZX$).

\begin{definition}[Symmetry degree of a permutation~\cite{ShalabyWoods}] \label{def:symdeg}
	For any circular permutation $\pi$, we say $n \in \mathbb{N}$ is a symmetry degree of $\pi$ if $\pi = y^n$ for some $y$, a prefix of $\pi$. 
\end{definition}

\begin{example}
	The set of  symmetry degrees of permutation $\pi = XZXZXZXZ$ is 
	$\{1,2,4\}$, since $\pi = (XZXZXZXZ)^1 = (XZXZ)^2 = (XZ)^4$.
\end{example}

For any circular permutation $\pi$, 
its maximum symmetry degree is denoted $v(\pi)$, and the corresponding repeating prefix $x$, such that $x^{v(\pi)} = \pi$, is the \emph{fundamental component} of $\pi$. 
It can be seen that $x$ is the smallest prefix that repeats over $\pi$. 
Indeed, $v(\pi)$ is the number of cyclic permutations that map each strand to a strand of the same type. 
Any repeating prefix of $\pi$ must be a multiple of its fundamental component~\cite{ShalabyWoods}.

\begin{remark}[Notation: $X_m^n$]\label[remark]{rem:FC notation}
	For any circular permutation $\pi$, 
	its \emph{augmented version}  gives  full ordering information for each fundamental component. 
	For example, 
	if $\pi = XYXZ \, XYXZ$, then its fundamental component is $XYXZ$, and  
	$ X_1^1 Y_1^1 X_2^1 Z_1^1  \; X_1^2 Y_1^2 X_2^2 Z_1^2$ is augmented version of $\pi$, such that  $X_m^n$, means the $m$th strand of type $X$ in the $n$th fundamental component of $\pi$.   
\end{remark}

We can visualize any ordering $\pi$ by representing it as a \emph{regular $v(\pi)$-gon} with each of its $v(\pi)$ vertices   representing a fundamental component. 
Let $\rho = (1 \ 2 \ 3 \ ... \ v(\pi))$,\footnote{Here we use algebraic cycle notation.  
	The order of $\rho$, denoted by $o(\rho)$, is the length of $\rho$ which is $v(\pi)$.} and consider the cyclic group\footnote{$G^\pi$ is isomorphic to $C_{v(\pi)}$, cyclic group of order $v(\pi)$.} $G^\pi$ generated by $\rho$. Intuitively, $G^\pi$ is the group of all $v(\pi)$ rotational motions in plane of the regular $v(\pi)$-gon that give the same $v(\pi)$-gon. 
We can represent  $G^\pi$ as follows:  $G^\pi = \{\rho^0, \rho^1 ...,\rho^{v(\pi)-1}\}$, where $\rho^i$ represents rotation of the regular $v(\pi)$-gon by the angle of  
$i \times {\frac{360^{\circ}}{v(\pi)}}$, 
where  $|G^\pi| = v(\pi)$.

Intuitively, the rotational symmetry of a secondary structure $S$ with strand ordering $\pi$ is the number of rotations of its polymer graph $\PolySpi$ that give the same polymer graph, as shown in \cref{fig:1}.

\begin{definition}[$R$-fold rotationally symmetric structure~\cite{ShalabyWoods}]  \label{def:sym}
	A connected unpseudoknotted secondary structure $S$ and strand ordering $\pi$ 
	(and thus polymer graph $\PolySpi$) 
	is $R$-fold rotationally symmetric,  
	if for any base pair $(i,j)$ in the polymer graph $\PolySpi$
	the rotation of that base pair by multiples of $(360^\circ/R)$ is also in $\PolySpi$. 
	More formally:   
	$(i_{X^l_k},j_{Y^n_m}) \in \PolySpi$,   
	iff 
	$(i_{X_k^{a(l)}}, j_{Y_m^{a(n)}}) \in \PolySpi$ for all $a \in H \! \leq \! G^\pi$, where $H$ is any subgroup\footnote{$H \leq G$, notationally means $H$ is a subgroup of $G$. Every subgroup of cyclic group is also cyclic.} satisfying the condition, and if $|H| = R$.  
\end{definition}

\begin{remark}[$R$ vs $R_\text{max}$]\label[remark]{rem:VS}
	In \cref{def:sym},  a symmetric secondary structure $S$ can have more than one degree of symmetry. For example, if $S$ is $4$-fold rotationally symmetric, $S$ is also $2$-fold rotationally symmetric. The highest degree of symmetry that $S$ can obtain is denoted by $R_\text{max}(S)$, see \cref{def:NN}. In general, any $R$-fold rotationally symmetric secondary structure is also $R'$-fold rotationally symmetric if $R'$ divides $R$, denoted by $R'|R$. As a consequence any secondary structure can be considered a $1$-fold rotationally symmetric structure (or in other words, asymmetric).         
\end{remark}

\subsection{PF, MFE, \#SSEL problems, and magnification}
Now, we define the main computational problems for which we are going to establish algorithmic relationships in the nearest neighbour model, through the lens of energy model magnification concept.

\begin{definition}[Partition Function: $Q$]~
	\begin{description}
		\item \textbf{Input:}  Multi-stranded system $\mathcal{S}$ with $c\geq 1$ strands, $N$ bases, and a temperature $T \geq 0$.
		\item \textbf{Output:} The partition function $Q = \sum_{S \in \Omega} e^{-{\Delta G(S)}/{k_{\mathrm{B}} T}}$. 
	\end{description}
\end{definition}

\begin{definition2}[Minimum Free Energy: MFE]~
	\begin{description}
		\item \textbf{Input:}  Multi-stranded system $\mathcal{S}$ with $c\geq 1$ strands,  $N$ bases, and a temperature $T \geq 0$.
		\item \textbf{Output:} The minimum free energy $\mathrm{MFE} = \min\Big\{\Delta G(S) \mid S \in \Omega \Big\}$. 
	\end{description}
\end{definition2}

\begin{definition2}[Density of states: \ssELtext]~
	\begin{description}
		\item \textbf{Input:}  Multi-stranded system $\mathcal{S}$ with $c\geq 1$ strands, $N$ bases, and a temperature $T \geq 0$, and an energy value $L \in \mathbb{R}$.
		
		\item \textbf{Output:} $\ssEL(L)$: number of structures $S \in \Omega$ such that $\Delta G(S)=L$.
	\end{description}
\end{definition2}

\begin{note}[Energy model magnification~\cite{ducloz}] \label{note:magnification}
	For a positive real number $\alpha\in\mathbb{R}^+$ the notation $\alpha \cdot \Delta G(S)$ simply means to multiply the free energy $\Delta G(S)$ by $\alpha$. Whenever we apply magnification in this paper, we do so uniformly over all secondary structures $S \in \Omega$, which does not change their relative free energy ordering nor distribution per energy levels, see \cref{fig:mag}. 
\end{note}

\begin{definition}[Magnifiable energy model~\cite{ducloz}]\label{def:mag}
	The nearest neighbour is \textbf{magnifiable}
	if there exists a polynomial time algorithm that for any $\alpha \in \mathrm{poly}(N)$ computes the $\alpha$-magnified partition function:
	$$Q_\alpha = \sum_{S \in \Omega} e^{-\alpha \cdot \Delta G(S)/ k_BT}$$
\end{definition}

This definition captures the main question that we positively answer in this paper: \emph{Is the nearest neighbour energy model magnifiable}?


\section{Distinguishability complications: $\OmegaD$ versus $\OmegaI$}\label{sec:dis}

\subsection{The algorithmic ensemble $\OmegaD$ and the symmetry-naive nearest neighbour model}\label{sec:distinguishableSS}

Consider a multi-stranded system $\mathcal{S}$ of $c \geq 1$ distinct strands (or if we pretend they are), 
then say we are in the {\em symmetry-naive nearest neighbour model}, we set 
$R_{\text{max}}(S)=1$, hence $k_\mathrm{B} T \log R_{\text{max}}(S) = 0$ in \cref{eq:NNDG}, meaning 
the free energy of a secondary structure $S$  simplifies to: 
\begin{equation}
	\DeltaGD(S) = \sum_{l\in S} \Delta G(l) + (c-1)\Delta G^{\textrm{assoc}}
\end{equation}
Then, let $\PiD$ be the set of all $(c-1)!$ circular permutations of the strands of $\mathcal{S}$, and 
let $\OmegaDS$ be the set (or {\em algorithmic ensemble}) of all connected unpseudoknotted secondary structures over $\mathcal{S}$.

For any circular permutation $\pi \in \PiD$, let $\OmegaDpiS \subseteq \OmegaDS$ be the subset of $\OmegaDS$ such that each connected unpseudoknotted secondary structure $S\in \OmegaDpiS$ is representable as a crossing-free polymer graph with circular permutation $\pi$. 
The representation theorem of~\cite{dirks} (Theorem 2.1) shows that the sets $\OmegaDpiS$, for all $\pi \in \PiD$, form a partition of $\OmegaD$; in other words, for each connected unpseudoknotted secondary structure $S$, there is exactly one permutation $\pi \in \PiD$ such that $S \in \OmegaDpiS$.

To compute the partition function  $\QD$ over $\OmegaD$, a known
strategy is to first compute the partition function for each of the $(c-1)!$ circular permutations $\pi$ separately, denoted $\QD(\pi)$:
\begin{equation}\label{eq:PFdistpi}
	\QD(\pi) =  \sum_{S \in \OmegaDpi} \mathrm{e}^{- \DeltaGD(S) / k_\mathrm{B} T}    
\end{equation} 

Then $\QD$ is simply a sum \QD = $\sum_{\pi \in \PiD} \QD(\pi)$ since the representation theorem~\cite{dirks} guarantees  the computed partition function considers every secondary structure $S \in \OmegaDS$ exactly once, more precisely it gives the second equality: 
\begin{equation}\label{eq:PFdist}
	\QD = \sum_{S \in \OmegaD} \mathrm{e}^{- \DeltaGD(S) / k_\mathrm{B} T} = \sum_{\pi \in \PiD} ~\sum_{S \in \OmegaDpi} \mathrm{e}^{- \DeltaGD(S) / k_\mathrm{B} T} = \sum_{\pi \in \PiD} \QD(\pi)   
\end{equation}  
We let snMFE, and snMFE($\pi$) denote the minimum free energy over $\OmegaD$ and $\OmegaDpi$ respectively, and we call it the symmetry naive MFE: 
\begin{equation}\label{eq:MFEdist}
	\textrm{snMFE} = \min_{S \in \OmegaD} \DeltaGD(S)  =\min\limits_{\pi \in \PiD} \left\{ \min_{S \in \OmegaD(\pi)} \DeltaGD(S) \right\} = \min\limits_{\pi \in \PiD} \left\{ \textrm{snMFE}(\pi) \right\}  
\end{equation}

\subsection{The physical ensemble $\Omega$ and the symmetry-aware nearest neighbour model}
Here we consider the more realistic, and challenging, {\em symmetry-aware nearest neighbour model}, or simply the nearest neighbour model see \cref{def:NN}, 
where for a  multi-stranded system $\mathcal{S}$ of $c$ strands, 
we treat strands with the same DNA/RNA sequence as indistinguishable from each other, 
which aligns with physical and experimental realities. 
The energy model is that from \cref{eq:NNDG}, and to emphasise the relationship between the symmetry-naive (\cref{sec:distinguishableSS}) and the symmetry-aware models we note: 
$\DeltaGI(S) = \DeltaGD(S) + k_\mathrm{B} T \log R_\text{max}(S)$. 

Let $\OmegaI \subseteq \OmegaD$, be  a maximal subset of {\em distinct} secondary structures. Specifically, $\OmegaIS$ denotes the set (called the {\em physical ensemble}) of all \textit{distinct} connected unpseudoknotted secondary structures over $\mathcal{S}$, 
and $\PiI$ to denote the set of {\em distinct}$\,$\footnote{For example if we have three strands $X$, $X$ and $Y$, i.e.~$\mathcal{S} = \{ (X,2) , (Y,1) \} $, then $\Pi = \{XXY \}$ and  $\PiD = \{X_1 X_2 Y , X_2 X_1 Y \}$.  In general, $\PiI \subseteq \PiD$  hence $|\PiI| \leq |\PiD|$, a fact which becomes important when computing $\QI$ and while avoiding unintended algorithmic overcounting~\cite{dirks}.} circular permutations of $\mathcal{S}$. And for any circular permutation $\pi \in \PiI$, we use $\OmegaIpiS \subseteq \OmegaIS$ to be the subset of $\OmegaIS$ such that each connected unpseudoknotted secondary structure $S\in \OmegaIpiS$ is representable as a crossing-free polymer graph with circular permutation $\pi$.  
Analogously to the algorithmic ensemble, the partition function $Q$ is: 
\begin{equation}
	Q = \sum_{S \in \OmegaI} \mathrm{e}^{- \DeltaGI(S) / k_\mathrm{B} T} 
	= \sum_{\pi \in \PiI} \sum_{S \in \OmegaIpi} \mathrm{e}^{- \DeltaGI(S) / k_\mathrm{B} T} = \sum_{\pi \in \PiI} \QI(\pi)   
\end{equation}  
The representation theorem~\cite{dirks} gives the second equality since it  holds in the physical ensemble.  
The MFE over the physical ensemble $\Omega$ is:  
\begin{equation}\label{eq:MFE}
	\textrm{MFE} = \min_{S \in \Omega} \Delta G(S)  =\min\limits_{\pi \in \Pi} \left\{ \min_{S \in \Omega(\pi)} \Delta G(S) \right\} =  \min\limits_{\pi \in \Pi} \left\{ \textrm{MFE}(\pi) \right\}   
\end{equation}

\subsection{Why consider the algorithmic ensemble  at all?}\label{sec:why}

If we threw up our hands, and only used explicit, exponential time, enumeration of all secondary structures, then we would drop the algorithmic ensemble altogether and only consider the physical ensemble $\Omega$ to predict physical systems: indistinguishable strands with the same DNA/RNA sequences are considered to be physically identical~\cite{bailey1978statistical,dirks}.
So why do we care about the algorithmic ensemble at all?  
Pretending strands are distinct is a matter of algorithmic necessity~\cite{NUPACK}, 
since all of the state-of-the-art algorithms use a fast dynamic programming paradigm that operates on the algorithmic ensemble. Dynamic programming is not obviously applicable to the physical ensemble: 
the algorithms act locally on matrix entries, unaware of the globally context, meaning they treat strands as distinct even if they have the same DNA/RNA sequence.

For example in the context of the partition function, since it is computed over $\overline{\Omega}$ rather than $\Omega$. If the strand ordering can be mapped onto itself by a cyclic permutation~\cite{dirks}, some
indistinguishable secondary structures are treated as distinct by the algorithm (i.e. see \cref{fig:counting} for an example),
resulting in an \textbf{overcounting} of the corresponding partition function contributions. This algorithmic approach implies that all computed physical quantities over $\OmegaD$ (algorithmic ensemble) need to be post-processed~\cite{dirks,ShalabyWoods} to obtain the corresponding quantities for $\OmegaI$ (physical ensemble). 

Specifically, for partition function, the algorithmic approach of~\cite{dirks} to compute $\QI(\pi)$ for indistinguishable strands, 
is to first compute $\QD(\pi)$ (\cref{eq:PFdistpi}), then post-process $\QD(\pi)$ by dividing by $v(\pi)$ to simultaneously cancel out overcounting and to correct the energy model from $\DeltaGD(S)$ to $\DeltaGI(S)$ to include the entropic penalty for rotational symmetry~\cite{bailey1978statistical}, resulting in $\QI(\pi)$ at the end, see \cref{app:dirks} for more details. 
And for MFE, our algorithmic approach~\cite{ShalabyWoods} for indistinguishable strands, is to first find the snMFE($\pi$)  (\cref{eq:MFEdist}), then post-process by backtracking over a polynomial set of secondary structures near the snMFE($\pi$) to find the true MFE($\pi$).

\subsection{The algebraic relationship between $\OmegaIpi$ and $\OmegaDpiS$}\label{sec:algebra}

Fully understanding the algebraic relationship between $\Omega(\pi)$ and $\overline{\Omega}(\pi)$, plays a significant role in our magnifiable partition function algorithm, since it sheds a light on what exactly should be done in our post-processing stage to compute the correct physical quantities, which is the magnifiable partition function in our case. Let's start with an example to pave the road for what we want to formalize.

\begin{figure}[t]
	\centering\includegraphics[width=0.8\textwidth]{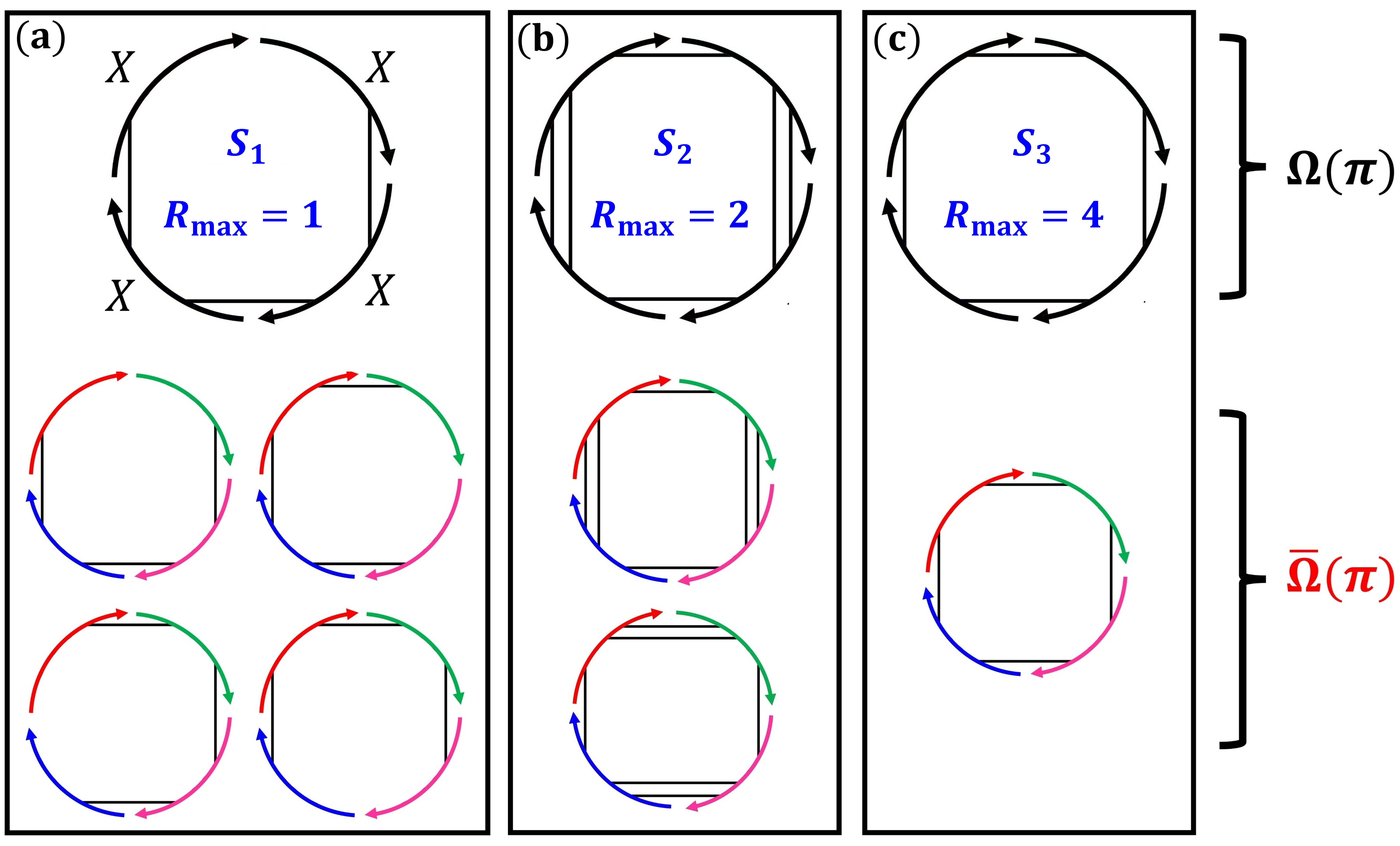}
	
	\caption{The symmetry-overcounting relationship between $\OmegaIpi$ and $\OmegaDpiS$, see \cref{ex:1}}\label{fig:counting}.
\end{figure}

\begin{example}\label{ex:1}
	In \cref{fig:counting}, since the state-of-the-art dynamic programming partition function algorithm~\cite{dirks} to compute $Q(\pi)$ runs over $\overline{\Omega}(\pi)$ instead $\Omega(\pi)$ (followed by a post-processing step), secondary structure $S_1 \in \Omega(\pi)$ with $R_\text{max}(S_1) =1$, will be encountered
	four times during the recursions, \cref{fig:counting} (a). Hence, the partition function contribution of $S_1$ will be overcounted by a factor of four. Comparing to structure $S_2$ where $R_\text{max}(S_2) = 2$, which will be encountered two times and its partition function
	contribution overcounted by a factor of two, see \cref{fig:counting} (b). With respect to $S_3$ where $R_\text{max}(S_3) = 4$, its contribution
	will be counted only once to the partition function, see \cref{fig:counting} (c). That's one of the reasons why the algorithm of~\cite{dirks} requires a post-processing step, see \cref{app:dirks}.
\end{example}

The main intuition we get from \cref{ex:1} is that: the lower the maximum degree of symmetry a secondary structure $S \in \Omega(\pi)$ has, the more $S$ will be overcounted during the partition function algorithm calculation over $\OmegaDpi$. Now, we algebraically formalize this intuition.      

Since $\PiI \subseteq \PiD$, for any $\pi \in \Pi$, 
consider secondary structure $S\in \overline{\Omega}(\pi)$. A permutation $g \in G^\pi$ acts on $S$ by relabeling strand identifiers: $g(S) = \{ (i_{X_k^{a(l)}}, j_{Y_m^{a(n)}}) ~|~ (i_{X^l_k},j_{Y^n_m}) \in S \}$. The stabilizer of $S$, denoted by $\text{Stab}(S) = \{g \in G^\pi ~|~ g(S) = S\}$, is the set of rotations of the polymer graph that map $S$ onto
itself. Note that $|\text{Stab}(S)|=R_\text{max}(S)$.

The orbit of $S$, denoted by $\text{Orb}(S)= \{g(S) ~|~ g \in G^\pi\}$, is the subset of $\overline{\Omega}(\pi)$ corresponding
to the images of $S$ under all permutations of $G^\pi$. Note that the members of $\text{Orb}(S)$ represent indistinguishable secondary structures~\cite{dirks} within $\overline{\Omega}(\pi)$ from the perspective of $\Omega(\pi)$ (see \cref{fig:counting}), and therefore they are represented by exactly one secondary structure in $\Omega(\pi)$. Hence
the partition function contribution of secondary structure $S \in \Omega(\pi)$ is overcounted by a factor of $|\text{Orb}(S)|$ (i.e. when algorithms run over $\overline{\Omega}(\pi)$ followed by a post-processing step), since the recursive scheme treats elements of
the orbit as algorithmically distinct even though they are physically indistinguishable~\cite{dirks}.
The orbit-stabilizer theorem from group theory~\cite{dummitfoote1999} provides the following relationship~\cite{dirks,bailey1978statistical}:
\begin{align}\label{eq:orbit}
	|\text{Orb}(S)| \cdot |\text{Stab}(S)| = |G^\pi| = v(\pi) \quad \forall S \in \overline{\Omega}(\pi) 
\end{align}

Since $|\text{Stab}(S)|=R_\text{max}(S)$, hence we get the symmetry-overcounting relationship (see \cref{fig:counting}) for any secondary structure $S$ as follows:
\begin{align}\label{eq:overcounting}
	|\text{Orb}(S)| = \frac{v(\pi)}{R_\text{max}(S)}
\end{align}

Since any secondary structure $S \in \Omega(\pi)$ may have different degrees of symmetries, see \cref{rem:VS}, but a unique maximum symmetry degree $R_\text{max}(S)$. Let's partition $\Omega(\pi)$ based on $R_\text{max}$. Let $\Omega_\text{max}(\pi,R)$ denote the set of all secondary structures $S \in \Omega(\pi)$, such that $R_\text{max}(S)=R$, and let $\Omega(\pi,R)$ denote the set of all $R$-fold rotationally symmetric secondary structures (not necessarily the maximum), hence we get the following identities: 
\begin{align}\label{identity1}
	\Omega(\pi) = \bigcup\limits_{R|v(\pi)} \Omega_\text{max}(\pi,R)
\end{align}
\begin{align}\label{identity2}
	\Omega(\pi,R) = \bigcup\limits_{R'| v(\pi) \text{ and } R|R'} \Omega_\text{max}(\pi,R')
\end{align}
\cref{identity1,identity2} could be extended directly to the algorithmic ensemble as follows:
\begin{align}\label{identity1'}
	\overline{\Omega}(\pi) = \bigcup\limits_{R|v(\pi)} \overline{\Omega}_\text{max}(\pi,R)
\end{align}
\begin{align}\label{identity2'}
	\overline\Omega(\pi,R) = \bigcup\limits_{R'|v(\pi) \text{ and } R|R'} \overline\Omega_\text{max}(\pi,R')
\end{align}
The relationship between $\Omega_\text{max}(\pi,R)$ and $\overline\Omega_\text{max}(\pi,R)$ could be achieved now using the symmetry-overcounting relationship in \cref{eq:overcounting}: 
\begin{align}\label{identity:relation}
	|\overline\Omega_\text{max}(\pi,R)| = \frac{v(\pi)}{R} \cdot |\Omega_\text{max}(\pi,R)|
\end{align}


\section{Strategy: Computing partition function by pizza slices} \label{sec:chunks}

Our goal is to  compute $Q_\alpha(\pi)$, the magnifiable partition function, over $\Omega(\pi)$, see \cref{def:mag}. 

Our algorithm will operate on symmetric ``slices'', meaning a $N/R$ size segment of a polymer graph \PolySpi,   
for any $R$ such that $R|v(\pi)$  (in the maximum $R_\text{max}$ and the non-maximum $R < R_\text{max}$ sense, see \cref{rem:VS}).  
Namely, $\Omega_\text{max}(\pi,R),~ \overline\Omega_\text{max}(\pi,R),~ \Omega(\pi,R)$, and $\overline{\Omega}(\pi,R)$, see \cref{sec:algebra}. We denote their magnifiable partition functions by $Q_{\alpha,\text{max}}(\pi,R),~ 
\overline Q_{\alpha,\text{max}}(\pi,R),~ Q_{\alpha}(\pi,R)$, and $\overline Q_{\alpha}(\pi,R)$ respectively.
\begin{align}
	Q_{\alpha,\text{max}}(\pi,R) = \sum_{S \in \Omega_\text{max}(\pi,R)} e^{- \alpha \Delta G(S)/k_B T}
\end{align}
\begin{align}\label{eq:disToindis}
	\overline Q_{\alpha,\text{max}}(\pi,R) = \sum_{S \in \overline\Omega_\text{max}(\pi,R)} e^{- \alpha \overline{\Delta G}(S)/k_B T} = \sum_{S \in \Omega_\text{max}(\pi,R)} \frac{v(\pi)}{R} \cdot e^{- \alpha \overline{\Delta G}(S)/k_B T} 
\end{align}

Note that, in \cref{eq:disToindis}, in the second equality we changed the underlying state space from $\overline\Omega_\text{max}(\pi,R)$ to $\Omega_\text{max}(\pi,R)$ but overcounted the contribution of every $S \in \Omega_\text{max}(\pi,R)$ by $\frac{v(\pi)}{R}$ using the symmetry-overcounting relationship \cref{eq:overcounting}, to balance the contributions.
\begin{align}
	Q_{\alpha}(\pi,R) = \sum_{S \in \Omega(\pi,R)} e^{- \alpha \Delta G(S)/k_B T} &= \sum_	{{R'| v(\pi) } \text{ and } {R|R'}}  Q_{\alpha, \text{max}}(\pi,R') & \text{Using \cref{identity2}}
\end{align}	
\begin{align}
	\overline{Q}_{\alpha}(\pi,R) = \sum_{S \in \overline{\Omega}(\pi,R)} e^{- \alpha \overline{\Delta G}(S)/k_B T} &= \sum_	{{R'| v(\pi) } \text{ and } {R|R'}}  \overline{Q}_{\alpha, \text{max}}(\pi,R') & \text{Using \cref{identity2'}}
\end{align}	
It's straightforward to show that: 
\begin{align}\label{eq:sumR}
	Q_\alpha(\pi) & = \sum_{R|v(\pi)} Q_{\alpha,\text{max}}(\pi,R)   & \text{Using  \cref{identity1}}
\end{align}
\begin{align}
	Q_{\alpha, \text{max}}(\pi,R) &=  Q_{\alpha}(\pi,R) -\sum_{R < R' \leq v(\pi) }  Q_{\alpha, \text{max}}(\pi,R')  & \text{Such that } R'|v(\pi) \textbf{ and } R|R'  
\end{align}	
\begin{align}\label{eq:Qmax subtract}
	\overline Q_{\alpha, \text{max}}(\pi,R) &=  \overline Q_{\alpha}(\pi,R) -\sum_{R < R' \leq v(\pi) }  \overline Q_{\alpha, \text{max}}(\pi,R')  & \text{Such that } R'|v(\pi) \textbf{ and } R|R'  
\end{align}

\subsection{Computing partition function via distinct maximum symmetry degrees: Overview and lemmas}\label{sec:strategy mag PF}

For any multi-stranded system $\mathcal{S}$ with strand ordering $\pi$, 
our approach to design a magnifiable partition function algorithm for $\QalphaI(\pi)$ is to algorithmically separate  the two problems of (a) secondary structures overcounting over the algorithmic ensemble $\OmegaDpi$, and (b) the energy model correction from $\DeltaGD(S)$ to $\DeltaGI(S)$ for the conversion from the algorithmic to the physical ensemble, as follows:
\begin{strategybox}{Computing Partition Function by breaking it into separate maximum symmetry degrees classes (Proof overview of \cref{thm:main})}{strategy1}
	\begin{enumerate}
		\item \label{step:1}
		For all $R$ such that $R|v(\pi)$, compute $\overline Q_{\alpha}(\pi,R)$, the magnified partition function over the algorithmic ensemble $\overline\Omega(\pi,R)$, the set of all secondary structures with symmetry degree $R$ (not necessarily the maximum $R$) in the {\bf algorithmic} ensemble. \cref{strategy2} has an overview and \Cref{sec:PFMAalgo} achieves this task.
		
		\item\label{step:2} Use the computed $\overline Q_{\alpha}(\pi,R)$, 
		for all $R|v(\pi)$, 
		to separate and compute $\overline Q_{\alpha,\text{max}}(\pi,R)$, the magnified partition function over $\overline\Omega_\text{max}(\pi,R)$, the set of all secondary structures that have the same \textbf{maximum} symmetry degree $R$ in the {\bf algorithmic} ensemble. \Cref{lem:conv1} and the ``conversion'' \cref{algo:conv1} achieve this separation.  
		
		\item\label{step:3} Use $\overline Q_{\alpha,\text{max}}(\pi,R)$ to compute its parallel magnified partition function $Q_{\alpha,\text{max}}(\pi,R)$ over $\Omega_\text{max}(\pi,R)$, the set of all secondary structures that have the same maximum symmetry degree $R$ in the {\bf physical} ensemble. \Cref{lem:conv2} and the ``conversion''  \cref{algo:conv2} achieve this conversion from the algorithmic ensemble to the physical ensemble.
		
		\item\label{step:4} Sum the contribution of $Q_{\alpha,\text{max}}(\pi,R)$ over every symmetry degree $R$ such that $R|v(\pi)$ to compute the desired $\QalphaI(\pi)$ at the end, see \cref{eq:sumR}.    
	\end{enumerate}
\end{strategybox}

\begin{restatable}[Algorithm $\overline Q_{\alpha}(\pi,R) \Longrightarrow$ algorithm $\overline Q_{\alpha,\text{max}}(\pi,R)$]{lemma}{conversionA} 
	\label{lem:conv1}
	
	Given any multi-stranded system  $\mathcal{S}$ with strand ordering $\pi$, and magnification factor $\alpha$. Assume the existence of a magnifiable partition function algorithm $A$ that computes $\overline Q_{\alpha}(\pi,R)$ for any symmetry degree $R$ such that $R|v(\pi)$, then there exists another magnifiable partition function algorithm $B$ that computes $\overline Q_{\alpha,\text{max}}(\pi,R)$. And $B$ has the same time and space complexity of $A$.    
\end{restatable}

\begin{proof}
	Assume the existence of such an algorithm $A$ that computes $\QalphaD(\pi,R)$ for any symmetry degree $R$ such that $R|v(\pi)$. Recall \cref{eq:Qmax subtract},we claim that \cref{algo:conv1} implements the following recursive equation for all $1 \leq R \leq v(\pi)$ such that $R|v(\pi)$:
	\begin{equation}\label{eq:rec-proof}
		\overline Z_{\alpha, \text{max}}(\pi,R) = \overline Q_{\alpha}(\pi,R) - \sum_	{\substack{{R < R' \leq v(\pi) }\\ \text{s.t.}{R|R'}}} \overline Z_{\alpha, \text{max}}(\pi,R')
	\end{equation}
	We claim that the computed
	$\overline Z_{\alpha, \text{max}}(\pi,R) = \overline Q_{\alpha, \text{max}}(\pi,R)$, and we prove its correctness by mathematical induction on $R$.
	\begin{itemize}
		\item \textbf{Base case:} If $R = v(\pi)$, the highest degree of symmetry, there is no $v(\pi) < R' \leq v(\pi)$, hence $\overline Z_{\alpha, \text{max}}(\pi,v(\pi)) = \overline Q_{\alpha}(\pi,v(\pi))$. Since $\overline Q_{\alpha}(\pi,v(\pi)) = \overline Q_{\alpha,\text{max}}(\pi,v(\pi))$, see \cref{eq:Qmax subtract}, this implies that $\overline Z_{\alpha, \text{max}}(\pi,v(\pi)) = \overline Q_{\alpha, \text{max}}(\pi,v(\pi))$, proving the base case.
		\item \textbf{Induction Hypothesis:} For all $k < R \leq v(\pi)$ such that $R|v(\pi)$, assume $\overline Z_{\alpha, \text{max}}(\pi,R) = \overline Q_{\alpha, \text{max}}(\pi,R)$. 	
		\item \textbf{Induction step:} If $R = k$ such that $k|v(\pi)$.
		Apply the induction hypothesis on \cref{eq:rec-proof}, by replacing $\overline Z_{\alpha, \text{max}}(\pi,R')$ with $\overline Q_{\alpha, \text{max}}(\pi,R')$, we get the following using \cref{eq:Qmax subtract}, proving the induction step.
		\begin{align*}
			\overline Z_{\alpha, \text{max}}(\pi,R) = \overline Q_{\alpha}(\pi,R) -\sum_	{\substack{{R < R' \leq v(\pi) }\\ \text{s.t.}{R|R'}}} \overline Q_{\alpha, \text{max}}(\pi,R')
			= 	\overline Q_{\alpha, \text{max}}(\pi,R))
		\end{align*}	
	\end{itemize}
	Since the set of divisors of $v(\pi)$ is of $\mathcal{O}(1)$ size (since $c=O(1)$), then algorithm B computes $\overline{Q}_{\alpha, \text{max}}(\pi, R)$ in the same time and space complexity of algorithm $A$. 
\end{proof}

\begin{algorithm}[H] 
	\caption{Conversion algorithm B, that computes $\overline Q_{\alpha,\text{max}}(\pi,R)$ for any symmetry degree $R$ such that $R|v(\pi)$, over the distinguishable ensemble $\overline\Omega_\text{max}(\pi,R)$, given a multi-stranded system  $\mathcal{S}$ with strand ordering $\pi$, magnification factor $\alpha$, and magnifiable algorithm $A$ that computes $\overline Q_{\alpha}(\pi,R)$ for any symmetry degree $R$ over the distinguishable ensemble $\overline\Omega(\pi,R)$.
	} \label{algo:conv1}
	\begin{algorithmic}[1]
		\For{$R \gets 1 \ldots v(\pi)$ \textbf{and} $R|v(\pi)$ } \Comment{$O(1)$ of R's since $v(\pi) = \mathcal{O}(1)$}
		\State $\overline Z_{\alpha, \text{max}}(\pi,R) = 0$
		\State Compute $\overline Q_{\alpha}(\pi,R)$ \Comment{Using algorithm A}
		\EndFor
		
		\vspace{0.3em}

		\State $\overline Z_{\alpha, \text{max}}\big(\pi,v(\pi)\big)  = \overline Q_{\alpha} \big(  \pi,v(\pi)\big)  $  \Comment{Base case}
		
		\vspace{0.3em}
		
		\For{$R \gets (v(\pi)-1) \ldots 1$ \textbf{and} $R|v(\pi)$} \Comment{$1 \leq R < v(\pi)$ descending}
		
		\State $Z = 0$
		\For{$R' \gets R+1 \ldots v(\pi)$ \textbf{and} $R|R'$} \Comment{Summing $O(1)$ of them}
		
		\State $Z += \overline Z_{\alpha, \text{max}}(\pi,R')$
		
		\EndFor
		
		\State 
		$\overline Z_{\alpha, \text{max}}(\pi,R) = \overline Q_{\alpha}(\pi,R) - Z$
		\EndFor
		
		\State \textbf{Return } $\overline Z_{\alpha, \text{max}}(\pi,R)$ \textbf{ for all $R|v(\pi)$}

	\end{algorithmic}
\end{algorithm}

\begin{restatable}[Algorithm $\overline Q_{\alpha,\text{max}}(\pi,R) \Longrightarrow $ algorithm $Q_{\alpha,\text{max}}(\pi,R)$]{lemma}{conversionB} 
	\label[lemma]{lem:conv2}
	
	Given any multi-stranded system  $\mathcal{S}$ with strand ordering $\pi$, symmetry degree $R$ such that $R|v(\pi)$, and magnification factor~$\alpha$. 
	Assume the existence of a magnifiable partition function algorithm $B$ that computes $\overline Q_{\alpha,\text{max}}(\pi,R)$, then there exists another magnifiable partition function algorithm $C$ that computes $ Q_{\alpha,\text{max}}(\pi,R)$, such that $C$ has the same asymptotic time and space complexity as $B$.
\end{restatable}

\begin{proof}
	Assume the existence of algorithm $B$ that computes $\overline Q_{\alpha,\text{max}}(\pi,R)$ for any symmetry degree $R$ such that $R|v(\pi)$. 
	\cref{algo:conv2} represents the magnifiable algorithm $C$ that computes $ Q_{\alpha,\text{max}}(\pi,R)$. 
	
	\begin{algorithm}[H] 
		\caption{Conversion algorithm C, that computes $ Q_{\alpha,\text{max}}(\pi,R)$ over the indistinguishable ensemble $\Omega_\text{max}(\pi,R)$, given a multi-stranded system  $\mathcal{S}$ with strand ordering $\pi$, symmetry degree $R$ such that $R|v(\pi)$, magnification factor $\alpha$, and a magnifiable algorithm $B$ that computes $\overline Q_{\alpha,\text{max}}(\pi,R)$ over the distinguishable ensemble $\overline\Omega_\text{max}(\pi,R)$.   
		} \label{algo:conv2}
		\begin{algorithmic}[1]
			\State  \textbf{Let }  $x = \overline Q_{\alpha,\text{max}}(\pi,R)$ 
			\Comment{Compute \cref{eq:disToindis} using algorithm $B$} 	
			
			\State  $x = x \cdot R/v(\pi)$ 
			\Comment{Getting rid of algorithmic overcounting} 	
			
			\State  $x = x \cdot e^{-\alpha \log R} $ 
			\Comment{Free energy model correction from $\DeltaGD(S)$ to $\DeltaGI(S)$} 	
			
			\State \textbf{Return } $x$
			
		\end{algorithmic}
	\end{algorithm}
	
	\begin{itemize}
		\item Step 2 in	\cref{algo:conv2} gets rid of the algorithmic overcounting problem, since multiplying \cref{eq:disToindis} by $R/v(\pi)$ gives the following:
		\begin{align}\label{eq:prev}
			\overline Q_{\alpha,\text{max}}(\pi,R) \cdot \frac{R}{v(\pi)} =  \sum_{S \in \Omega_\text{max}(\pi,R)} e^{-\alpha \DeltaGD(S) /{k_\mathrm{B}T}}
		\end{align} 
		\item Step 3 in	\cref{algo:conv2}, the underlying energy model correction from $\DeltaGD(S)$ to $\DeltaGI(S)$ is achieved while applying the magnification factor $\alpha$ to the rotational symmetry penalty term. Multiplying \cref{eq:prev} by $e^{-\alpha \log R}$ gives the following: 
		\begin{align*}
			\overline Q_{\alpha,\text{max}}(\pi,R) \cdot \frac{R}{v(\pi)} \cdot e^{-\alpha \log R} &=  \sum_{S \in \Omega_\text{max}(\pi,R)} e^{-\alpha \big[ \DeltaGD(S) + k_\mathrm{B}T \log R \big]/{k_\mathrm{B}T}} \notag \\ 
			&=  \sum_{S \in \Omega_\text{max}(\pi,R)} e^{-\alpha \DeltaGI(S)/{k_\mathrm{B}T}} = Q_{\alpha,\text{max}}(\pi,R)
		\end{align*}
	\end{itemize} 
	Hence, these two steps of multiplication gives the desired algorithm $C$ that has the same asymptotic time and space complexity as $B$.      
\end{proof}

One might ask, why  not skip Step \ref{step:1} in \cref{strategy1}, and design an algorithm that computes $\overline{Q}_{\alpha, \text{max}} (\pi,R)$ directly instead of $\overline{Q}_{\alpha}(\pi,R)$? 
The answer is that we still use  McCaskill's~\cite{mccaskill1990equilibrium} recursive scheme, but as a sub-module 
with 
modifications in our algorithm, see \cref{sec:assistant} for full details. 
It turns out to be tricky, without sacrificing  efficiency, to convince McCaskill's recursive scheme to work strictly over $\overline\Omega_\text{max}(\pi,R)$, 
the set of all structures with  maximum symmetry degree $R$ in the algorithmic ensemble. 
This happens since any $R$-fold rotationally symmetric secondary structure is also $R'$-fold rotationally symmetric if $R'|R$, see \cref{rem:VS}. 
From the algorithmic perspective, the difficulty of separating $\overline\Omega_\text{max}(\pi,R)$ classes for different values of $R$,  forces us to work over $\overline\Omega(\pi,R)$ instead.

\subsection{Empty pizza slicing and forming strategy}

Step \ref{step:1} in \cref{strategy1} is the only missing step, hence our goal now is to design a magnifiable partition function algorithm that computes $\overline Q_{\alpha}(\pi,R)$, for any symmetry degree $R$ such that $R|v(\pi)$.
To motivate our \textbf{slice-based} strategy to achieve step \ref{step:1}, in solving the MFE problem for multiple strands including rotational symmetries, we~\cite{ShalabyWoods} start with a pizza \pizza (symmetric secondary structure) and slice it into isomorphic slices \slice.\footnote{They go on to use this slicing strategy in a technical lemma that puts a quadratic bound on the number of structures  a backtracking algorithm is required to search to find the MFE value.}

Intuitively, in this paper, our goal is the complete opposite, instead of cutting pizzas into slices, we wish to build pizzas from scratch by forming slices (segments with base pairs) and repeating them around the disk to compute the  magnifiable partition function, as follows: 
\begin{strategybox}{Empty Pizza Slicing and Forming Strategy \faPizzaSlice~ (Step 1 of \cref{strategy1})}{strategy2}
	\begin{itemize}
		\item We start with an empty pizza \emptypizza, a secondary structure with no base pairs (hence disconnected in the case of multiple strands) and slice it into $R$ empty slices \emptyslice for any valid symmetry degree $R$.
		
		\item Then we take one of these $R$ empty slices \emptyslice, which we call the \textbf{template slice} \templateslice, 
		and analyse 
		all connected, unpseudoknotted slices \slice induced by adding all possible combinations of base pairs within \templateslice.
		
		\item We repeat that slice \slice $R$ times, joining each to the next by a backbone bond, to form an $R$-fold rotationally symmetric pizza \pizza.    
	\end{itemize}
\end{strategybox}

Note that, in \cref{strategy2} having each derived slice \slice be connected is sufficient for connectedness of each derived pizza \pizza, since~\cite{ShalabyWoods} showed that a symmetric secondary structure \pizza is connected if and only if each of its isomorphic slices \slice is also connected.  Also, during the strategy, as combinations of structures are formed, their ``partial'' partition functions are computed and summed into larger partial values, until the final \pf is obtained. Finally, some technical lemmas are employed to prevent unintended over- or under-counting of structures, see \cref{sec:overcounting}. 

\section{Structural properties of symmetric pizzas} \label{sec:prosym}

From now on we assume a global indexing of all bases (over all strands of any multi-stranded system $\mathcal{S}$ given a strand ordering $\pi$) from $1$ to $N$, and use the square brackets notation, $[i, i+1]$, to  denote the covalent bond\footref{foot:cov} connecting bases $i$ and $i +1$, implying that both bases belong to the same strand. This notation also helps to differentiate the covalent bond notation $[i, i+1]$ from the base pair $(j,k)$ (hydrogen bond) notation. 
We define an ordering relation $\prec$ on the set of all covalent bonds $E_v$, such that $[i,i+1] \prec [k,k+1]$ if $i<k$.

For any connected secondary structure $S$ with ordering $\pi$, 
there are two different routes from any base $i$ to any base $j$ around the backbone, or circumference, of $\PolySpi$ (clockwise or anticlockwise). We define the \emph{length} function $l\llbracket i,j \rrbracket$ to be the length of the shorter path, including both $i$ and $j$, as follows: 
\begin{align*}
	l\llbracket i,j \rrbracket = \min \{|i-j|+1,N-|i-j|+1\}
\end{align*} 
where $\llbracket i,j \rrbracket$  is used to denote that shorter {\em segment} of length $l\llbracket i,j \rrbracket$. If the segment $\llbracket k,l \rrbracket$ is contained in the segment $\llbracket i,j \rrbracket$, we denote that containment by  $\llbracket k,l \rrbracket \subseteq \llbracket i, j \rrbracket$. And if covalent bond $[k, k+1]$ is contained in  segment $\llbracket i, j \rrbracket$, we denote that by $[k, k+1] \in \llbracket i, j \rrbracket$.

\subsection{Pizza Cuts}

\begin{definition} [Empty secondary structure \protect \emptypizza] Given a multi-stranded system of strands with ordering $\pi$, the empty secondary structure, denoted $\emptypizza$, is the secondary structure with no base pairs, hence $\emptypizza = \phi$, and the polymer graph representation of $\emptypizza$ is  denoted  $\Polypi$. \end{definition}

Intuitively, an $R$-symmetric backbone cut slices an empty secondary structure  \emptypizza into $R$ empty slices \emptyslice or segments, defined via  a set of $R$ evenly spaced covalent bonds\footref{foot:cov} along the circumference of $\Polypi$.
Since the covalent bonds are evenly spaced, there is exactly one covalent bond $b$ belonging to the segment $\llbracket 1, \frac{N}{R} \rrbracket$, and hence $b$ and $R$ are sufficient to define the cut~\cite{ShalabyWoods}:

\begin{figure}[t]
	\centering\includegraphics[width=0.7\textwidth]{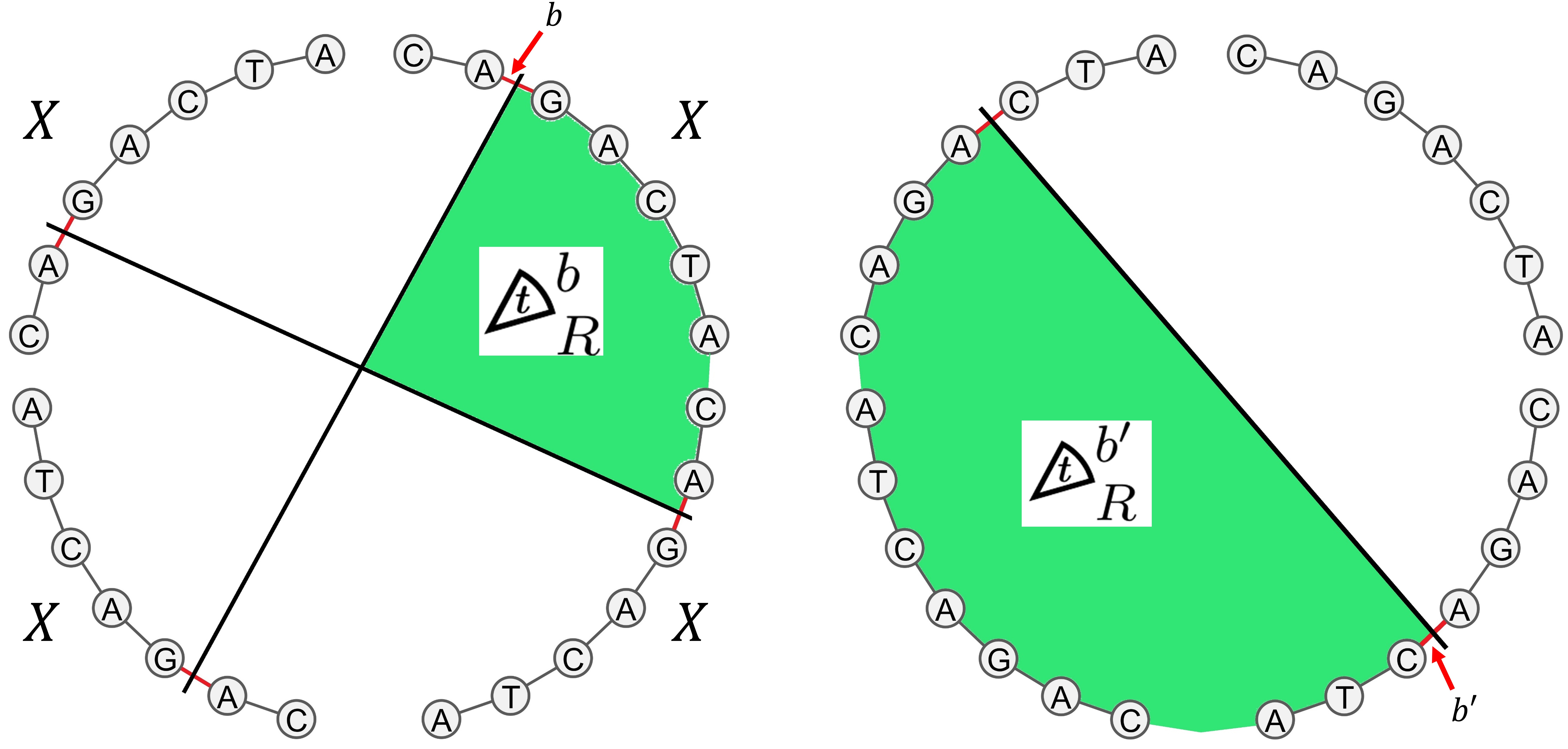}
	
	\caption{Empty secondary structure of multi-stranded system of $4$ strands of the same type $X$ with strand ordering $XXXX$; two different symmetric backbone cuts each associated with its template slices are shaded with green generated by $b$ and $b'$ respectively. (a) $4$-symmetric backbone cut, which is simply the four red covalent bonds where the cut is made(b) $2$-symmetric backbone cut which is the two red covalent bonds. If $b = f$, we get the prime template slice $\templateslice_R^f$. 
	}\label{fig:empty}
\end{figure}

\begin{definition}[$R$-symmetric backbone cut generated by a covalent bond]\label{def:cut}
	Given an empty secondary structure \emptypizza with strand ordering $\pi$, and a valid symmetry degree $R \geq 2$, the $R$-symmetric backbone cut generated by covalent bond $b=[i_{A_m^n}, (i+1)_{A_m^n}]$ such that $b \in \llbracket 1, \frac{N}{R} \rrbracket$, is $\bigotimes_R^b = \{ [i_{A_m^{a(n)}}, ({i+1})_{A_{m}^{a(n)}}]: \textrm{ for all } a \in H \!\!\leq \!\! \!\ G^\pi \textrm{ such that } |H| = R \}$. We call $b$ a {\em symmetric backbone cut generator}. \cref{fig:empty} shows an example, and \cref{rem:FC notation} and \cref{def:sym} explains the notation $A_m^{a(n)}$ which is the $m^\mathrm{th}$ instance of strand $A$ in  fundamental component number $a(n)$ where $n \in \{1, \ldots, v(\pi)/ R\}$.
\end{definition}

\begin{remark} [$R$-symmetric backbone cuts ordering]\label[remark]{rem:num}
	The number of all different $R$-symmetric backbone cuts is $\frac{N-c}{R}$, as each covalent bond $[i,i+1] \in \llbracket 1, \frac{N}{R} \rrbracket$ can be used as a cut generator~\cite{ShalabyWoods}, excluding nicks. We can extend the ordering relation $\prec$ defined on covalent bonds to the set of all $R$-symmetric backbone cuts based on the order of their generator bonds, formally $\bigotimes_R^{b_1}  \prec \bigotimes_R^{b_2}$, if $b_1 \prec b_2$.  
\end{remark}

Slicing  an empty pizza \emptypizza using an $R$-symmetric backbone cut $\bigotimes_R^{b}$ with generator $b = [i,i+1]$ leads to $R$ empty slices \emptyslice, we call the \textit{first empty slice} of them a \textit{template slice} (Intuitively the slice that just starts in the middle of the covalent bond $b$ with base $i+1$, see \cref{fig:empty}). The concept of template slices of a cut plays a major role in our algorithm as we intuitively explained in \cref{strategy2}.

\begin{definition} [Template slice]
	Given an empty secondary structure \emptypizza with strand ordering~$\pi$, symmetry degree $R \geq 2$, and covalent bond $b = [i,i+1] \in \llbracket 1, \frac{N}{R} \rrbracket$,
	then the graph~$G$ obtained from $\Polypi$ by removing the covalent bonds of the $R$-symmetric backbone cut~$\bigotimes_R^b$, formally: $G=(V(\Polypi), E(\Polypi)\setminus\bigotimes_R^b)$, consists exactly of $R$ isomorphic segments, called \textbf{empty slices}. We call the first one of these segments, the  \textbf{$R$-template slice} denoted by $\templateslice_R^b$, more formally: For covalent bond $b = [i,i+1]$, $\templateslice_R^b =\llbracket i+1, i+\frac{N}{R} \rrbracket$ is the $R$-template slice of the $R$-symmetric backbone cut generated by $b$, with base $i+1$ (the first index of $\templateslice_R^b$) as its \textbf{first border}, and we denote it by $\mathcal{B}_1(\templateslice_R^b) = i+1$, and with base $i+\frac{N}{R}$ (the second index of $\templateslice_R^b$) as its \textbf{second border}, and we denote it by $\mathcal{B}_2(\templateslice_R^b) = i+\frac{N}{R}$. Note that $\mathcal{B}_1(\templateslice_R^b) < \mathcal{B}_2(\templateslice_R^b)$. See \cref{fig:empty} for an example. 
\end{definition}

Given a strand ordering $\pi$, we denote the set of all $R$-template slices of all $R$-symmetric backbone cuts by $\mathcal{T}_R$,
where $|\mathcal{T}_R| = \frac{N-c}{R}$, since there is a template slice per cut. For the cut generated by the first covalent bond of the system, namely the covalent bond $f = [1,2]$, we call its corresponding $R$-template slice $\templateslice_R^f = \llbracket 2, 1+\frac{N}{R} \rrbracket$, {\em the  prime template slice}. For each possible $R$ and $\pi$ there is exactly one such prime template slice.  We extend the ordering relation $\prec$ to the set $\mathcal{T}_R$ of all template slices of $R$-symmetric backbone cuts: Formally  $\templateslice_R^{b_1}  \prec  \templateslice_R^{b_2}$, if $b_1 \prec b_2$.

\begin{definition}[Nest]\label{def:nest}
	For any connected unpseudoknotted secondary structure $S$ with strand ordering $\pi$, we call any subset of base pairs $A \subseteq S$ a \textbf {nest} if there exists a base pair $(i,j) \in A$ such that for all base pairs $(k,l) \in A$,  $\llbracket k,l \rrbracket \subseteq \llbracket i,j \rrbracket$ and there is no base pair $(m,n) \in S$ such that $\llbracket i,j \rrbracket \subset \llbracket m,n \rrbracket$. We call $(i,j)$ the \textbf {exterior} base pair of $A$. And we use $\# \mathrm{nest}(S)$ to denote the number of nests in $S$, see \cref{fig:1}.
\end{definition}

\begin{remark}\label[remark]{rem:nest}
	For any ($R \geq 2$)-fold rotationally symmetric secondary structure $S$, 
	note that each nick must be part of a nest, otherwise $S$ would be disconnected or asymmetric (a contradiction), this implies that there is a connected path (see \cref{fig:1}) of covalent bonds along the backbone (circumference) of $\PolySpi$ between any two consecutive nests  
\end{remark}

Transitioning from an empty pizza \emptypizza to an actual pizza \pizza, the following definition defines a special kind of pizza cuts, 
that can be thought of as slicing a pizza \pizza through its centre into $R$ slices
and through $R$ backbone bonds along the circumference, but  
without intersecting any formed base pair.

\begin{definition}[Admissible $R$-symmetric backbone cut] \label{def:admissible}
	For any connected unpseudoknotted secondary structure $S$ with strand ordering $\pi$, and a valid symmetry degree $R \geq 2$, the $R$-symmetric backbone cut $\bigotimes_R^b$ generated by $b$ is  admissible, 
	if for all covalent bonds $x \in \bigotimes_R^b$,  
	$x$ is not ``enclosed'' by any nest with exterior base pair $(i,j) \in \PolySpi$, more formally:  $x \notin  \llbracket i,j \rrbracket$. 
\end{definition}

Lemma $15$ of~\cite{ShalabyWoods} proves 
that for any $(R \geq 2)$-fold rotationally symmetric secondary structure $S$,
there exists at least one admissible $R$-symmetric backbone cut. 
Here, we give a stronger result based on nesting that we need to highlight in order to avoid potential overcounting that may happen during the recursions of our algorithm. The proof of \cref{lem:cutexist} is in \cref{app:omitted:prosym}.   

\begin{restatable}{lemma}{cutexist} 
	\label[lemma]{lem:cutexist}
	For any $(R \geq 2)$-fold rotationally symmetric secondary structure $S$, for any two consecutive nests $A$ and $B$ with exteriors $(i,j)$ and $(k,l)$ respectively, any covalent bond $[a,a+1]$ on the path of the circumference of $\PolySpi$ from $A$ to $B$
	is an admissible $R$-symmetric backbone cut generator.    
\end{restatable}

\subsection{Note on Central Loops and Multiloops}\label{sec:loops}

Given any multi-stranded system $\mathcal{S}$ with strand ordering $\pi$, a \textit{substructure} $t$ is a set of base pairs only over a subsequence $[i,j]$ of $\mathcal{S}$. For a substructure $t$, the free energy of $t$ is the summation of the loop energies of its constituent loops:  
$\Delta G(t) = \sum_{l\in t} \Delta G(l)$. Note that helix association penalty  $\Delta G^{\textrm{assoc}}$, and rotational symmetry penalty are not included in this definition; as they are global properties of the full secondary structure not a substructure of it. 

For any $(R \geq 2)$-fold  symmetric secondary structure $S$, Lemma $16$ of~\cite{ShalabyWoods} showed that the graph $G$ obtained from $\PolySpi$ by removing all covalent bonds of its admissible $R$-symmetric backbone cut $\bigotimes_R^b$, formally $G=(V(\PolySpi), E(\PolySpi)\setminus\bigotimes_R^b)$, is disconnected and consists exactly of $R$ connected isomorphic substructures.
Each of these substructures is simply called a {\em slice} of $\PolySpi$,  denoted by $\slice^{S,b}_R$, and we define the loop free energy of a slice as a separate substructure  as the summation of the loop energies of its constituent loops as follows: 
\begin{equation}
	\Delta G(\slice^{S,b}_R) = \sum_{l\in \slice^{S,b}_R} \Delta G(l)
\end{equation}

\begin{remark}[Central loop $\mathcal{C}(S)$]\label[remark]{rem:central}
	For any $(R \geq 2)$-fold rotationally symmetric secondary structure $S$, lemma $20$ of~\cite{ShalabyWoods} shows the necessary existence of a unique loop in the center of $\PolySpi$  surrounded by the outer base pairs of its $R$ slices $\slice^{S,b}_R$, called the {\em central loop} of $S$, and denoted by $\mathcal{C}(S)$. If $R_\text{max}(S)>2$, then $\mathcal{C}(S)$ is a multiloop, and if $R_\text{max}(S)=2$ then $\mathcal{C}(S)$ is either a multiloop (if each slice has more than one nest), stack, or an internal loop (if each slice has exactly one nest). Hence, the free energy of $S$ can equivalently~\cite{ShalabyWoods} be defined as follows: 
	\begin{equation}
		\DeltaGI(S) = R_\text{max}(S) *  \Delta G(\slice^{S,b}_{R_{\text{max}(S)}}) + \Delta G(\mathcal{C}(S)) + (c-1) \Delta G ^\text{assoc} +
		k_\mathrm{B} T \log R_{\text{max}}(S)
	\end{equation}\label{eq:NNcentral1}
	\vspace{-20pt}
	\begin{equation}\label{eq:NNcentral2}
		\DeltaGD(S) = R_\text{max}(S) *  \Delta G(\slice^{S,b}_{R_{\text{max}(S)}}) + \Delta G(\mathcal{C}(S)) + (c-1) \Delta G ^\text{assoc}
	\end{equation}
\end{remark}

Because of the crucial importance of multiloops for our algorithm, we highlight the multiloop energy model that is used in the standard dynamic programming algorithms. The free energy of a multiloop $m$ has the following linear form~\cite{dirks}:
\begin{equation} \label{eq:multi}
	\Delta G(m)  = \Delta G_\textrm{init}^\textrm{multi} + n_1 \cdot \Delta G_\textrm{bp}^\textrm{multi} + n_2 \cdot \Delta G_\textrm{nt}^\textrm{multi},
\end{equation}
where $\Delta G_\textrm{init}^\textrm{multi}$ is called the penalty for   formation of the multiloop, $\Delta G_\textrm{bp}^\textrm{multi}$ is called the penalty for each of its $n_1$ base pairs that border\footnote{Every base pair that borders the interior of a multiloop $m$ is called a \textit{branch} of $m$.\label{ft:branch}} the interior of the multiloop, and $\Delta G_\textrm{nt}^\textrm{multi}$ is called the penalty for each of the $n_2$ free bases inside the multiloop. For any $(R \geq 2)$-fold rotationally symmetric secondary structure~$S$ such that $\mathcal{C}(S)$ is a multiloop $m$, the contribution of the two terms
$n_1 \cdot \Delta G_\textrm{bp}^\textrm{multi}$ and $n_2 \cdot \Delta G_\textrm{nt}^\textrm{multi}$ is divided equally between the $R$ slices $\slice^{S,b}_R$ of $\PolySpi$.


\section{Computing $\overline Q_{\alpha}(\pi,R)$: The final piece of our magnifiable partition function algorithm}\label{sec:PFMAalgo}

\cref{strategy1}, in \cref{sec:chunks} described our overall strategy to prove \cref{thm:main}.  
Steps  \ref{step:2}, \ref{step:3} and \ref{step:4} of \cref{strategy1} were already shown, and it only remains  to  prove step \ref{step:1} which we do here. 
In this section, in formal notation, we give a magnifiable partition function algorithm that computes $\overline Q_{\alpha}(\pi,R)$, for any symmetry degree $R$ such that $R|v(\pi)$. 

Metaphorically, we can imagine the whole process of making any $R$-fold rotationally symmetric pizza \pizza as an ongoing process between two chefs:
\begin{itemize}
	\item \textbf{The Master Chef:} The ``thinking brain'' behind the whole process of making all possible pizzas, and ensuring that there are no undercounting or overcounting of pizzas (secondary structures)  in the process.
	The master chef goes through each template slice $\templateslice$ from the set of all template slices $\mathcal{T}_R$, and decides what toppings (base pairs) should be formed there. \Cref{sec:master} achieves this step.
	
	\item \textbf{The Assistant Chef:} Responsible for implementing what the master chef is asking in each template slice \templateslice, but repeated $R$ times, one per each empty slice \emptyslice. They also magnify  the energy contribution of every formed loop by $\alpha$. \Cref{sec:assistant} achieves this step.
\end{itemize}

The master and assistant chef algorithms constitute step  \ref{step:1} of  \cref{strategy1}, and the proof is in the remainder of this section. 
Together with steps  \ref{step:2}, \ref{step:3} and \ref{step:4} of \cref{strategy1}, this gives our main theorem (restated here), with the corresponding algorithm pseudocode being \cref{algo:PFMA}:

\begingroup
\renewcommand\footnote[1]{}
\main*
\endgroup

\begin{algorithm}[H] 
\caption{A  magnifiable partition function algorithm for the nearest neighbour energy model (see \cref{thm:main}), that takes as input: $c=\mathcal{O}(1)$ strands with total number of bases (length) $N$, and magnification factor $\alpha \in \mathrm{poly}(N)$.
	Runs in  $\mathcal{O}(N^3(c-1)!)$ time and $\mathcal{O}(N^2)$ space.
} 
\label{algo:PFMA}
\begin{algorithmic}[1]
	\State \textbf{Let $Q = 0$}
	\For{$\pi \in \Pi$}
	\For{$R$ such that $R|v(\pi)$}
	\State \parbox[t]{0.9\linewidth}{\textbf{-} Run the Assistant Chef \Cref{algo:assischef}  (\cref{lem:assistant})   to pre-compute the subsequence partition functions $Q_{i,j}, Q^{\text{bp}}_{i,j}$, and $Q^{m}_{i,j} $ that will be used by the Master Chef algorithm.}
	\Statex 
	
	\If{$R > 1$}
	
	\State \parbox[t]{0.87\linewidth}{\textbf{-} Run the Master Chef \cref{algo:master}  (\cref{lem:master}) to compute  the magnified partition function $\overline Q_{\alpha}(\pi,R)$ over
		$\overline\Omega(\pi,R)$  ensemble of all secondary structures with symmetry degree $R$ (\textbf{not} necessarily the maximum).}
	\Statex
	\EndIf
	\EndFor
	
	\State \parbox[t]{0.95\linewidth}{\textbf{-} Run Conversion \cref{algo:conv1} to separate and compute  the magnified partition function $\overline Q_{\alpha,\text{max}}(\pi,R)$ over 
		$\overline\Omega_\text{max}(\pi,R)$, the algorithmic ensemble of all secondary structures that have the same \textbf{maximum} symmetry degree $R$.}
	\Statex
	
	\State \parbox[t]{0.95\linewidth}{\textbf{-} Run Conversion \cref{algo:conv2} to compute the magnified partition function $Q_{\alpha,\text{max}}(\pi,R)$ over 
		$\Omega_\text{max}(\pi,R)$, the physical ensemble of all secondary structures that have the same \textbf{maximum} symmetry degree $R$.}
	\Statex
	
	\State $Q(\pi) = \sum\limits_{R|v(\pi)} Q_{\alpha,\text{max}}(\pi,R)$.
	
	\State $Q = Q + Q(\pi)$
	\EndFor
	\State \textbf{Return $Q$}
\end{algorithmic}
\end{algorithm}

\subsection{The Assistant Chef: Magnifies loop contributions by $\alpha$ and repeats every loop $R$ times} \label{sec:assistant}

\begin{figure}
\vspace{-2em}
\begin{minipage}[b]{1\textwidth}
	\centering
	\includegraphics[width=0.95\linewidth]{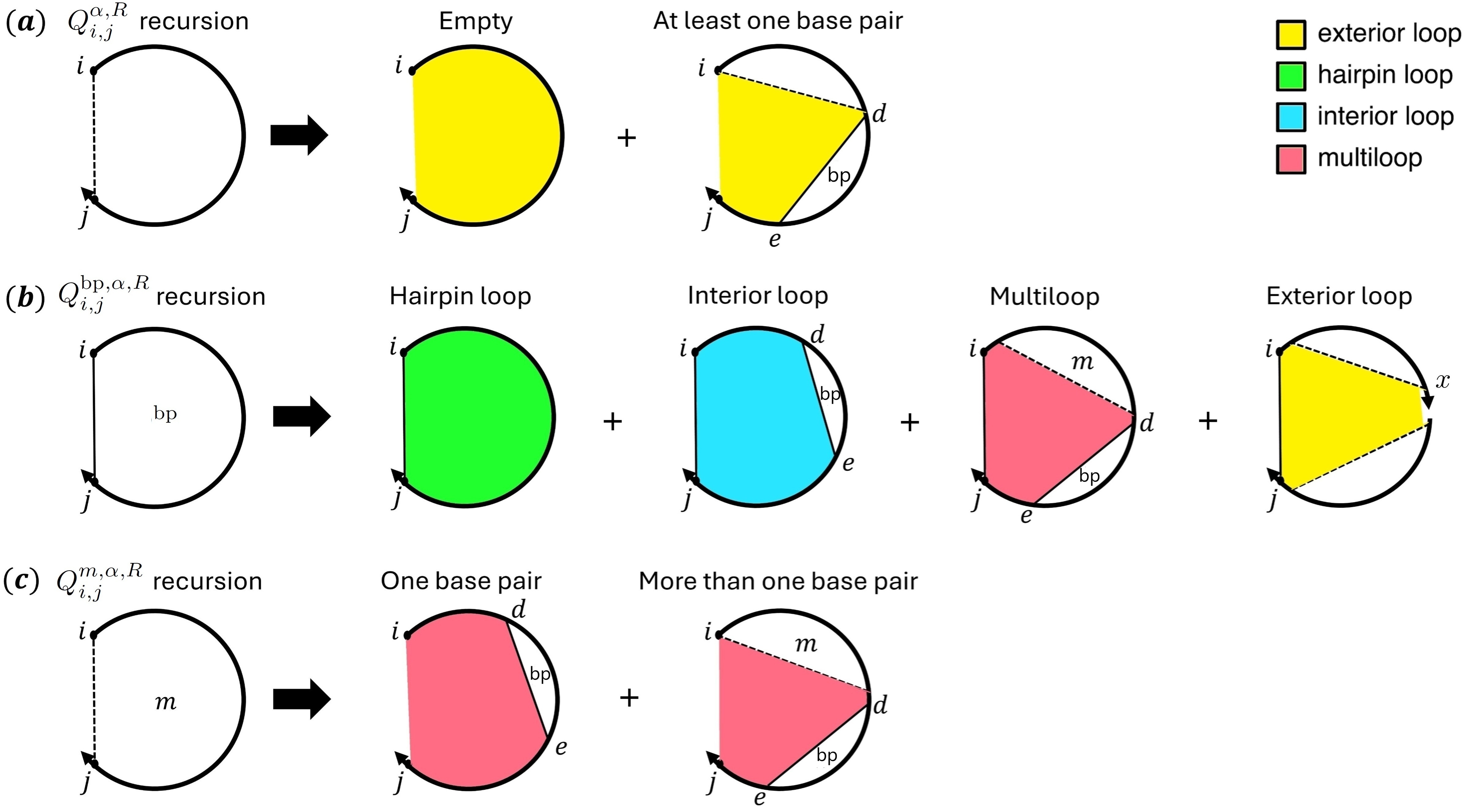} 
\end{minipage}

\begin{minipage}[b]{1\textwidth}
	
	\begin{align}
		Q^{\alpha,R}_{i,j} &= 1 + \sum_{i \leq d < e \leq j} Q^{\alpha,R}_{i,d-1} Q^{\text{bp},\alpha,R}_{d,e} \label{eq:pf1}
		\\[10pt] 
		Q^{\text{bp},\alpha,R}_{i,j} &= \exp{\{- \alpha R   \cdot \Delta G^\mathrm{hairpin}_{i,j}/k_\mathrm{B}T\}} \notag \\ 
		&+ \sum_{i < d < e \leq j} \Bigg[ Q^{\text{bp}, \alpha,R}_{d,e} \exp{\{- \alpha R \cdot \Delta G^\mathrm{interior}_{i,d,e,j}/k_\mathrm{B}T\}} \notag \\
		&+ Q^{m,\alpha,R}_{i+1,d-1} Q^{\text{bp},\alpha,R}_{d,e} \exp{\bigg\{- \alpha R \cdot \big[\Delta G^\mathrm{multi}_\mathrm{init} + 2 \Delta G^\mathrm{multi}_\mathrm{bp} + (j-e-1)\Delta G^\mathrm{multi}_\mathrm{nt}\big]/k_\mathrm{B}T\bigg\}} 
		\Bigg] \notag \\
		&+ \sum_{\substack{{i \leq c < j} \\ \text{s.t. $c+ \frac{1}{2}$ is a nick}}} Q_{i+1,c}^{\alpha,R} ~  Q_{c+1,j-1}^{\alpha,R} \label{eq:pf2}
		\\[6pt]
		Q^{m,\alpha,R}_{i,j}  &=  \sum_{i \leq d < e \leq j}   \Bigg[ Q^{\text{bp},\alpha,R}_{d,e} \exp{\bigg\{- \alpha R \cdot \big[\Delta G^\mathrm{multi}_\mathrm{bp} + (d-i+j-e) \Delta G^\mathrm{multi}_\mathrm{nt}\big] /k_\mathrm{B}T\bigg\}} \notag \\
		& \quad  \quad  \quad +  Q^{m,\alpha,R}_{i+1,d-1} Q^{\text{bp},\alpha,R}_{d,e} \exp{\bigg\{- \alpha R \cdot \big[\Delta G^\mathrm{multi}_\mathrm{bp} + (j-e)\Delta G^\mathrm{multi}_\mathrm{nt}\big]/k_\mathrm{B}T\bigg\}} 
		\Bigg] \label{eq:pf3}
	\end{align}
\end{minipage}
\vspace{-2em}
\caption{The Assistant Chef algorithm. 
	By convention~\cite{dirks,fornace2020unified}, a solid straight line between two different bases indicates a base pair formed between them, and
	a dashed line demarcates a region without implying that the connected bases are paired.
	Shaded regions correspond to loop free energies that are explicitly incorporated at the current level of recursion. 	
	The algorithm is described by recursion diagrams (top) and recursion equations (bottom). (a) $Q^{\alpha,R}_{i,j}$ (see \cref{eq:pf1}) represents the partition function for
	all substructures over the subsequence $[i, j]$, magnified by $\alpha$ and repeated $R$ times, such that the base pairing between $i$ and $j$ is undetermined.
	(b) $Q^{\text{bp},\alpha,R}_{i,j}$ (see \cref{eq:pf2}) is the partition function for
	subsequence $[i, j]$ with the restriction that bases $i$ and $j$ are paired. (c) $Q^{m,\alpha,R}_{i,j}$ (see \cref{eq:pf3}) is the partition function for
	subsequence $[i, j]$ with the restrictions that the subsequence
	is inside a multiloop and contains at least one base pair. For more details see~\cite{dirks,fornace2020unified}.  Pseudocode in \cref{algo:assischef} has some implementation details.}
\label{fig:assischef}
\end{figure}

In this subsection we describe the assistant chef algorithm, which is an algorithm on template slices. 
Specifically, the subsection contains a proof sketch of the following lemma, and the pseudocode implementation of the assistant chef algorithm, is in \cref{app:omitted:PFMAalgo}:

\begin{restatable}[Assistant Chef (correctness of \cref{algo:assischef})]{lemma}{assistant} 
\label{lem:assistant}
For any multi-stranded system $\mathcal{S}$ with strand ordering $\pi$, symmetry degree $R$, magnification factor $\alpha$, and given any subsequence $[i, j]$ of $\mathcal{S}$, 
the assistant chef algorithm described in recursive \cref{eq:pf1,eq:pf2,eq:pf3} 
(\cref{fig:assischef}), 
or in \cref{algo:assischef},
correctly computes  \cref{eq:spf1,eq:spf2,eq:spf3} in $\mathcal{O}(N^3)$  time and $\mathcal{O}(N^2)$ space.
\end{restatable}

\begin{proof}[Proof sketch]
The algorithm is a straightforward generalisation  
the partition function  recursive scheme developed by McCaskill~\cite{mccaskill1990equilibrium} for single strands and extended by Dirks et al.~\cite{dirks} to multiple strands. 
\cref{eq:pf1,eq:pf2,eq:pf3}, in \cref{fig:assischef}, represent our generalization which includes two new parameters, the repetition (or symmetry) parameter $R \in \mathbb{N}$, and the magnification parameter $\alpha \in \mathbb{Q}$. Previous work~\cite{mccaskill1990equilibrium,dirks} is the special case of $R=1$ and $\alpha = 1$, since that work had no repetition nor magnification.

Given any multi-stranded system $\mathcal{S}$ with strand ordering $\pi$, for any two bases $i$ and $j$ such that $i<j$, let $\Omega_{i,j}$ denote the set of all connected substructures over the subsequence $[i, j]$, such that  it is not yet determined whether $i,j$ are base-paired, and $i,j$ may be on the same or different strands.
And let $\Omega_{i,j}^{\text{bp}}$ denote the set of all connected substructures over the subsequence $[i, j]$, with the restriction that there is a base pair formed between base $i$ and base $j$.

For any substructure $t \in \Omega_{i,j}$ such $|t|\geq1$, in the case when the subsequence $[i, j]$ is supposed to be inside a multiloop $m$, let $\# \text{bp}(t)$ denote the number of base pairs of $t$ that border the interior\footnote{Note that $\# \text{bp}(t) \leq |t|$.} of $m$ and contribute to the free energy of $m$, and let $\# \text{nt}(t)$ denote the number of free bases of $t$ inside $m$ that contribute to the free energy of $m$. Formally speaking, for any connected substructure $t \in \Omega_{i,j}$, $|t| \geq 1$:
\begin{align}
	\# \text{bp}(t) = \# \text{nest}(t),  &&   \text{(as each nest is a branch\footref{ft:branch} of $m$ (\cref{def:nest}))}
\end{align} 
\begin{align}
	\# \text{nt}(t) = (j-i+1) - \sum_{\substack{{(d,e) \in t} \\ \text{}}} (e-d+1), &&  \text{such that $(d,e)$ is an exterior of a nest}
\end{align} 
For the case with no magnification nor repetition  (i.e.~$\alpha = R = 1$), the recursive scheme is simply that of McCaskill and Dirks et al~\cite{mccaskill1990equilibrium,dirks} which is the subsequence partition functions $Q^{\alpha,R}_{i,j}$, $Q^{\text{bp},\alpha,R}_{i,j}$, and $Q^{m,\alpha,R}_{i,j}$ in \cref{eq:pf1,eq:pf2,eq:pf3} in \cref{fig:assischef} but with $\alpha = R = 1$.  Hence  if we simply remove the superscripts $\alpha$ and $R$ from those recursions we get the following quantities from~ \cite{mccaskill1990equilibrium,dirks}:
\begin{align*}
	Q_{i,j} = \sum_{t \in \Omega_{i,j}} e^{-\Delta G(t)/k_\mathrm{B}T} 
\end{align*}
\begin{align*}
	Q^{\text{bp}}_{i,j} = \sum_{t \in \Omega_{i,j}^{\text{bp}}} e^{-\Delta G(t)/k_\mathrm{B}T} 
\end{align*}
\begin{align*}
	Q^{m}_{i,j} = \sum_{\substack{{t \in \Omega_{i,j}} \\  {\text{s.t. } |t| \geq 1}}} e^{-\big[\Delta G(t) ~ + ~ \# \text{bp}(t) \cdot \Delta G^\mathrm{multi}_\mathrm{bp} ~ + ~\# \text{nt}(t) \cdot \Delta G^\mathrm{multi}_\mathrm{nt} \big] /k_B T}
\end{align*}
The only modification in our generalized recursive equations over~\cite{mccaskill1990equilibrium,dirks} is multiplying every loop contribution\footnote{Note that the $1$ on the right hand side of \cref{eq:pf1} is explained as follows~\cite{mccaskill1990equilibrium,dirks}: $1 =  e^{-\Delta G^{\text{ext}}/k_B T}$, where $\Delta G^{\text{ext}} = 0$ for the energy contribution of an exterior loop or equivalently in the case of having empty substructure. Hence even after magnifying the exterior loop by $\alpha$ and repeating it $R$ times, their contribution to the partition function remains $1$, as shown in \cref{eq:pf1}.} by $\alpha R$, which intuitively means that whenever the assistant chef forms a loop $l$ in a given template slice $\templateslice$, he also magnifies its contribution by factor $\alpha$ and repeats it for $R$ (all) empty slices.

Like McCaskill's~\cite{mccaskill1990equilibrium} (and hence our) recursive scheme computes partition function by combining previously computed subsequence partition functions.
By the distributive law, multiplication of subsequence partition functions (each represents a sum over a set of substructures) implicitly sums over all pairwise combinations of substructures~\cite{dirks}. 
The independence of the loop contributions in the substructure \cref{sec:loops} energy model, $\Delta G(t) = \sum_{l \in t} \Delta G (l)$, implies that these products appropriately add the free energies in the partition function exponents. 
Hence multiplying each loop energy contribution, $\Delta G(l)$, by $\alpha R$ in \cref{eq:pf1,eq:pf2,eq:pf3} ends up adding their magnified and repeated version, $\alpha R \Delta G (l)$, in the exponents. As a direct consequence, the free energy  of any substructure $t$ is multiplied by $\alpha R$; since $\sum_{l \in t} \alpha R \Delta G (l) = \alpha R \sum_{l \in t}  \Delta G (l) = \alpha R \Delta G (t)$.

Note that the linear form of the free energy of multiloops (\cref{eq:multi}) is crucial here as the multiplication by $\alpha R$ is simply distributed over its three penalty components ($\Delta G_\textrm{init}^\textrm{multi}, \Delta G_\textrm{bp}^\textrm{multi},$ and $\Delta G_\textrm{nt}^\textrm{multi}$), in our generalized recursion scheme, see \cref{eq:pf2,eq:pf3}.

Hence formally speaking given any subsequence $[i, j]$: the subsequence partition functions $Q^{\alpha,R}_{i,j}$, $Q^{\text{bp},\alpha,R}_{i,j}$, and $Q^{m,\alpha,R}_{i,j}$ in \cref{eq:pf1,eq:pf2,eq:pf3}, for any $\alpha$ and $R $, compute the following three quantities correctly, in particular without overcounting any substructure\footnote{Note that, the connectedness of any substructure $t$ is crucial for correctness. For the simplicity of our generalized recursion scheme, the connectedness is only established within the pseudocode of \cref{algo:assischef}, we follow the approach of Dirks et al.~\cite{dirks} approach with respect to connectedness.}:
\begin{align}\label{eq:spf1}
	Q^{\alpha,R}_{i,j} =\sum_{t \in \Omega_{i,j}} e^{- \alpha R \Delta G(t)/k_\mathrm{B}T}
\end{align}
\begin{align}\label{eq:spf2}
	Q^{\text{bp},\alpha,R}_{i,j} = \sum_{t \in \Omega_{i,j}^{\text{bp}}} e^{- \alpha R  \Delta G(t)/k_\mathrm{B}T} 
\end{align}
\begin{align}\label{eq:spf3}
	Q^{m,\alpha,R}_{i,j} = \sum_{\substack{{t \in \Omega_{i,j}} \\  {\text{s.t. } |t| \geq 1}}} e^{- \alpha R \big[\Delta G(t) ~ + ~ \# \text{bp}(t) \cdot \Delta G^\mathrm{multi}_\mathrm{bp} ~ + ~ \# \text{nt}(t) \cdot \Delta G^\mathrm{multi}_\mathrm{nt} \big] /k_B T}
\end{align} 
Since the time and space complexity of McCaskill algorithm~\cite{mccaskill1990equilibrium} is $\mathcal{O}(N^3)$ and $\mathcal{O}(N^2)$ respectively, hence also our assistant chef \cref{algo:assischef}. 
This completes the proof sketch of \cref{lem:assistant}.
\end{proof}


\subsection{Avoiding the same pizza more than once} \label{sec:overcounting}

With respect to our empty pizza slicing and forming \cref{strategy2}, our goal established in \cref{sec:chunks},
of finding a magnifiable partition function  algorithm that computes $\QalphaD(\pi,R)$ has one technical problem. Since \cref{lem:cutexist} shows that any $R$-fold rotationally symmetric secondary structure \pizza may (or may not) have more than one admissible $R$-symmetric backbone cut. The main problem of our \Cref{strategy2} is the possibility of considering the same secondary structure \pizza more than once through two recursive trajectories of the DP starting with two different template slices $\templateslice_R^b$ and $\templateslice_R^{b'}$. This form of algorithmic overcounting leads to compute an incorrect and overcounted partition function $\QalphaD(\pi,R)$.

Now, our goal is to reach a criterion that avoids that form of overcounting, and fully separates the classes of secondary structures that could be formed starting from any two different $R$-template slices $\templateslice_R^b$ and $\templateslice_R^{b'}$. 
For the master chef to achieve such goal, let $\mathcal{P}: \mathcal{T}_R \times E_h \rightarrow \OmegaDRpi $ be a function mapping an $R$-template slice $\templateslice_R^b$ (for any valid $R$ and $b$) with exactly one single base pair $(i,j)$ that will be added to it (i.e. $\mathcal{B}_1(\templateslice_R^b)\leq i<j \leq\ \mathcal{B}_2(\templateslice_R^b)$, and hence a possibly disconnected structure) to
the set of all valid $R$-fold rotationally symmetric secondary structures $S$ that have that base pair, $(i,j) \in S$, and share the admissible $R$-symmetric backbone cut $\bigotimes_R^b$.
If $S \in \mathcal{P}(\templateslice_R^b, (i,j))$ for some $b,R$ and base pair $(i,j)$, we say $S$ is \textit{producible}  from $(\templateslice_R^b, (i,j))$.
\cref{lem:criterion} establishes such criterion.

\begin{figure}[t]
	\centering\includegraphics[width=1\textwidth]{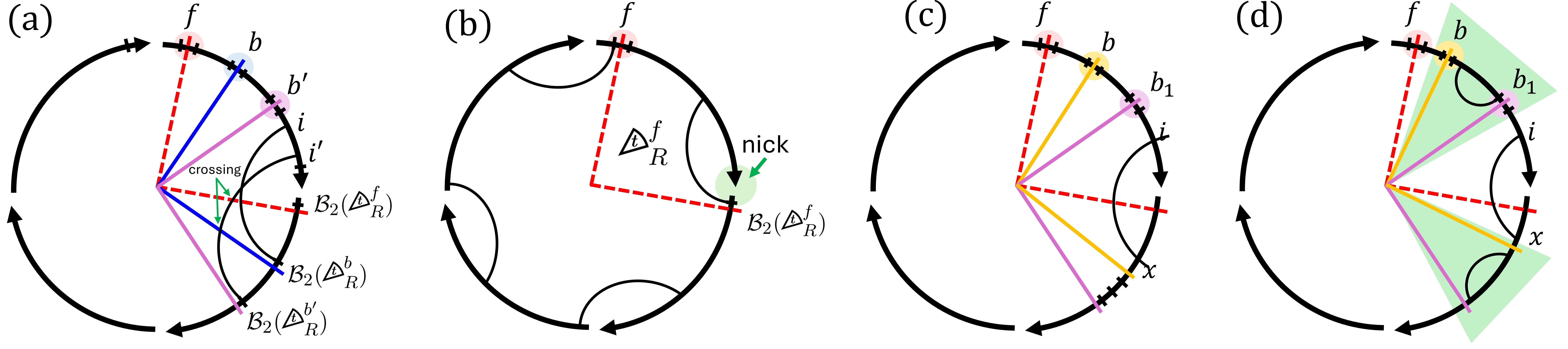}
	
	\caption{Crossing the prime template slice's second border $\mathcal{B}_2(\templateslice_R^{f})$ criterion. Dashed red lines indicates the prime template slice borders $\templateslice_R^{f}$.
	}\label{fig:criterion}
\end{figure}

\begin{lemma}[Criterion: Nest crosses the prime template slice's second border $\mathcal{B}_2(\templateslice_R^{f})$]\label[lemma]{lem:criterion}
	Let $R$ be the symmetry degree, hence  $\templateslice_R^f = \llbracket 2, 1+\frac{N}{R} \rrbracket$  is the prime template slice.  
	Let $b \text{ and } b' \in \llbracket 2, \frac{N}{R} \rrbracket$ to be two covalent bonds such that $b \prec b'$. We claim the following:
	\begin{itemize}
		\item First, 
		$\mathcal{P}(\templateslice_R^b,(i,x)) \cap \mathcal{P}(\templateslice_R^{b'},(i',y)) = \phi$, 
		where $x = \mathcal{B}_2(\templateslice_R^b)$ and $y = \mathcal{B}_2(\templateslice_R^{b'})$ (i.e. $x$ and $y$ equal the second borders of $\templateslice_R^b$ and $\templateslice_R^{b'}$ respectively), and $i, i' < \mathcal{B}_2(\templateslice_R^f) $ (i.e. base pair $(i',y)$ crosses the second border of the prime template slice $\templateslice_R^f$).\footnote{Note that $x$ and $y$ are constants and only $i$ and $i'$ are variables here. Also, note there must be a nick between bases $\frac{N}{R}$ and $1+ \frac{N}{R} = \mathcal{B}_2(\templateslice_R^{f})$, see \cref{fig:criterion} (b), and \cref{rem:nest}, that's why in our crossing criterion $i, i' < \mathcal{B}_2(\templateslice_R^f)$ to ensure connectedness, since if $i, i' = \mathcal{B}_2(\templateslice_R^f)$ they are still crossing the prime template second border but introducing disconnectedness of the formed secondary structure $S$ due to the existence of a nick.  \label{foot:range}} 
		
		\item
		\begin{minipage}{1\linewidth}
			\begin{equation}\label{eq:criterion}
				\bigcup_{\templateslice_R^b \in \mathcal{T}_R} \bigcup_{(i,x)}
				\mathcal{P}(\templateslice_R^b,(i,x))
				= \OmegaDRpi
			\end{equation}
		\end{minipage}
	
		where $(i,x)$ is the exterior base pair of a nest such that  $x = \mathcal{B}_2(\templateslice_R^b)$ (i.e. $x$ equals the second border of $\templateslice_R^b$), and 
		$\mathcal{B}_1(\templateslice_R^b)\leq i< \mathcal{B}_2(\templateslice_R^f)$ (i.e.~$(i,x)$ crosses the second border of the prime template slice $\templateslice_R^f$ if $f \prec b$).
	\end{itemize} 
\end{lemma}
\begin{proof}
	
	\begin{itemize}
		
		\item 	  
		Since $b \prec b'$, then $f \prec b'$.
		Since the base pair $(i',y)$ added to template slice $\templateslice^{b'}_R$ crosses the second border of the prime template slice $\templateslice^f_R$, then it also crosses the second border of $\templateslice^b_R$ since $\mathcal{B}_2(\templateslice_R^f) \leq \mathcal{B}_2(\templateslice_R^b)$, see \cref{fig:criterion} (a). For any secondary structures $S \in \mathcal{P}(\templateslice^{b}_R, (i,x))$, $(i',y) \notin S$ or equivalently $S \notin \mathcal{P}(\templateslice^{b'}_R, (i',y))$, otherwise it contradicts the fact that $S$ has the admissible $R$-symmetric backbone cut $\bigotimes_R^{b}$ by the definition of $\mathcal{P}$, proving the first claim.

		\item First direction ($\Rightarrow$): If a secondary structure $S$ belongs to the left hand side of \cref{eq:criterion} (i.e. $S$ is an element of the range of $\mathcal{P}$), then $S \in \OmegaDRpi$ by the definition of $\mathcal{P}$. 
		\vspace{0.1cm} 
		
		Second direction ($\Leftarrow$): Suppose $S \in \OmegaDRpi$.
		If $S$ has the admissible $R$-symmetric backbone cut $\bigotimes_R^{f}$ with generator $f=[1,2]$, we are done, since we have $b=f$, and by the definition of $f$, there must be nest with exterior base pair $(i,x)$ where $x = \mathcal{B}_2(\templateslice_R^f) = 1+ \frac{N}{R}$ (the second border of the prime template slice $\templateslice_R^f$), otherwise $S$ is disconnected since there is a nick between bases $\frac{N}{R}$ and $1+ \frac{N}{R} = \mathcal{B}_2(\templateslice_R^{f})$ (see \cref{rem:nest}). See \cref{fig:criterion} (b). Hence $S \in \mathcal{P}(\templateslice_R^f,(i,x))$.
		
		Otherwise, since $S$ is an $R$-fold rotationally symmetric secondary structure, $S$ may have more than one admissible $R$-symmetric backbone cut (see \cref{lem:cutexist}),  $\bigotimes_R^{b_1}, \bigotimes_R^{b_2}, \ldots \bigotimes_R^{b_k}$, such that $f \prec b_1 \prec b_2 \prec  \ldots b_k $. We claim that $S$ is producible from $\templateslice^{b_1}_R$, with an exterior base pair $(i,x)$ of some nest where $i< \mathcal{B}_2(\templateslice_R^f)$ and $x = \mathcal{B}_2(\templateslice_R^{b_1})$ (the second border of $\templateslice_R^{b_1}$). 
		
		By the definition of $\mathcal{P}$, all secondary structures $S'$ that are producible from the prime template slice $\templateslice^f_R$ are all of those that do not have a base pair crossing its own second border $\mathcal{B}_2(\templateslice_R^{f})$.
		Hence, there is a nest with an exterior base pair $(i,x)$ added to $\templateslice^{b_1}_R$ crosses the second border of $\templateslice^f_R$, in other words $i < \mathcal{B}_2(\templateslice_R^{f})$. We have two cases: 
		\begin{enumerate}
			\item If $x = \mathcal{B}_2(\templateslice_R^{b_1})$, we are done since we have satisfied the requirement of $\mathcal{P}$ in \cref{eq:criterion}.
			\item  Otherwise $x < \mathcal{B}_2(\templateslice_R^{b_1})$, there are two cases:  
			\begin{enumerate}
				\item All bases $d$ in $S$, such that $x <d \leq \mathcal{B}_2(\templateslice_R^{b_1})$ are unbound bases.  
				Hence by the $R$-fold rotational symmetry of $S$, there is another cut generator $b \prec b_1$, where $b = [x-\frac{N}{R},x-\frac{N}{R}+1]$, and hence template slice $\templateslice^{b}_R$ that satisfies the requirement of $\mathcal{P}$ in \cref{eq:criterion} such that $S \in \mathcal{P}(\templateslice_R^b,(i,x))$, contradicting the fact that $b_1$ is the least possible $R$-admissible cut generator of $S$. See \cref{fig:criterion} (c).
				
				\item There is another nest with an exterior $(i',y)$ such $x<i'<y \leq \mathcal{B}_2(\templateslice_R^{b_1})$, otherwise we violate the fact that $(i,x)$ being an exterior base pair of nest in $S$. 
				Again by the $R$-fold rotational symmetry of $S$, there is another cut generator $b \prec b_1$, where $b = [x-\frac{N}{R},x-\frac{N}{R}+1]$, and hence template slice $\templateslice^{b}_R$ that satisfies the requirement of $\mathcal{P}$ in \cref{eq:criterion} such that $S \in \mathcal{P}(\templateslice_R^b,(i,x))$, contradicting the fact that $b_1$ is the least possible $R$-admissible cut generator of $S$. See \cref{fig:criterion} (d).
			\end{enumerate} 
		\end{enumerate} 
	\end{itemize}
\end{proof}
\begin{note}[$\mathcal{P}(\templateslice_R^{b},(i,x))$ partitioning based on the number of nests]\label{note:numnest}
	For any template slice $\templateslice_R^{b}$ and base pair $(i,j)$ such that $\mathcal{B}_1(\templateslice_R^b)\leq i<j \leq\ \mathcal{B}_2(\templateslice_R^b)$, the set of secondary structures $\mathcal{P}(\templateslice_R^{b},(i,j))$ can be logically partitioned into two disjoint sets based on the number of nests in each slice $\slice^{S,b}_R$ of a secondary structure $S \in \mathcal{P}(\templateslice_R^{b},(i,j))$. 
	(1) The first denoted by $\mathcal{P}(\templateslice_R^{b},(i,j), \# \text{nest} =1)$ is the set of all secondary structures such that each has exactly one nest per slice   $\slice^{S,b}_R$.
	(2) The second denoted by $\mathcal{P}(\templateslice_R^{b},(i,j), \# \text{nest} > 1)$  is the set of all secondary structures such that each has more than one nest per slice   $\slice^{S,b}_R$.w
	Note that if $S$ has zero nests, it's disconnected, and hence discarded. 
\end{note}

\subsection{The Master Chef}\label{sec:master}

Now, we are ready to introduce our master chef algorithm, the last remaining sub-piece of our magnifiable partition function algorithm. The main job of the master chef ensures that there is no under or overcounting happening during our new proposed recursive scheme for template slices \templateslice. The master chef achieves this goal by incorporating our criterion of having one nest crosses the prime template slice's second border $\mathcal{B}_2(\templateslice_R^f)$, see \cref{lem:criterion}. 

\begin{figure}
	\centering
	\begin{minipage}[b]{1\textwidth}
		\centering
		\includegraphics[width=0.9\linewidth]{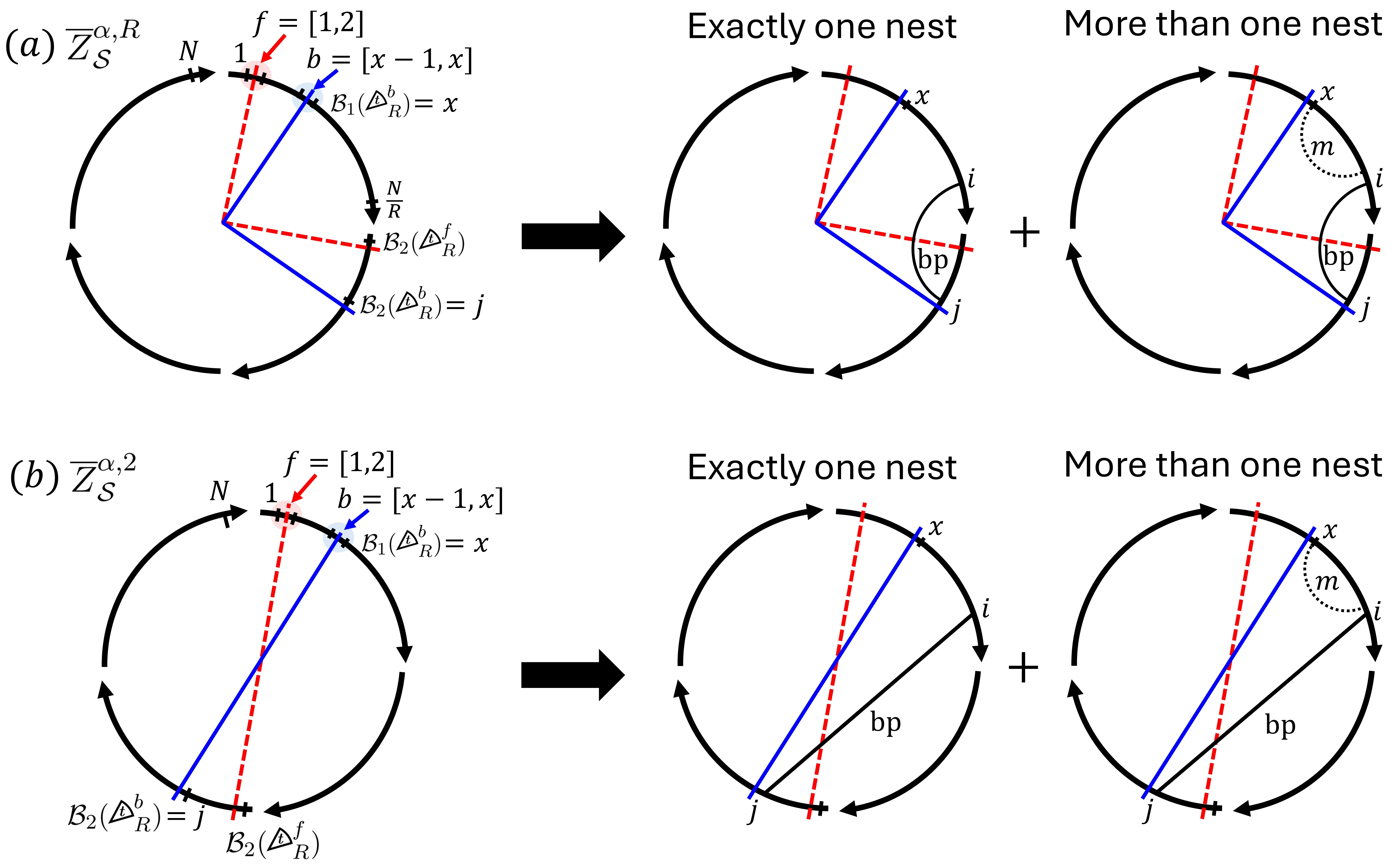} 
	\end{minipage}

	\begin{minipage}[b]{1\textwidth}

		\begin{align}
			\overline{Z}^{\alpha,R}_{\mathcal{S}} =  \sum_{\templateslice_R^b \in \mathcal{T}_R}~    \sum_{(i,j)} &\Big[ \mathds{1}_{x,i} \hspace{0.1cm} Q^{\text{bp},\alpha,R}_{i,j} \hspace{0.1cm} 		
			e^{-\alpha \big[ \Delta G^\mathrm{multi}_\mathrm{init} + R \cdot \Delta G^\mathrm{multi}_\mathrm{bp} + R \cdot (i-x)\Delta G^\mathrm{multi}_{\mathrm{nt}} + \beta \big]/k_\mathrm{B}T} & \notag \\ 
			&+  Q^{m,\alpha,R}_{x,i-1} ~ \mathds{1}_{i} ~ Q^{\text{bp},\alpha,R}_{i,j} ~ e^{- \alpha \big[\Delta G^\mathrm{multi}_\mathrm{init} + R \cdot \Delta G^\mathrm{multi}_\mathrm{bp} + \beta \big]/k_\mathrm{B}T}\Big]
			\label{eq:master1}
			\\[10pt] 
			\overline{Z}^{\alpha,2}_{\mathcal{S}} = \sum_{\templateslice_2^b \in \mathcal{T}_2}   \sum_{(i,j)}  & \Big[ \mathds{1}_{x,i} ~ Q^{\text{bp},\alpha,2}_{i,j}\hspace{0.15cm} e^{-\alpha \big[ \Delta G^\mathrm{interior}_{i,j,i',y} + \beta \big]/k_\mathrm{B}T} & \notag\\ 
			&+ 
			Q^{m,\alpha,2}_{x,i-1}
			\hspace{0.1cm}
			\mathds{1}_{i} 
			\hspace{0.1cm} 
			Q^{\text{bp},\alpha,2}_{i,j}
			\hspace{0.15cm}
			e^{- \alpha \big[ \Delta G^\mathrm{multi}_\mathrm{init} + 2 \Delta G^\mathrm{multi}_\mathrm{bp} + \beta \big]/k_\mathrm{B}T}\Big] \label{eq:master2}
		\end{align}
		
		\vspace{0.2cm}
		
		Such that $\beta = (c-1) \Delta G^\mathrm{assoc}$, $x = \mathcal{B}_1(\templateslice_R^b)$,  $j =\mathcal{B}_2(\templateslice_R^b)$, $i' = i+ \frac{N}{2}$, and $y = (j+ \frac{N}{2})\mod N$. Base $i$ of base pair $(i,j)$ ranges\footref{foot:range}  as follows: $\mathcal{B}_1 (\templateslice_R^b)\leq i < \mathcal{B}_2(\templateslice_R^f)$. 
		And let the following two indicator variables  $\mathds{1}_{a,b}$ and $\mathds{1}_a$ to be defined as follow:
		\begin{align} \label{eq:indicator1}
			\mathds{1}_{a,b} = \begin{cases}
				1 & \text{if the interval $[a,b]$  contains zero nicks, (i.e. hence connected)} \\
				0 & \text{otherwise} 
			\end{cases}
		\end{align}
		\vspace{-0.5cm}
		\begin{align}\label{eq:indicator2}
			\mathds{1}_{a} = \begin{cases}
				1 & \text{if $[a-1,a]$  is a covalent bond, (i.e. not a nick)} \\
				0 & \text{otherwise} 
			\end{cases}
		\end{align}
		\vspace{-0.5cm}

	\end{minipage}
	\caption{The Master Chef algorithm, described by recursion diagrams (top) and recursion equations (bottom),  for computing the partition function for multi-stranded system $\mathcal{S}$ of $c = O(1)$ strands, given a strand ordering $\pi$, symmetry degree $R$, and magnification factor $\alpha$.  Dashed red lines indicates the prime template slice $\templateslice_R^{f}$ borders.  
	By convention~\cite{dirks,fornace2020unified}, a solid straight line between two different bases indicates a base pair formed between them, and a dashed line demarcates a region without implying that the connected bases are paired. (a) When $R>2$, (b) when $R=2$. Pseudocode implementation details are provided in \cref{algo:master}.}
	\label{fig:master}
\end{figure}

The master chef algorithm, in its new recursive scheme (\cref{fig:master}), incorporates results from our MFE algorithm~\cite{ShalabyWoods} related to central loops of symmetric secondary structures, see \cref{sec:loops}. This allows to separate different classes of secondary structures based on the type of their central loop, and how many nests per slice each of them has.
\Cref{fig:master} contains the  recursion diagrams, and their corresponding recursion equations, see \cref{eq:master1,eq:master2}. The proof of the following lemma and the pseudocode implementation of the master chef \cref{algo:master} are in \cref{app:omitted:PFMAalgo}.

\begin{restatable}[Master Chef]{lemma}{master} 
	\label{lem:master}
	
	Given any multi-stranded system of strands $\mathcal{S}$ with strand ordering $\pi$, magnification factor $\alpha$, and valid symmetry degree $R \geq 2$ such that $R|v(\pi)$:
	\begin{itemize}
		\item  if $R > 2$, \cref{eq:master1} computes  
		$\QalphaD(\pi,R)$ correctly. Formally speaking:
		\begin{align*}
			\overline{Z}^{\alpha,R}_{\mathcal{S}} = \sum\limits_{S \in \OmegaDRpi} e^{- \alpha  \DeltaGD(S)/k_\mathrm{B}T} = \QalphaD(\pi,R). 
		\end{align*}
		\item  if $R = 2$, \cref{eq:master2} computes  
		$\QalphaD(\pi,2)$ correctly. Formally speaking:
		\begin{align*}
			\overline{Z}^{\alpha,2}_{\mathcal{S}} = \sum\limits_{S \in \OmegaDtwopi} e^{- \alpha  \DeltaGD(S)/k_\mathrm{B}T} = \QalphaD(\pi,2).
		\end{align*} 
	\end{itemize}
	
	Furthermore, \cref{eq:master1,eq:master2} are implemented in the master chef \cref{algo:master}, that has asymptotic complexity of $\mathcal{O}(N^2)$ time, and $\mathcal{O}(1)$ space. 
	
\end{restatable}


\section{Peeling onions: Applications of the magnifiable partition function algorithm}\label{sec:Applications}

Beside computing the partition function (i.e. set $\alpha = 1$), in this section we explore wide range of applications of our magnifiable partition function algorithm in the nearest neighbour model, from both theoretical and practical perspectives. These applications range from purely combinatorial and counting applications to the ability to efficiently compute the equilibrium landscape using just a single call to our algorithm. In \cref{sec:app:count}, we prove that our magnifiable PF algorithm can be used to answer the fundamental counting questions~\cite{distinguishable} about the algorithmic and the physical ensembles $\overline{\Omega}$, $\Omega$ respectively. In \cref{sec:app:land}, we prove the two remaining main results of this work, \cref{thm:MFE} and \cref{thm:SSEL}.

\subsection{Combinatorial applications ($\alpha = 0$)}\label{sec:app:count}

\begin{theorem}\label[theorem]{thm:counting}
	Given any multi-stranded system $\mathcal{S}$ with strand ordering $\pi$, and symmetry degree $R$ such that $R|v(\pi)$. A single call to our magnification adaptable partition function \cref{algo:PFMA}, with $\alpha = 0$, can be used to compute all the following quantities correctly, in $\mathcal{O}(N^3)$ time and $\mathcal{O}(N^2)$ space:
	\begin{itemize}
		\item $|\Omega(\pi)|$, the number of secondary structures in the physical ensemble (i.e. distinguishable secondary structures). 
		
		\item $|\Omega_\text{max}(\pi,R)|$, the number of secondary structures with maximum degree of symmetry $R$ such that $R|v(\pi)$ in the physical ensemble. 
		
		\item $|\Omega(\pi,R)|$, the number of secondary structures with symmetry degree $R$, not necessarily the maximum, such that $R|v(\pi)$ in the physical ensemble. 
	\end{itemize}
	
	The corresponding quantities $|\overline{\Omega}(\pi)|,|\overline{\Omega}_\text{max}(\pi,R)|$, and $|\overline{\Omega}(\pi,R)|$ in the algorithmic ensemble could be computed also within the same call.
	
\end{theorem}

\begin{proof}
	The intuition here is simple: we call our magnifiable \pf  algorithm with  magnification factor $\alpha = 0$.  
	Hence, the energy contribution of each secondary structure $S \in \Omega(\pi)$ becomes $0$, and, since $e^0=1$, our partition function algorithm becomes a counting algorithm, such that $Q_0(\pi) = |\Omega(\pi)|$. 
	
	Since our algorithm is slice-based, it computes $Q_{\alpha, \text{max}}(\pi,R)$ for every $R|v(\pi)$, as a sub-module within the algorithm, see step 8 in \cref{algo:PFMA}, all we need is just extract such information. Hence $Q_{0, \text{max}}(\pi,R) = |\Omega_\text{max}(\pi,R)|$. Using \cref{identity2,identity1',identity2',identity:relation} to compute $|\Omega(\pi,R)|$, $|\overline{\Omega}(\pi)|$, $|\overline{\Omega}(\pi,R)|$, and $|\overline{\Omega}_\text{max}(\pi,R)|$ respectively.  
\end{proof}

\subsection{Taking a limit: as $\alpha \rightarrow \infty$, the magnified \pf $Q_\alpha\rightarrow $ MFE}\label{sec:lim alpha is MFE}

The next theorem\footnote{The first bullet point of \cref{thm:lim} is already known in statistical thermodynamics and is widely known as the zero or low temperature limit, which is less general than magnification since it  considers $T \rightarrow 0$ assuming temperature-independent energy models. The main proof technique used in \cref{thm:lim} is called `factoring out the ground state' (MFE in our case). This technique is also commonplace in statistical thermodynamics and optimization, i.e. Log-Sum-Exp operation (see section 6.2 of~\cite{yeomans1992statistical} or~\cite{vandenberghe2004convex,pathria2017statistical,tong2011statistical}).  GPT-4o mini was used to source the related literature, and formed a variant of this proof in late April 2026. \label{ft:gpt}}  states that when the parameter $\alpha \rightarrow \infty$, the contribution of MFE structures 
dominate the $\alpha$-magnified partition function. Later in \cref{sec:app:land} we will use  a bounded version of this limit to extract both the MFE value, and the number of MFE structures, from a single \pf call.  

\begin{theorem}\label{thm:lim}
	Given any multi-stranded system $\mathcal{S}$ with strand ordering $\pi$: 
	
	\begin{itemize}
		\item \refstepcounter{equation}\label{eq:MFElim} 
		$ \lim\limits_{\alpha \rightarrow \infty} \frac{-k_B T ~\log Q_\alpha(\pi)}{\alpha} = \mathrm{MFE}(\pi)$, {\normalfont i.e. the MFE value, and} \hfill {\normalfont (\theequation)}
		
		\item \refstepcounter{equation}\label{eq:SSELlim} 
		$ \lim\limits_{\alpha \rightarrow \infty}  \frac{Q_\alpha(\pi)}{\mathrm{e}^{-\alpha \cdot \mathrm{MFE}(\pi)/ k_\mathrm{B} T}}  = \mathrm{\#SSEL}\big(\mathrm{MFE}(\pi)\big)$,  {\normalfont i.e. the number of MFE structures.} \hfill {\normalfont (\theequation)}
	\end{itemize} 
\end{theorem}

\begin{proof}	
	Let $S^* \in \Omega(\pi)$ denote any structure with minimum free energy, i.e. $\Delta G(S^*) = \mathrm{MFE}(\pi)$.  
	We define a difference function $\delta: \Omega(\pi) \rightarrow \mathbb{R}_{\geq 0}$ so that for any $S \in \Omega(\pi)$, 
	$\Delta G(S) = \mathrm{MFE}(\pi) + \delta(S)$, where $\delta(S) \geq 0$. 
	Hence  $Q_\alpha(\pi)$ can be expressed as follows:  
	\begin{align}\label{eq:lim1}	
		Q_\alpha(\pi) =  \sum_{S \in \Omega(\pi)} \mathrm{e}^{- \alpha \big[\mathrm{MFE}(\pi) ~+~ \delta(S)\big] / k_\mathrm{B} T} = \mathrm{e}^{-\alpha \cdot \mathrm{MFE}(\pi)/ k_\mathrm{B} T} \sum_{S \in \Omega(\pi)} \mathrm{e}^{-\alpha \cdot \delta(S)/ k_\mathrm{B} T}
	\end{align}
	
	\noindent By applying the natural logarithm, and multiplying \cref{eq:lim1} by $\frac{- k_\mathrm{B} T}{\alpha}$, we get: 
	\begin{align}
		\frac{- k_\mathrm{B} T ~\log\left(Q_\alpha(\pi)\right)}{\alpha}
		= \mathrm{MFE}(\pi) - \frac{k_\mathrm{B} T ~\log\left( \sum_{S \in \Omega(\pi)} \mathrm{e}^{-\alpha \cdot \delta(S)/ k_\mathrm{B} T} \right)}{\alpha}
		\label{eq:MFE approx}
	\end{align}
	
	\noindent and taking the limit $\alpha \rightarrow \infty$:  
	\begin{align*}
		\displaystyle \lim\limits_{\alpha \rightarrow \infty} \frac{- k_\mathrm{B} T ~\log \left(Q_\alpha(\pi)\right)}{\alpha}
		=  \mathrm{MFE}(\pi) - k_\mathrm{B} T  \displaystyle \lim\limits_{\alpha \rightarrow \infty} \frac{\log \left( \sum_{S \in \Omega(\pi)} \mathrm{e}^{-\alpha \cdot \delta(S)/ k_\mathrm{B} T} \right)}{\alpha}
	\end{align*}
	
	Our goal is to sandwich the right-hand side limit  between lower and upper bounds that converge to $0$ as $\alpha \rightarrow \infty$, yielding the first claim of the theorem statement. 
	For the lower bound:  $S^* \in \Omega(\pi)$, hence $\delta(S^*) = 0$ and $\mathrm{e}^{-\alpha \cdot \delta(S^*)/ k_\mathrm{B} T} = 1$. 
	The exponential function is strictly positive for all real numbers, hence $\sum_{S \in \Omega(\pi)} \mathrm{e}^{-\alpha \cdot \delta(S)/ k_\mathrm{B} T} \geq 1$, and hence:  $0 \leq \log \left( \sum_{S \in \Omega(\pi)} \mathrm{e}^{-\alpha \cdot \delta(S)/ k_\mathrm{B} T} \right)$.
	For the upper bound: $\delta(S) \geq 0$, and the constants $ k_\mathrm{B}$ and $T$ (in Kelvin) are both positive, hence $ \mathrm{e}^{-\alpha \cdot \delta(S)/ k_\mathrm{B} T} \leq 1$ and $\sum_{S \in \Omega(\pi)} \mathrm{e}^{-\alpha \cdot \delta(S)/ k_\mathrm{B} T} \leq |\Omega(\pi)| $, hence we get our upper bound:  $ \log \left( \sum_{S \in \Omega(\pi)} \mathrm{e}^{-\alpha \cdot \delta(S)/ k_\mathrm{B} T} \right) \leq \log |\Omega(\pi)|$.
	Since $\lim\limits_{\alpha \rightarrow \infty} \log |\Omega(\pi)| / \alpha = 0$, and  applying the sandwich theorem~\cite{bartle2000introduction,rudin1976principles} implies that: $ \lim\limits_{\alpha \rightarrow \infty} \frac{\log \left( \sum_{S \in \Omega(\pi)} \mathrm{e}^{-\alpha \cdot \delta(S)/ k_\mathrm{B} T} \right)}{\alpha} = 0$, proving our first claim: $ \lim\limits_{\alpha \rightarrow \infty} \frac{-k_B T ~\log \left(Q_\alpha(\pi)\right)}{\alpha} = \mathrm{MFE}(\pi)$.

	For our second claim, let's partition $\Omega(\pi)$ into two disjoint sets:
	\begin{itemize}
		\item The set $\mathcal{F}$ of all secondary structures $S$ such that $\Delta G(S) = \mathrm{MFE}(\pi)$. Note, $|\mathcal{F}| = \mathrm{\#SSEL}\big(\mathrm{MFE}(\pi)\big)$.
		\item The set $\mathcal{F}'$ of all other secondary structures $S$ such that $\Delta G(S) > \mathrm{MFE}(\pi)$.
	\end{itemize}
	$Q_\alpha(\pi)$ can be written as follows:
	\begin{align*}
		Q_\alpha(\pi) = \mathrm{\#SSEL}\big(\mathrm{MFE}(\pi)\big) \cdot \mathrm{e}^{- \alpha \cdot \mathrm{MFE}(\pi)  / k_\mathrm{B} T} + \sum_{S \in {\mathcal{F}'} }  \mathrm{e}^{- \alpha \big[\mathrm{MFE}(\pi) + \delta(S)\big] / k_\mathrm{B} T}, \text{ hence:}
	\end{align*}
	\begin{align} 
		\frac{Q_\alpha(\pi)}{\mathrm{e}^{-\alpha \cdot \mathrm{MFE}(\pi)/ k_\mathrm{B} T}}  = \mathrm{\#SSEL}\big(\mathrm{MFE}(\pi)\big) +  \sum_{S \in \mathcal{F}'}  \mathrm{e}^{- \alpha \cdot \delta(S) / k_\mathrm{B} T}.
		\label{eq:SSEL}
	\end{align}
	\noindent By taking the limit, when $\alpha \rightarrow \infty$, of \cref{eq:SSEL}, we get the following:
	
	\begin{align*} 
		\displaystyle \lim\limits_{\alpha \rightarrow \infty}  \frac{Q_\alpha(\pi)}{\mathrm{e}^{-\alpha \cdot \mathrm{MFE}(\pi)/ k_\mathrm{B} T}}  = \mathrm{\#SSEL}\big(\mathrm{MFE}(\pi)\big) +  \sum_{S \in \mathcal{F}'} \displaystyle \lim\limits_{\alpha \rightarrow \infty} \mathrm{e}^{- \alpha \cdot \delta(S) / k_\mathrm{B} T}.
	\end{align*}
	
	Since $|\mathcal{F}'| < |\Omega(\pi)| =  \mathcal{O}(1)$, and $\delta(S) \geq 0$ by definition, then $\sum_{S \in \mathcal{F}'} \displaystyle \lim\limits_{\alpha \rightarrow \infty} \mathrm{e}^{- \alpha \cdot \delta(S) / k_\mathrm{B} T} = 0$, proving our second claim: $\displaystyle \lim\limits_{\alpha \rightarrow \infty}  \frac{Q_\alpha(\pi)}{\mathrm{e}^{-\alpha \cdot \mathrm{MFE}(\pi)/ k_\mathrm{B} T}}  = \mathrm{\#SSEL}\big(\mathrm{MFE}(\pi)\big)$.
\end{proof}

\subsection{Numerical Considerations} 

Throughout algorithmic work on the nearest neighbour model, there is seldom discussion about precision of real numbers in algorithms, perhaps for two reasons: 
(a) The free energies for some hardcoded features (e.g. stacks) in the nearest neighbour model are experimentally given to one or two decimal places~\cite{santa,santalucia1998unified,mathews1999expanded,jaeger1989improved,walter1994coaxial}, and although single-stranded loop energy scales logarithmically, precision beyond a few decimal places is somewhat meaningless experimentally.
And (b) many algorithmic results have the flavour that better precision in the algorithm's variables simply means a smooth, quantitive improvement in accuracy, i.e. no qualitative change to the result.

However,  software implementations have specific precision~\cite{viennaRNA,NUPACK}. 
And, in this section we will assume constant, but arbitrary, precision, independent of system length (number of bases) $N$.

\begin{remark}[Precision assumption]\label[remark]{rem:precision}\d
	Throughout this work we assume the existence of a minimum gap of $\varepsilon > 0$ between any two energy levels in the nearest neighbour model. 
	Specifically, $\varepsilon = \left(\frac{1}{10}\right)^k$ for some constant natural number $k \geq 1$, meaning $k$ is independent of the number of bases (system size) $N$. In other words, for any two secondary structures $S_i, S_j \in \Omega$, $|\Delta G(S_i) - \Delta G(S_j)| = n \cdot \varepsilon$ for some $n \geq 0$. 
\end{remark}

This assumption is used in previous works~\cite{cupal1996dynamic,viennaRNA,ducloz,NUPACK,cupal1997density,lou2010thermodynamics}. For example, ViennaRNA\footnote{ViennaRNA~\cite{viennaRNA,flamm2000rna,flamm2002barrier,wolfinger2004efficient,wuchty1999complete,cupal1997density} is a widely used open-source software package designed for predicting and analyzing RNA secondary structures. It features algorithms for computing the MFE, partition function, density of states, folding kinetics, and sequence design.} assumes $\varepsilon = 0.01$ kcal/mol~\cite{viennaRNA}.
Using this assumption, we get the following lemma: 
\begin{lemma}\label[lemma]{lem:NNlinear}
	In the nearest neighbour energy model, for any multi-stranded system $\mathcal{S}$ of total length $N$. Let $\mathcal{G} = \{\Delta G(S) \mid  S \in \Omega\}$, the set of {\em occupied} real valued free energies. We claim that $|\mathcal{G}| = \mathcal{O}(N)$, i.e. linear number of energy levels.
\end{lemma}

\begin{proof}
	In the worst case, any secondary structure free energy is a sum of $\lfloor \frac{N}{2} \rfloor = \mathcal{O}(N)$ loop free energies. 
	The contribution of any loop is
	$\mathcal{O}(1)$ or $\mathcal{O}(N)$ size using the precision assumption and loop type. In the worst case, any secondary structure is composed of $\mathcal{O}(N)$ loops of $\mathcal{O}(1)$ size and $\mathcal{O}(1)$ loops of $\mathcal{O}(N)$ size.	
	This implies that for any secondary structure $S \in \Omega$,  the free energy $\Delta G(S) \in \mathcal{G}$ grows linearly, i.e. $\Delta G(S) = \mathcal{O}(N)$. Hence, $|\mathcal{G}| = \mathcal{O}(N)$.
\end{proof}

\subsection{Even more efficient MFE, \#SSEL, and occupied energy levels algorithms}\label{sec:app:land}

Having these numerical considerations in mind, instead of having $\alpha \rightarrow \infty$ as in \cref{thm:lim}, the goal of \cref{lem:dom} is to 
choose a finite $\alpha$ to achieve similar results to \cref{thm:lim}. 

\begin{note}\label[note]{note:num structures}
	For a single RNA/DNA strand, Stein and Waterman~\cite{stein1979some} showed that the asymptotic number of RNA/DNA secondary structures\footnote{Note that the upper bound $\mathcal{U}_N$ is with the restriction that each hairpin loop has at least one free base.} is upper bounded by $\mathcal{U}_N = 1.104366 \cdot N^{-3/2} \cdot 2.618034^{N}$, where $N$ is the number of bases. The same upper bound holds for multiple strands given a strand ordering $\pi$ (in fact, a slightly lower one as multi-stranded systems require at least some base pairs to maintain connected structures).   
	For the sake of simplicity of our analysis, we use $e^N$ as an upper bound, since $\mathcal{U}_N < e^N$ for all $N \geq 2$.
	Note that, these upper bounds are system-independent, hence not tight. 
	One of our results is an algorithm that gives the the exact number of structures $|\Omega(\pi)|$ for a multi-stranded system, 
	achieved simply by setting $\alpha=0$, call our magnifiable \pf algorithm to get $Q_0(\pi)$, and noting that $|\Omega(\pi)| = Q_0(\pi)$, 
	see \cref{thm:counting}. Note that $e^N > |\Omega(\pi)|$ for all $N$.
\end{note}

The goal of \cref{lem:dom} is to magnify secondary structure energies so those at the any energy level $L$ strictly dominate over all at energy levels $L'>L$, in the contribution to the $\alpha$-magnified partition function, regardless of the secondary structures distribution over the energy landscape.

\begin{lemma}[Hierarchical dominance of energy levels]\label[lemma]{lem:dom}
	Let $\mathcal{S}$ be a multi-stranded system  of length $N$, 
	strand ordering $\pi$ and gaps  $\geq \varepsilon = (\frac{1}{10})^k$ between energy levels  (\cref{rem:precision}). 
	For any $\alpha \geq k_\mathrm{B} T N / \varepsilon$, and for any occupied energy level $L$, we claim the following:  
	\begin{align}
		\sum_{ S \text{ s.t. } \Delta G(S) = L}    e^{-\alpha \Delta G(S)/k_\mathrm{B}T} > 
		\sum_{L'>L}~\sum_{ S' \text{ s.t. } \Delta G(S') = L'} e^{-\alpha \Delta G(S')/k_\mathrm{B}T}
	\end{align}   
\end{lemma}

\begin{proof}
	
	To prove our claim, we satisfy the worst case scenario of the distribution of secondary structures over the energy landscape which is as follows: 
	\begin{itemize}
		\item There exist exactly one secondary structure $S$ belonging to energy level $L$.
		\vspace{0.25em}
		\item Every other secondary structure of the system $\mathcal{S}$, belongs to the closest possible energy level $L'$ above $L$, i.e. $L'= L + \varepsilon$. 
		Even worse, we assume that $e^N$ secondary structures belong to $L'$ ($e^N$ is a loose upper bound on the total number of secondary structures, see~\cref{note:num structures}).   
	\end{itemize}
	\noindent Assume $\alpha = k_\mathrm{B}TN/\varepsilon$, and consider $L$:
	\begin{align*}
		\sum_{ S \text{ s.t. } \Delta G(S) = L} e^{-\alpha \Delta G(S)/k_\mathrm{B}T} & = \sum_{ S \text{ s.t. } \Delta G(S) = L} e^{- N  L/\varepsilon} & \\
		& \geq e^{- N L/\varepsilon} & \text{ $\geq 1$  structure at level $L$}
		\\
		& = e^{- N (L+\varepsilon- \varepsilon)/\varepsilon}
		\\
		&=  e^{ N} e^{- N (L+\varepsilon) / \varepsilon}   & 
		\\
		& > \sum_{L'>L}~\sum_{ S' \text{ s.t. } \Delta G(S') = L'} e^{-\alpha \Delta G(S')/k_\mathrm{B}T} 
	\end{align*}
	since $< e^N$ secondary structures have energy  $L'$ since $e^N$ is an upper bound on the number of secondary structures, 
	which proves our claim.
\end{proof}

\Cref{lem:dom} will allow us to unpeel the entire energy landscape (see \cref{thm:SSEL,algo:onion}) like an \textit{onion}, where we can peel  layers, one by one starting at the MFE level/layer and working upwards. 

\subsubsection{MFE}\label{sec:MFE from finite alpha theorem and algorithm}

\MFE*

\begin{proof}
	For any $\pi \in \Pi$, recall \cref{eq:MFE approx}  derived in the proof of \cref{thm:lim}:      
	\begin{align*}
		\frac{- k_\mathrm{B} T ~\ln\left(Q_\alpha(\pi)\right)}{\alpha}
		= \mathrm{MFE}(\pi) - \frac{k_\mathrm{B} T ~\ln\left( \sum_{S \in \Omega(\pi)} \mathrm{e}^{-\alpha \cdot \delta(S)/ k_\mathrm{B} T} \right)}{\alpha}
		\tag*{(\ref{eq:MFE approx})}
	\end{align*}
	
	\noindent Here, we call $0 \leq \theta(\alpha) =\frac{k_\mathrm{B} T ~\ln\left( \sum_{S \in \Omega(\pi)} \mathrm{e}^{-\alpha \cdot \delta(S)/ k_\mathrm{B} T} \right)}{\alpha}$,  the $\mathrm{MFE}(\pi)$ approximation error. Note that $\theta(\alpha)$ is a monotonically decreasing function (a fact derived in the proof of \cref{thm:lim} but straightforward to derive). 
	
	Our goal  is to show  $\theta(\alpha) < \varepsilon$ when $\alpha \geq k_B T N / \varepsilon$. In other words, the $\mathrm{MFE}(\pi)$ approximation error is strictly less than the minimum  gap between any two energy levels, and hence we can extract $\mathrm{MFE}(\pi)$  from $Q_\alpha(\pi)$.
	\vspace{0.5em}
	
	\noindent Assume $\alpha = k_B T N / \varepsilon$: 
	\begin{align*}
		\theta(\alpha) & =  \frac{\varepsilon}{N} ~\log \sum_{S \in \Omega(\pi)} \mathrm{e}^{- N \cdot \delta(S)/ \varepsilon}
		\\
		&= 	\frac{\varepsilon}{N} ~\log \Big[ \sum_{S \in \mathcal{F}} \mathrm{e}^{- N\cdot \delta(S)/ \varepsilon} + \sum_{S' \in \mathcal{F}'} \mathrm{e}^{- N\cdot \delta(S')/ \varepsilon} \Big]  &  \text{partitioning of } \Omega(\pi) \text{ in \cref{thm:lim} proof}\\
		&= \frac{\varepsilon}{N} ~\log \Big[ \mathrm{\#SSEL}\big(\mathrm{MFE}(\pi)\big) + \sum_{S' \in \mathcal{F}'} \mathrm{e}^{- N\cdot \delta(S')/ \varepsilon} \Big]  & \text{$\delta(S) = 0$} \\
		& \leq \frac{\varepsilon}{N} ~\log \Big[ \mathrm{\#SSEL}\big(\mathrm{MFE}(\pi)\big) + \sum_{S' \in \mathcal{F}'} \mathrm{e}^{- N} \Big] &  \delta(S') = \varepsilon, \text{ the worst candidate} \\
		& < \frac{\varepsilon}{N} ~\log \Big[ \mathrm{\#SSEL}\big(\mathrm{MFE}(\pi)\big) + \mathrm{e}^{N} \mathrm{e}^{- N} \Big] & \text{since } |\mathcal{F}'| < e^N \text{, \cref{note:num structures}} \\
		& < \frac{\varepsilon}{N} ~\log(\mathrm{e}^{N}) & \text{since } \mathrm{\#SSEL}\big(\mathrm{MFE}(\pi)\big) + 1 < e^N \\
		& = \varepsilon	
	\end{align*}
	
	This proves that $0 \leq \theta(\alpha) < \varepsilon$ when $\alpha = k_B T N / \varepsilon$. Hence the following equation represents an $\mathrm{MFE}(\pi)$ prediction algorithm that runs in $\mathcal{O}(N^3)$ time and $\mathcal{O}(N^2)$ space, where rescaling by $1/\varepsilon$ and the ceiling operator remove the approximation error which we know to be $< \varepsilon$:  
	\begin{align}
		\mathrm{MFE}(\pi) = \bigg\lceil \frac{- k_\mathrm{B} T ~\log Q_\alpha(\pi)}{\alpha} \cdot \frac{1}{\varepsilon} \bigg\rceil \cdot \varepsilon & \hspace{8em} \text{where } \alpha \geq k_B T N / \varepsilon
	\end{align} 
	
	Since $|\Pi| \leq (c-1)!$, \cref{algo:MFE} predicts the MFE value in $\mathcal{O}(N^3 (c-1)!)$ time and $\mathcal{O}(N^2)$ space, proving our claim.
\end{proof}

\begin{algorithm}[H] 
	\caption{MFE algorithm for the nearest neighbour energy model, that takes as input: $c=\mathcal{O}(1)$ strands with total number of bases (length) $N$. 
		Runs in  $\mathcal{O}(N^3(c-1)!)$ time and $\mathcal{O}(N^2)$ space.} 
	\label{algo:MFE}
	\begin{algorithmic}[1]
		\State \textbf{Let } $\text{MFE} = +\infty$
		\For{$\pi \in \Pi$}

		\State \textbf{Let  $Z = Q_\alpha(\pi)$ for  $\alpha = k_B T N / \varepsilon$}		
		\Comment{Using \cref{algo:PFMA}}
		
		\State $
		\mathrm{MFE}(\pi) = \bigg\lceil \frac{- k_\mathrm{B} T ~\log Z}{\alpha} \cdot \frac{1}{\varepsilon} \bigg\rceil \cdot \varepsilon$
		
		\If{$\text{MFE}(\pi) < \text{MFE}$}
		\State $\text{MFE} = \text{MFE}(\pi)$
		\EndIf

		\EndFor
		\State \textbf{Return } MFE
	\end{algorithmic}
\end{algorithm}

\begin{remark}\label[remark]{remark:MFE value}
	The main limitation of this new efficient algorithm for MFE is that it computes the MFE value but not the corresponding MFE secondary structure.
\end{remark}

\subsubsection{\#SSEL and the set of occupied energy levels}\label{sec:SSEL occupied}

Next, for a $\pi \in \Pi$, we seek an algorithm that  returns the  number of secondary structures that have MFE($\pi$) value. 

\begin{theorem}[\#SSEL(MFE($\pi$)) Algorithm] \label{th:SSELMFEalgo}
	Let $\mathcal{S}$ be a multi-stranded system of length $N$, 
	strand ordering $\pi$ 
	and a minimum  gap of $\geq \varepsilon > 0$ between energy levels  (\cref{rem:precision}).
	There is an $\mathcal{O}(N^3)$ time, $\mathcal{O}(N^2)$ space algorithm to predict $\mathrm{\#SSEL}\big(\mathrm{MFE}(\pi)\big)$, i.e. the number of {\normalfont MFE} structures with ordering $\pi$, for the nearest neighbour model. 
\end{theorem}

\begin{proof}
	For any $\pi \in \Pi$, recall \cref{eq:SSEL}  derived in the proof of \cref{thm:lim}:           
	\begin{align*}
		\frac{Q_\alpha(\pi)}{\mathrm{e}^{-\alpha \cdot \mathrm{MFE}(\pi)/ k_\mathrm{B} T}}  = \mathrm{\#SSEL}\big(\mathrm{MFE}(\pi)\big) +  \sum_{S \in \mathcal{F}'}  \mathrm{e}^{- \alpha \cdot \delta(S) / k_\mathrm{B} T}
		\tag*{(\ref{eq:SSEL})}
	\end{align*}

	\noindent where $0 \leq \vartheta(\alpha) = \sum_{S \in \mathcal{F}'}  \mathrm{e}^{- \alpha \cdot \delta(S) / k_\mathrm{B} T}$, is called the approximation error on the  number of {\normalfont MFE}$(\pi)$ structures. Note that $\vartheta(\alpha)$ is a monotonically decreasing function.

	Our goal is to show that $\vartheta(\alpha) < 1$ when $\alpha = k_B T N / \varepsilon$. In other words,  the approximation error on the  number of {\normalfont MFE}$(\pi)$ structures is strictly less than one, and hence we can use a simple floor function to obtain $\mathrm{\#SSEL}\big(\mathrm{MFE}(\pi)\big)$.
	\vspace{0.5em}
	
	\noindent Assume $\alpha = k_B T N / \varepsilon$: 
	\begin{align*}
		\vartheta(\alpha) & =  \sum_{S \in \mathcal{F}'}  \mathrm{e}^{- N \cdot \delta(S) / \varepsilon} \\
		& \leq \sum_{S \in \mathcal{F}'}  \mathrm{e}^{- N}  &  \delta(S) = \varepsilon, \text{ the worst candidate} \\
		& < e^N \cdot e^{-N} =  1 & \text{since } |\mathcal{F}'| < e^N \text{, \cref{note:num structures}}
	\end{align*}
	
	This proves that $0 \leq \vartheta(\alpha) < 1$ when $\alpha = k_B T N / \varepsilon$. Hence the following equation represents an algorithm, that calls our magnifiable \pf once (with a single $\alpha$ value) and uses a simple floor, that computes  the number of {\normalfont MFE}$(\pi)$ secondary structures, i.e. $\mathrm{\#SSEL}\big(\mathrm{MFE}(\pi)\big)$, and runs in $\mathcal{O}(N^3)$ time and $\mathcal{O}(N^2)$ space.  
	\begin{align}
		\mathrm{\#SSEL}\big(\mathrm{MFE}(\pi)\big) = \bigg\lfloor 	\frac{Q_\alpha(\pi)}{\mathrm{e}^{-\alpha \cdot \mathrm{MFE}(\pi)/ k_\mathrm{B} T}}   \bigg\rfloor, & \hspace{8em} \text{where } \alpha = k_B T N / \varepsilon
	\end{align}
\end{proof}

For a strand ordering $\pi \in \Pi$, in the previous proof we peeled away the $\text{MFE}(\pi)$  energy level. Next, we move on and peel the next occupied energy level above MFE($\pi$), and by keeping iterating this process we efficiently compute the full energy landscape in \cref{algo:onion}. The correctness proof of \cref{thm:SSEL} and \cref{algo:onion} are in \cref{app:omitted:app}.

\SSEL*

\begin{note}
	We should note that \cref{algo:onion} peels energy levels in ascending order starting from the optimal MFE$(\pi)$ level, 
	but also could be adapted to run in descending order by using $\alpha = -k_\mathrm{B}TN/ \varepsilon$, where the negative sign reverses the energy landscape without affecting the relative energies nor its distribution.  
\end{note}

\subsubsection{Revealing the energy landscape at a higher resolution (rotational symmetry)}
Since our magnifiable partition function algorithm has a slice-based approach, and it computes $Q_{\alpha, \text{max}}(\pi,R)$ for every $R$ such that $R|v(\pi)$, as a sub-module within the algorithm, see step 8 in \cref{algo:PFMA}. 
We can again benefit from our algorithmic slice-based construction, and run the onion peeling \cref{algo:onion} on each of $Q_{\alpha, \text{max}}(\pi,R)$ for all $R$'s separately. This leads to  higher resolution information about the energy landscape of any system $\mathcal{S}$. 
For example, for any $\pi \in \Pi$, with a single call to our magnifiable algorithm we can answer a question like:
\begin{itemize}
	\item How many structures $S \in \Omega(\pi)$ that belong to energy level $L$, such that $R_\text{max}(S) = R$? 
\end{itemize}
This leads to \cref{lem:high}, the higher resolution version of \cref{thm:SSEL}:
\begin{theorem}[Onion Peeling: High Resolution Version]\label[theorem]{lem:high}
	Given any multi-stranded system $\mathcal{S}$ of total length $N$, with strand ordering $\pi$, symmetry degree $R$ such that $R|v(\pi)$, energy level $L$.  There exists an $\mathcal{O}(N^3)$ time, $\mathcal{O}(N^2)$ space algorithm to predict $\mathrm{\#SSEL}(L,\pi,R)$, i.e. the number of secondary structures $S \in \Omega(\pi)$ at energy level $L$ with maximum symmetry degree $R$, i.e. $\Delta G(S) = L$, and $R_\text{max}(S) = R$. 
\end{theorem}

\subsection{One partition function call to rule them all}\label{sec:one call}

We showed different useful applications for our new magnifiable partition function algorithm. These applications range from (usual) partition function calculation, using $\alpha = 1$, counting applications, using $\alpha = 0$, or computing important properties of the equilibrium energy landscape, using $\alpha = k_\mathrm{B}TN/\varepsilon$. Here we go one step further and show that a single call with $\alpha = k_\mathrm{B}TN/\varepsilon = \Theta(N)$, gives all these results.  

We already showed one call to our magnifiable partition function algorithm $Q_\alpha(\pi)$, with $\alpha \geq k_\mathrm{B}TN/\varepsilon$,  
computes the set of occupied energy levels $\mathcal{G}$ and with the number of secondary structures per level. 
We can compute the (usual) partition function $Q=Q_1(\pi)$ from that information, since:
\begin{align*}
	Q_1(\pi) = \sum_{1 \leq i \leq |\mathcal{G}|} \mathrm{\#SSEL}(g_i) \cdot e^{- g_i/k_\mathrm{B} T}
\end{align*}  
where $g_i \in \mathbb{R}$ is the free energy of the $i^\mathrm{th}$ energy level that is occupied by at least one  secondary structure. 
In other words, the partition function can be computed as a sum of products, 
of an energy $g_i$ multiplied by the number of structures $\mathrm{\#SSEL}(g_i)$ with energy $g_i$.

And since we established a high resolution version of our onion peeling \cref{algo:onion}, where we have the full equilibrium landscape over $\Omega_\text{max}(\pi,R)$, we can achieve all combinatorial quantities that we achieved in \cref{thm:counting}.
Hence, just one magnifiable partition function call with $\alpha = k_\mathrm{B}TN/\varepsilon$ is enough.

\subsubsection*{Declaration of AI Use}
None of the theorems or results were proven, and none of the text was written or edited, by generative AI/large language models (LLMs). 
In one instance an LLM (GPT‑4o mini) gave, in April 2026, the proof intuition of Case 1 of \cref{thm:lim}, a technique  already known in the statistical thermodynamics literature as documented in Footnote 22 in \cref{sec:Applications}.  LLMs were used to help LaTeX code for the blue boxes of \cref{strategy1} and \cref{strategy2}, and to draw tasty geometric shapes, specifically: \pizza, \slice, \templateslice, and \emptyslice.

\subsubsection*{Acknowledgments}
We thank Constantine Evans for his helpful comments on the statistical mechanics origin of the rotational symmetry penalty.
We dedicate this paper to the memory of Robert Dirks. 

\bibliography{bib}

\newpage
\appendix

\section{Appendix: Distinguishability complications: the partition function algorithm of Dirks et al.}\label{app:dirks}

As briefly mentioned earlier in \cref{sec:why}, the Dirks et al.\ partition function algorithm~\cite{dirks} computes $\overline{Q}(\pi)$ first, then post-process $\overline{Q}(\pi)$ by dividing by $v(\pi)$ as follows:

\begin{align*}
	\frac{\QD(\pi)}{v(\pi)}  &=  \sum_{S \in \OmegaDpi} \frac{1}{	v(\pi)} ~ \mathrm{e}^{- \DeltaGD(S) / k_\mathrm{B} T}  \\
	& =   \sum_{S \in \OmegaDpi} \frac{1}{ 	|\text{Orb}(S)| \cdot |\text{Stab}(S)|} ~ \mathrm{e}^{- \DeltaGD(S) / k_\mathrm{B} T} &\text{Using \cref{eq:orbit}}  \\
	& =  \sum_{S \in \OmegaIpi} \mathrm{e}^{- \log R_\text{max}(S)} ~  \mathrm{e}^{- \DeltaGD(S) / k_\mathrm{B} T}  &\text{Since $|\text{Stab}(S)| = R_\text{max}(S)$}  
	\\
	&=  \sum_{S \in \OmegaIpi}  \mathrm{e}^{-\left[ \DeltaGD(S) + k_\mathrm{B} T \log R_\text{max}(S) \right]/ k_\mathrm{B} T} \\
	&= \sum_{S \in \OmegaIpi}  \mathrm{e}^{- \DeltaGI(S) / k_\mathrm{B} T} = \QI(\pi)
\end{align*}

This division by $v(\pi) = 	|\text{Orb}(S)| \cdot R_\text{max}(S)$, achieves the following simultaneously:

\begin{itemize}
	\item Cancels out overcounting by correcting the underlying state space from $\OmegaDpi$ to $\OmegaIpi$. This is achieved when divided by $|\text{Orb}(S)|$, since there are $|\text{Orb}(S)|$ indistinguishable secondary structures within $\overline{\Omega}(\pi)$ that are represented as a single structure in $\Omega(\pi)$, see \cref{sec:algebra}.
	
	\item Corrects the underlying free energy model from $\DeltaGD(S)$ to $\DeltaGI(S)$, this is achieved when divided by $|\text{Stab}(S)| = R_\text{max}(S)$. As a consequence the entropic penalty for rotational symmetry~\cite{bailey1978statistical}, got included, leading to $\QI(\pi)$.
\end{itemize}

Dirks at al.'s algorithm benefits hugely from two things: (1) The algebraic relationship between $\OmegaDpi$ and $\OmegaIpi$, see \cref{sec:algebra}, and (2) The natural exponential nature of the weighted Boltzmann distribution of the partition function, otherwise the previous post-processing algebraic link will never work.    We believe the simultaneous handling of Dirks at al.'s~\cite{dirks} algorithm of those two problems of turns out to be its main obstacle for being a magnifiable partition function algorithm.

\newpage
\section{Appendix: Omitted proofs and algorithms}\label{app:proofs}

\subsection{Proofs omitted from \cref{sec:prosym}}
\label{app:omitted:prosym}  

\cutexist*
\begin{proof}
	Given any covalent bond $b = [a,a+1]$ on the path of the circumference of $\PolySpi$ from  $A$ to $B$, we claim $b$ is an admissible $R$-symmetric backbone cut generator, 
	otherwise there exists a base pair $(m,n) \in \PolySpi$ such that $b \in \llbracket m,n \rrbracket$, 
	giving a contradiction by any of the following cases: 
	\begin{itemize}%
		\item $\llbracket m,n \rrbracket$ must contain either $\llbracket i,j \rrbracket$ or $\llbracket k,l \rrbracket$ or both which contradicts the fact that $(i,j)$ and $(k,l)$ are the exteriors of $A$ and $B$ respectively.
		
		\item $( m,n )$ intersect with either $( i,j )$ or $( k,l )$ or both  forming a pseudoknot.
		
		\item $m$ and $n$ bases belong to the path of the circumference of $\PolySpi$ between $A$ and $B$ contradicting the fact that $A$ and $B$ are consecutive nests. 
	\end{itemize}
	All covalent bonds in $\bigotimes_R^{b}$  have the same situation because of the $R$-fold symmetry of $S$, 
	which implies that $\bigotimes_R^{b}$ is an admissible $R$-symmetric backbone cut of $S$.
\end{proof}

\subsection{Proofs and algorithms omitted from \cref{sec:PFMAalgo}}
\label{app:omitted:PFMAalgo}

\begin{algorithm}[H] 

	\caption{\small The assistant chef algorithm pseudocode (see \cref{lem:assistant}), that  
		runs in  $\mathcal{O}(N^3)$ time and $\mathcal{O}(N^2)$ space with
		recursive calls illustrated in \cref{fig:assischef}. Following the same notations of~\cite{dirks}, nicks between strands are denoted by half indices (e.g.~$x+ \frac{1}{2}$). 
		The function $\eta[i+ \frac{1}{2}, j+\frac{1}{2}]$ returns the number of nicks in the interval $[i+ \frac{1}{2}, j+\frac{1}{2}]$. 
		The shorthand $\eta[i+ \frac{1}{2}]$ is equivalent to $\eta[i+ \frac{1}{2}, i+\frac{1}{2}]$ and by convention, $\eta[i+ \frac{1}{2}, i-\frac{1}{2}] =0$.
	} \label{algo:assischef}
	\begin{algorithmic}[1]
		\footnotesize
		\State Initialize $Q^{\alpha,R}, Q^{\text{bp},\alpha,R}, Q^{m,\alpha,R}$ by setting all values to $0$, except $Q^{\alpha,R}_{i,i-1} = Q^{\alpha,R}_{i,i} = 1$  for all $i=1,\ldots,N$
		
		\For{$l \gets 1 \ldots N/R$}
		\For{$i \gets 1 \ldots  \frac{2N}{R}-l+1$}\Comment{If $R = 1$, let $i \gets 1 \ldots  N-l+1$}
		\State $j = i+l-1$
		
		\Comment{$Q^{\text{bp},\alpha,R}$ recursion equations} 
		
		\Comment{hairpin loop requires no nicks} 
		\If{$\eta[i+\frac{1}{2}, j-\frac{1}{2}] ==0$}
		$Q_{i,j}^{\text{bp},\alpha,R} =    \exp \{\alpha R \cdot \Delta G_{i,j}^\text{hairpin} / k_B T\}$

		\EndIf 
		
		\For{$d \gets i+1 \ldots j-2$}
		\Comment{loop over all possible $3'$-most pairs $(d,e)$} 
		
		\For{$e \gets d+1 \ldots j-1$} 
		
		\If{$\eta[i+\frac{1}{2}, d-\frac{1}{2}] ==0$ and $\eta[e+\frac{1}{2}, j-\frac{1}{2}] ==0$}
		
		\State 	$Q_{i,j}^{\text{bp},\alpha,R} +=  \exp \{ \alpha R \cdot \Delta G^\mathrm{interior}_{i,d,e,j} /k_B T \} Q_{d,e}^{\text{bp},\alpha,R}$

		\EndIf 
		
		\Comment{multiloop: no nicks}
		\If{$\eta[e+\frac{1}{2}, j-\frac{1}{2}] ==0$ \textbf{and} $\eta[i+\frac{1}{2}] ==0$ \textbf{and} $\eta[d-\frac{1}{2}] ==0$}

		\State 		
		$
		\begin{aligned}
			Q_{i,j}^{\text{bp},\alpha,R} += Q_{i+1,d-1}^{m,\alpha,R}   Q_{d,e}^{\text{bp},\alpha,R}  \exp \big\{ &- \alpha R  [\Delta G^\mathrm{multi}_\mathrm{init} + 2 \Delta G^\mathrm{multi}_\mathrm{bp} \\
			&+ (j-e-1) \Delta G^\mathrm{multi}_\text{nt}] / k_B T \big\}
		\end{aligned}$

		\EndIf
		\EndFor		
		\EndFor

		\For{$x \in \{i, \ldots,j-1\}$ s.t. $\eta[x+\frac{1}{2}] = 1$} 
		\Comment{loop over all nicks $\in [i+\frac{1}{2}, j-\frac{1}{2}]$} 
		
		\If{($\eta[i+\frac{1}{2}] == 0$  \textbf{and} $\eta[j-\frac{1}{2}] == 0$)  \textbf{or} ($i==j-1$)  \textbf{or}  ($x==i$  \textbf{and} $\eta[j-\frac{1}{2}] == 0$)  \textbf{or}
			\par \hskip\algorithmicindent  
			\hspace{6 mm} ($x==j-1$  \textbf{and} $\eta[i+\frac{1}{2}] == 0$)}
		
		\State  $Q_{i,j}^{\text{bp},\alpha,R} += Q_{i+1,c}^{\alpha,R} ~  Q_{c+1,j-1}^{\alpha,R} $\Comment{exterior loops} 
		
		\EndIf
		\EndFor

		\Statex
		\Comment{$Q^{\alpha,R}, Q^{m,\alpha,R}$ recursion equations} 
		
		\If{$\eta[i+\frac{1}{2}, j-\frac{1}{2}] == 0$} 
		$Q_{i,j}^{\alpha,R} = 1$\Comment{empty substructure}
		\Else \hspace{0.1cm}{$Q_{i,j}^{\alpha,R} = 0$}\Comment{unconnected substructure}
		
		\EndIf
		
		\For{$d \gets i \ldots j-1$} \Comment{loop over all possible $3'$-most pairs $(d,e)$}
		
		\For{$e \gets d+1 \ldots j$}
		
		\If{$\eta[e+\frac{1}{2}, j-\frac{1}{2}] == 0$} 
		
		\If{$\eta[d-\frac{1}{2}] == 0$ \textbf{or} $d==i$}
		
		\State $Q_{i,j}^{\alpha,R} +=  Q_{i,d-1}^{\alpha,R}~ Q^{\text{bp},\alpha,R}_{d,e}$
		\EndIf
		
		\If{$\eta[i+\frac{1}{2}, d-\frac{1}{2}] == 0$}
		\Comment{single base pair}
		\State $Q_{i,j}^{m,\alpha,R} += 
		\exp \big\{- \alpha R  [\Delta G^\mathrm{multi}_\mathrm{bp} + (d-i + j-e) \Delta G^\mathrm{multi}_\text{nt}] /k_\mathrm{B}T \big\} Q^{\text{bp},\alpha,R}_{d,e} $

		\EndIf
		\If{$\eta[d-\frac{1}{2}] == 0$}
		\Comment{more than one base pair}
		\State $Q^{m,\alpha,R}_{i,j} +=	Q^{m,\alpha,R}_{i,d-1}~ Q^{\text{bp},\alpha,R}_{d,e} \exp \big\{ - \alpha R [\Delta G^\mathrm{multi}_\text{bp} + (j-e) \Delta G^\mathrm{multi}_\text{nt}]/k_\mathrm{B}T \big\} $

		\EndIf
		
		\EndIf
		
		\EndFor
		
		\EndFor
		
		\EndFor
		\EndFor 
		
	\end{algorithmic}
\end{algorithm}

\master*
\begin{proof}
	Let's start with \cref{eq:master1} for the case when  $R >2$. It's a double summation of two terms:
	\begin{itemize}
		\item $t_1 = \mathds{1}_{x,i} ~ Q^{\text{bp},\alpha,R}_{i,j} ~ 
		e^{-\alpha \big[\Delta G^\mathrm{multi}_\mathrm{init} + R \cdot \Delta G^\mathrm{multi}_\mathrm{bp} + R \cdot (i-x)\Delta G^\mathrm{multi}_{\mathrm{nt}} +  \beta \big]/k_\mathrm{B}T}$, that captures the case when there is exactly one nest will be formed in the template slice $\templateslice_R^b$. 
		
		\item $t_2 =Q^{m,\alpha,R}_{x,i-1} ~ \mathds{1}_{i} ~  Q^{\text{bp},\alpha,R}_{i,j} \hspace{0.1cm} e^{- \alpha \big[\Delta G^\mathrm{multi}_\mathrm{init} + R \cdot \Delta G^\mathrm{multi}_\mathrm{bp} + \beta \big]/k_\mathrm{B}T}$, that captures the only other case when there is more than one nest will be formed in $\templateslice_R^b$. 
	\end{itemize}
	
	Let's simplify the first term $t_1$ first, see \cref{fig:master} (a). By substituting $Q^{\text{bp},\alpha,R}_{i,j}$ using \cref{eq:spf2}, and reordering: 
	\begin{align}\label{eq:t1}
		t_1
		&= 
		e^{-\alpha \big[ \Delta G^\mathrm{multi}_\mathrm{init} + R \cdot \Delta G^\mathrm{multi}_\mathrm{bp} +  R \cdot (i-x)\Delta G^\mathrm{multi}_{\mathrm{nt}} + \beta \big]/k_\mathrm{B}T} 
		\hspace{0.1cm}
		\mathds{1}_{x,i} 
		\sum_{r \in \Omega_{i,j}^{\text{bp}}} e^{- \alpha R  \Delta G(r)/k_\mathrm{B}T} 
	\end{align} 
	If there is any nick in the interval $[x,i]$, where $x = \mathcal{B}_1(\templateslice_R^b)$, then the substructure that will be within the template slice $\templateslice_R^b$ is disconnected, and hence neglected in the partition function calculation by multiplication by $\mathds{1}_{x,i} = 0$. On the other hand, if $\mathds{1}_{x,i} = 1$, we end up concatenating the empty connected substructure over the interval $[x,i]$, with any connected substructure $r \in \Omega_{i,j}^{\text{bp}}$. This leads to form a connected slice. After repeating that formed slice $R$ times, joining each to the next by a covalent bond where our $R$-symmetric cut was initiated, we form an $R$-fold rotationally symmetric secondary structure $S$.

	Since the base pair $(i,j)$, where $j =\mathcal{B}_2(\templateslice_R^b)$, is chosen such that it crosses the second border of the prime template's slice, hence it  satisfies the requirement of $\mathcal{P}$ in our main criterion, see \cref{eq:criterion}. So, $S \in \mathcal{P}(\templateslice_R^b,(i,j), \# n = 1)$, see \cref{note:numnest}. 
	Now we can replace the dummy variable $r$, in \cref{eq:t1}, with the actual slice $\slice^{S,b}_R$ of $S$ since they are equivalent.
	\begin{align} \label{eq:t12}
		t_1 &= 
		e^{-\alpha \big[ \Delta G^\mathrm{multi}_\mathrm{init} + R \cdot \Delta G^\mathrm{multi}_\mathrm{bp} + R \cdot (i-x)\Delta G^\mathrm{multi}_{\mathrm{nt}} + \beta \big]/k_\mathrm{B}T} 
		\sum_{S \in \mathcal{P}(\templateslice_R^b,(i,j), \# n = 1)}  \hspace{-2.5em} e^{- \alpha R  \Delta G(\slice^{S,b}_R)/k_\mathrm{B}T} \nonumber \\
		&= 
		\sum_{S \in \mathcal{P}(\templateslice_R^b,(i,j), \# n = 1)} \hspace{-2.5em} 
		e^{-\alpha \Big[ R \Delta G(\slice^{S,b}_R) +\Delta G^\mathrm{multi}_\mathrm{init} + R \big[ \Delta G^\mathrm{multi}_\mathrm{bp} + (i-x)\Delta G^\mathrm{multi}_{\mathrm{nt}}\big] + \beta\Big]\big/ k_\mathrm{B}T}	 
	\end{align} 
	
	Since $R>2$, \cref{rem:central} shows that the central loop $\mathcal{C}(S)$ must be a multiloop. Since there is exactly one nest per slice $\slice^{S,b}_R$, therefore one exterior base pair per slice that borders the interior of $\mathcal{C}(S)$. Each slice contributes to the free energy of $\mathcal{C}(S)$ with exactly $\Delta G^\mathrm{multi}_\mathrm{bp} + (i-x)\Delta G^\mathrm{multi}_{\mathrm{nt}}$. The free energy of the multiloop $\Delta G(\mathcal{C}(S)) = \Delta G^\mathrm{multi}_\mathrm{init} + R \big[ \Delta G^\mathrm{multi}_\mathrm{bp} + (i-x)\Delta G^\mathrm{multi}_{\mathrm{nt}}\big]$. Substitute $\Delta G(\mathcal{C}(S))$ and $\beta$ in \cref{eq:t12}, we get the following:   
	\begin{align}\label{eq:t13}
		t_1 &= 
		\sum_{S \in \mathcal{P}(\templateslice_R^b,(i,j), \# n = 1)} 
		e^{-\alpha \big[ R  \Delta G(\slice^{S,b}_R) +\Delta G(\mathcal{C}(S)) + (c-1)\Delta G^\mathrm{assoc} \big]/k_\mathrm{B}T} \nonumber \\
		&=  
		\sum_{S \in \mathcal{P}(\templateslice_R^b,(i,j), \# n = 1)} 
		e^{-\alpha \DeltaGD(S) /k_\mathrm{B}T}, &\text{using \cref{eq:NNcentral2}}
	\end{align}
	
	With respect to the second term $t_2$, that captures the case when there is more than one nest will be formed in $\templateslice_R^b$,  see \cref{fig:master} (a).  By substituting $Q^{m,\alpha,R}_{i,j}$ and $Q^{\text{bp},\alpha,R}_{i,j}$  using \cref{eq:spf3,eq:spf2}, and reordering we get the following: 
	
	\begin{align}\label{eq:t2}
		t_2 &=
		\mathds{1}_{i} 
		\Bigg[\sum_{r \in \Omega_{x,i-1}} e^{- \alpha R \big[\Delta G(r) ~+~ \# \text{bp}(r) \cdot \Delta G^\mathrm{multi}_\mathrm{bp} ~+~ \# \text{nt}(r) \cdot \Delta G^\mathrm{multi}_\mathrm{nt} \big] /k_B T} \Bigg]
		\Bigg[\sum_{r' \in \Omega_{i,j}^{\text{bp}}}  e^{- \alpha R  \Delta G(r')/k_\mathrm{B}T} \Bigg] \nonumber \\ 
		&  \quad \cdot  e^{- \alpha \big[\Delta G^\mathrm{multi}_\mathrm{init} ~+~ R \cdot \Delta G^\mathrm{multi}_\mathrm{bp} ~+~ \beta  \big]/k_\mathrm{B}T} \hspace{0.1cm}
	\end{align}
	
	If $\mathds{1}_{i} = 0$, then there is a nick between bases $i-1$ and $i$. therefore the substructure that will be formed within the template slice $\templateslice_R^b$ is disconnected, and hence neglected in the partition function calculation using multiplication by $\mathds{1}_{i} = 0$. On the other hand, if $\mathds{1}_{i} = 1$, we end up concatenating two connected substructures $r \in \Omega_{x,i-1}$ and $r' \in \Omega_{i,j}^{\text{bp}}$ using the covalent bond $[i-1,i]$. Concatenating $r$ and $r'$ results in a new connected substructure $t \in \Omega_{x,j}$ such that $\Delta G(t) = \Delta G(r) + \Delta G(r')$. Since $R>2$, $t$ contributes to the central multiloop with $\# \text{nt}(t) = \# \text{nt}(r)$ free bases\footnote{Note that,  $r' \in \Omega_{i,j}^{\text{bp}}$ contributes with zero free bases to the central multiloop, since $r'$ is just one  nest with exterior base pair $(i,j)$.}, and  $\# \text{bp}(t) = \# \text{bp}(r) + 1$ base pairs that border the interior of the central multiloop. Note that the extra base pair comes from the nest with exterior base pair $(i,j)$ from substructure $r'$. Reflecting these changes in \cref{eq:t2}, we get the following:  
	\begin{align}\label{eq:t22}
		t_2
		& = 
		\sum_{r \in \Omega_{x,i-1}} \sum_{r' \in \Omega_{i,j}^{\text{bp}}}  e^{- \alpha \Big[R \big(  \Delta G(r) + \Delta G(r') \big)+ \Delta G^\mathrm{multi}_\mathrm{init} + R\big( \# \text{bp}(r) + 1 \big) \Delta G^\mathrm{multi}_\mathrm{bp} + R\cdot\# \text{nt}(r)  \Delta G^\mathrm{multi}_\mathrm{nt}  + \beta \Big] \big/k_B T} \nonumber \\
		&= \sum_{t \in \Omega_{x,j}} e^{- \alpha \Big[R \cdot \Delta G(t) + \Delta G^\mathrm{multi}_\mathrm{init} +  R \cdot\# \text{bp}(t) \cdot \Delta G^\mathrm{multi}_\mathrm{bp} + R \cdot \# \text{nt}(t) \cdot \Delta G^\mathrm{multi}_\mathrm{nt}  + \beta \Big] \big/ k_B T}
	\end{align}
	Since $t \in \Omega_{x,j}$ is a connected substructure over the interval $[x,j]$, where $x = \mathcal{B}_1(\templateslice_R^b)$,  $j =\mathcal{B}_2(\templateslice_R^b)$, in other words the template slice $\templateslice_R^B$. Therefore $t$ is a connected slice. After repeating that formed slice $R$ times, joining each to the next by a covalent bond where our cut was done, we form an $R$-fold rotationally symmetric secondary structure $S$. And since the base pair $(i,j)$ is chosen such that it crosses the second border of the prime template's slice, so it  satisfies the requirement of $\mathcal{P}$ in our main criterion, see \cref{eq:criterion}. So, $S \in \mathcal{P}(\templateslice_R^b,(i,j), \# n > 1)$, see \cref{note:numnest}. 
	We then can replace the dummy variable $t$, in \cref{eq:t22}, with the actual slice $\slice^{S,b}_R$ of $S$ since they are equivalent.
	\begin{align*}
		t_2
		&= \sum_{S \in \mathcal{P}(\templateslice_R^b,(i,j), \# n > 1)} \hspace{-2.2em} e^{- \alpha \Big[R \cdot \Delta G(\slice^{S,b}_R) ~+~ \Delta G^\mathrm{multi}_\mathrm{init} ~+~ R \cdot  \# \text{bp}(\slice^{S,b}_R) \cdot \Delta G^\mathrm{multi}_\mathrm{bp} ~+~ R \cdot \# \text{nt}(\slice^{S,b}_R) \cdot \Delta G^\mathrm{multi}_\mathrm{nt}  ~+~ \beta \Big] \big/ k_B T}
	\end{align*}

	Since $S$ has a central multiloop $\mathcal{C}(S)$, the free energy of $\mathcal{C}(S)$ is as follows: $\Delta G(\mathcal{C}(S)) = \Delta G^\mathrm{multi}_\mathrm{init} + R \big[ \# \text{bp}(\slice^{S,b}_R) \cdot \Delta G^\mathrm{multi}_\mathrm{bp} + \# \text{nt}(\slice^{S,b}_R) \cdot \Delta G^\mathrm{multi}_\mathrm{nt} \big]$. Substitute $\Delta G(\mathcal{C}(S))$ and $\beta$ in \cref{eq:t22}, we obtain the following:
	
	\begin{align}\label{eq:t24}
		t_2
		&= \sum_{S \in \mathcal{P}(\templateslice_R^b,(i,j), \# n > 1)} e^{- \alpha \Big[R \Delta G(\slice^{S,b}_R) + \Delta G(\mathcal{C}(S))  + (c-1)\Delta G^\mathrm{assoc} \Big] /k_B T}
		\nonumber \\
		&=  
		\sum_{S \in \mathcal{P}(\templateslice_R^b,(i,j), \# n > 1)} 
		e^{-\alpha \DeltaGD(S) /k_\mathrm{B}T}, & \text{using \cref{eq:NNcentral2}}
	\end{align}
	Now, by substituting $t_1$  and $t_2$ in \cref{eq:master1}, we get:
	\begin{align*}
		\overline{Z}^{\alpha,R}_{\mathcal{S}} &=  \sum_{\templateslice_R^b \in \mathcal{T}_R} \hspace{0.05cm} \sum_{(i,j)}  \Bigg[
		\sum_{S \in \mathcal{P}(\templateslice_R^{b},(i,j), \# n = 1)} e^{- \alpha  \DeltaGD(S)/k_\mathrm{B}T} 
		+
		\sum_{S \in \mathcal{P}(\templateslice_R^{b},(i,j), \# n > 1)} e^{- \alpha  \DeltaGD(S)/k_\mathrm{B}T}  
		\Bigg]
		\\
		& = \sum_{\templateslice_R^b \in \mathcal{T}_R} \hspace{0.05cm}   \sum_{(i,j)} \hspace{0.05cm}
		\sum_{S \in \mathcal{P}(\templateslice_R^{b},(i,j))} e^{- \alpha  \DeltaGD(S)/k_\mathrm{B}T} \quad \quad \quad \quad \quad \quad \quad\quad \quad \quad  \quad \quad \quad \text{using \cref{note:numnest}} \\
		&= \sum_{S \in \OmegaDRpi} e^{- \alpha  \DeltaGD(S)/k_\mathrm{B}T} \quad \quad \quad \quad \quad \quad \quad \text{by  \cref{eq:criterion} in our criterion, see \cref{lem:criterion}}\\
		& = \QalphaD(\pi,R) \quad \hspace{9.6cm} \text{by definition} 
	\end{align*}
	This proves our first case when $R>2$. 
	
	When $R=2$, see \cref{eq:master2}, the proof goes through the same lines except for the first case when there is only one nest per slice. In this case, the central loop $\mathcal{C}(S)$ of the formed secondary structure $S$ is internal loop or stack. $\mathcal{C}(S)$ is determined by its four bases $i,j,i'$, and $j'$, where $i'= i+\frac{N}{2}$ and $j'= (j+\frac{N}{2}) \mod N$. The free energy of $\mathcal{C}(S)$ is $\Delta G (\mathcal{C}(S)) = \Delta G^\mathrm{interior}_{i,j,i',j'}$. In conclusion this shows that $\overline{Z}^{\alpha,2}_{\mathcal{S}} = \sum\limits_{S \in \OmegaDtwopi} e^{- \alpha  \DeltaGD(S)/k_\mathrm{B}T} = \QalphaD(\pi,2)$. 
	
	With respect to time and space complexity, each of \cref{eq:master1,eq:master2} has double summation. The first summation ranges over $|\mathcal{T}_R| = \frac{N-c}{R} = \mathcal{O}(N)$, see \cref{sec:prosym}. And in the second summation, base $i$ of base pair $(i,j)$ ranges\footref{foot:range} as follows: $\mathcal{B}_1 (\templateslice_R^b)\leq i < \mathcal{B}_2(\templateslice_R^f)$, since  $\mathcal{B}_2(\templateslice_R^f) - \mathcal{B}_1 (\templateslice_R^b) < \frac{N}{R} = \mathcal{O}(N)$. Hence we have $\mathcal{O}(N^2)$ iterations, each iteration uses the already pre-computed values from the Assistant Chef \cref{algo:assischef}, and hence each is done in $\mathcal{O}(1)$ time and space. In total our Master Chef \cref{algo:master} runs in $\mathcal{O}(N^2)$ time, and $\mathcal{O}(1)$ space, proving our claim.
\end{proof}

\begin{algorithm}[H] 

	\caption{The master chef algorithm pseudocode (see \cref{lem:master}), that 
		runs in  $\mathcal{O}(N^2)$ time and $\mathcal{O}(1)$ space with
		recursive calls illustrated in \cref{fig:master}. 
		The indicator variables  $\mathds{1}_{a,b}$ and $\mathds{1}_a$ are defined in \cref{eq:indicator1,eq:indicator2} respectively. 
	} \label{algo:master}
	\begin{algorithmic}[1]
		
		\State \textbf{Initialize} $	\overline{Z}^{\alpha,R}_{\mathcal{S}} = 0$
		
		\State \textbf{Set} $ \beta = (c-1) \Delta G^\text{assoc}$ 
		
		\For{$a \gets 1 \ldots \frac{N}{R}-1$}
		\State $b = [a,a+1]$ \Comment{The covalent bond that will be uses as backbone cut generator}
		\State $x = a+1$
		\Comment{$x = \mathcal{B}_1(\templateslice_R^b) = a+1$}
		
		\State $j = a+\frac{N}{R}$
		\Comment{$j = \mathcal{B}_2(\templateslice_R^b) = a+\frac{N}{R}$} 
		
		\Statex
		\For{$i \gets x \ldots  1+\frac{N}{R}$}
		\Comment{$\mathcal{B}_1(\templateslice_R^b)\leq i < \mathcal{B}_2(\templateslice_R^f)$}
		
		\Statex
		\If{$\mathds{1}_{x,i} == 1$}
		\Comment{Case 1: Only one nest per slice} 	
		
		\If{$R == 2$}
		\State $ \begin{aligned}[t]
			\overline{Z}^{\alpha,R}_{\mathcal{S}} +=  Q^{\text{bp},\alpha,R}_{i,j} ~
			\exp\Big\{-\alpha \big[  \Delta G^\mathrm{interior}_{i,j,i',j'} ~+~ \beta \big]/k_\mathrm{B}T\Big\}
		\end{aligned}$
		
		\EndIf 
		
		\Statex
		\If{$R > 2$}
		\State $ \begin{aligned}[t]
			\overline{Z}^{\alpha,R}_{\mathcal{S}} +=  Q^{\text{bp},\alpha,R}_{i,j}
			\exp\Big\{-\alpha \big[ \Delta G^\mathrm{multi}_\mathrm{init} & + R \cdot \Delta G^\mathrm{multi}_\mathrm{bp} \\ 
			&+ (i-x)\Delta G^\mathrm{multi}_{\mathrm{nt}} + \beta \big]/k_\mathrm{B}T\Big\}
		\end{aligned}$
		
		\EndIf 
		\EndIf 
		
		\Statex
		
		\If{$\mathds{1}_i == 1$}
		\Comment{Case 2: More than one nest per slice}

		\State $\overline{Z}^{\alpha,R}_{\mathcal{S}} +=  Q^{m,\alpha,R}_{x,i-1} \hspace{0.1cm} Q^{\text{bp},\alpha,R}_{i,j} \hspace{0.1cm} \exp \Big\{- \alpha \big[\Delta G^\mathrm{multi}_\mathrm{init} + R \cdot \Delta G^\mathrm{multi}_\mathrm{bp} + \beta \big]/k_\mathrm{B}T\Big\}$
		\EndIf 
		
		\EndFor
		\EndFor
		
		\State \textbf{Return} $\overline{Z}^{\alpha,R}_{\mathcal{S}}$   
	\end{algorithmic}
\end{algorithm}

\subsection{Proofs and algorithms omitted from \cref{sec:Applications}}
\label{app:omitted:app}  

\SSEL*

\begin{proof}
	For any strand ordering $\pi \in \Pi$, 	
	let $\mathcal{G}_\pi = \{\Delta G(S)|~ S\in \Omega(\pi)\} = \{g_1 < g_2 < \ldots < g_{|\mathcal{G}_\pi|}\}$ denote the set of \textbf{occupied} energy levels of system $\mathcal{S}$. Let  $\mathrm{\#SSEL}(g_i, \pi)$ denote the number of secondary structures $S \in \Omega(\pi)$ such that $\Delta G(S) = g_i$. Note that for any energy level $g_i \in \mathcal{G}$, $\mathrm{\#SSEL}(g_i,\pi) > 0$, and $g_1 = \mathrm{MFE}(\pi)$.
	In \cref{algo:onion}, we claim the following about the returned arrays $\mathcal{L}$ and $\mathcal{N}$:
	\begin{itemize}
		\item For every $1 \leq i \leq |\mathcal{G}_\pi|$,  $\mathcal{L}[i] = g_i$ and $\mathcal{N}[i] = \mathrm{\#SSEL}(g_i,\pi)$.
	\end{itemize}  
	
	\noindent We prove this claim by mathematical induction on arrays index $i$.

	In \cref{algo:onion}, note that $\mathcal{R}$ is a loop variant that get updated at the end of every loop iteration. Let $\mathcal{R}_j$ denote the value of $\mathcal{R}$ after the $j^\text{th}$ iteration, the value of $\mathcal{R}$ is updated according to the following rule: $\mathcal{R}_j = \mathcal{R}_{j-1} -  \mathcal{N}[j] \cdot e^{-\alpha\mathcal{L}[j]/k_\mathrm{B} T}$, see line \ref{algo:onion:update}. Note that $\mathcal{R}_j$ is used during iteration $j+1$, and the initial value of $\mathcal{R}$ is $\mathcal{R}_0 = Q_\alpha(\pi)$, see line \ref{algo:onion:initial}. 
	
	\begin{itemize}
		\item \textbf{Base case:}  When $i = 1$ (just after the loop first iteration), since $\mathcal{R}_0 = Q_\alpha(\pi)$, \cref{thm:MFE} guarantees that $\mathcal{L}[1] = \mathrm{MFE}(\pi) = g_1$, and \cref{th:SSELMFEalgo} guarantees that $\mathcal{N}[1] = \mathrm{\#SSEL}(\mathrm{MFE}(\pi),\pi) = \mathrm{\#SSEL}(g_1)$, proving the base case. And $\mathcal{R}_0$ gets updated to $\mathcal{R}_1 = Q_\alpha(\pi) - \mathcal{N}[i] \cdot e^{- \alpha \mathcal{L}[i]/k_\mathrm{B} T}$ after the end of the first iteration.
		
		\item \textbf{Induction Hypothesis:} For all $1 \leq i < k$, assume $\mathcal{L}[i] = g_i$, and $\mathcal{N}[i] = \mathrm{\#SSEL}(g_i,\pi)$.
		
		\item \textbf{Induction step:}  
		When $i = k$ (just after the loop $k^\text{th}$ iteration), in \cref{algo:onion}, lines \ref{algo:onion:level} and \ref{algo:onion:number}, $\mathcal{L}[k] = \big\lceil -k_\mathrm{B} T ~\log \mathcal{R}_{k-1} \big/ (\alpha\cdot \varepsilon) \big\rceil \cdot \varepsilon$, and $\mathcal{N}[k] = \big\lfloor \mathcal{R}_{k-1} \big/ \mathrm{e}^{-\alpha \cdot \mathcal{L}[k]/ k_\mathrm{B} T}   \big\rfloor$. Applying the induction hypothesis on $\mathcal{R}_{k-1}$:
		\begin{align*}
			\mathcal{R}_{k-1} &  = \mathcal{R}_{k-2} -  \mathcal{N}[k-1] \cdot e^{-\alpha \cdot \mathcal{L}[k-1]/k_\mathrm{B} T}  \\ 
			&  = \mathcal{R}_{k-2} -  \mathrm{\#SSEL}(g_{k-1},\pi) \cdot e^{-\alpha \cdot g_{k-1}/k_\mathrm{B} T}  \\
			& = Q_\alpha(\pi) - \sum_{1 \leq i \leq k-1} \mathrm{\#SSEL}(g_{i},\pi) \cdot e^{-\alpha \cdot g_{i}/k_\mathrm{B} T} \\
			&= \sum_{1 \leq i \leq |\mathcal{G}_\pi|} \mathrm{\#SSEL}(g_i,\pi) \cdot e^{-\alpha \cdot g_i/k_\mathrm{B} T} -  \sum_{1 \leq i \leq k-1} \mathrm{\#SSEL}(g_{i},\pi) \cdot e^{-\alpha \cdot g_{i}/k_\mathrm{B} T} \\
			& =  \sum_{k \leq i \leq |\mathcal{G}|} \mathrm{\#SSEL}(g_i,\pi) \cdot e^{-\alpha \cdot g_i/k_\mathrm{B} T} \\
			& = \sum_{S \in B } e^{-\alpha \cdot \Delta G(S)/k_\text{B}T} \hspace{6em} \text{ such that } B = \{S \in \Omega(\pi)| ~\Delta G(S) \geq g_{k} \}
		\end{align*}

		In other words, $\mathcal{R}_{k-1}$ is the magnified partition function over $B$. Hence we can think of $\mathcal{R}_{k-1}$ as the magnified partition function of a new system $\mathcal{S}'$, where $B$ is the set of secondary structures under interest of $\mathcal{S}'$. Since $|B| < |\Omega(\pi)| $, the hierarchical dominance \cref{lem:dom}  applies on $\mathcal{S}'$.
		
		Hence, we now can apply the MFE$(\pi)$ algorithm, see \cref{thm:MFE}, on $\mathcal{S}'$, and it guarantees that $\mathcal{L}[k] = g_k$ (The minimum free energy level over $B$). And apply the \#SSEL$(\pi)$ algorithm, see \cref{th:SSELMFEalgo}, also guarantees that $\mathcal{N}[k] = \mathrm{\#SSEL}(g_k,\pi)$, proving the induction step.   
	\end{itemize}  
	
	Hence, \cref{algo:onion} computes the sorted set of occupied levels $\mathcal{G}_\pi$, along with the number of secondary structures per level $\mathrm{\#SSEL}(L,\pi)$ for all energy levels with respect to any strand ordering $\pi \in \Pi$. Since $|\mathcal{G}_\pi| = \mathcal{O}(N)$, see \cref{lem:NNlinear}, the asymptotic time and space complexity of \cref{algo:onion} is $\mathcal{O}(N^3)$ and $\mathcal{O}(N^2)$ respectively, as it involves exactly one call to our magnifiable partition function $Q_\alpha(\pi)$.
	
	Since $\text{\#SSEL}(L) = \sum_{\pi \in \Pi} \text{\#SSEL}(L,\pi)$, where $|\Pi| \leq (c-1)!$. 
	Since, different circular permutations $\pi$ may have overlapping energies, 
	a simple merge step (i.e. takes $\mathcal{O}(N)$ time and space) after each run of \cref{algo:onion} on each $\pi \in \Pi$, 
	we compute the sorted set of occupied levels $\mathcal{G}$, along with the number of secondary structures per level $\mathrm{\#SSEL}(L)$ for all energy levels of any system $\mathcal{S}$. 
	The asymptotic time and space complexity of our algorithm are $\mathcal{O}(N^3(c-1)!)$ and $\mathcal{O}(N^2)$ respectively. 
\end{proof}

\begin{algorithm}[H] 
	\caption{The Onion Peeling Algorithm for the nearest neighbour energy model, that takes as input: $c=\mathcal{O}(1)$ strands with total number of bases (length) $N$, and strand ordering $\pi$.  
		Runs in  $\mathcal{O}(N^3)$ time and $\mathcal{O}(N^2)$ space. Proof of correctness:  \cref{thm:SSEL}.
	} \label{algo:onion}
	\begin{algorithmic}[1]
		\State \textbf{Let $\mathcal{L}$ and  $\mathcal{N}$ be empty arrays}\Comment{Initialize $\mathcal{L}$ to be an empty array}
		
		\State \textbf{Let} $i =  0$ \Comment{Array counter used for our induction}
		
		\State \textbf{Let  $\mathcal{R} = Q_\alpha(\pi)$ for  $\alpha = k_B T N / \varepsilon$}\label{algo:onion:initial}		
		\Comment{Using \cref{algo:PFMA}}

		\Repeat
		
		\State $i = i+1$
		\Statex
		\State $\mathcal{L}[i] = \bigg\lceil - k_\mathrm{B} T ~\log( \mathcal{R}) \big/ (\alpha\cdot \varepsilon) \bigg\rceil \cdot \varepsilon$  \label{algo:onion:level}
		\Comment{See \cref{thm:MFE}}
		\Statex
		
		\State $\mathcal{N}[i] = \bigg\lfloor \mathcal{R} \big/ \mathrm{e}^{-\alpha \cdot \mathcal{L}[i]/ k_\mathrm{B} T}   \bigg\rfloor$
		\label{algo:onion:number}  \Comment{See \cref{th:SSELMFEalgo}}

		\Statex

		\State $\mathcal{R} = \mathcal{R} - \mathcal{N}[i] \cdot e^{- \alpha \mathcal{L}[i]/k_\mathrm{B} T} $\label{algo:onion:update}
		
		\Until{$\mathcal{R} > 0$}

		\State \textbf{Return} $\mathcal{L}, \mathcal{N}$

	\end{algorithmic}
\end{algorithm}

 \end{document}